\documentclass[10pt,a4paper]{article}
\usepackage[dvips]{color}
\usepackage{epsfig}
\usepackage{amsmath}
\usepackage{graphicx}
\usepackage{authblk}
\usepackage{amssymb,amsmath}
\usepackage{amsmath}

\begin{document}
\title{\bf Toy models of Universe with an Effective varying $\Lambda$-Term in Lyra Manifold}
\author{{Martiros Khurshudyan$^{a}$ \thanks{Email:
khurshudyan@yandex.ru}}\\
$^{a}${\small {\em Department of Theoretical Physics, Yerevan State
University, 1 Alex Manookian, 0025, Yerevan, Armenia}}\\}  \maketitle
\begin{abstract}
We are interested by the study of several toy models of the Universe in presence of interacting quintessence DE models. Models are considered in the cosmology with an Effective varying $\Lambda$-Term in Lyra Manifold. The motivation of the phenomenological models discussed in this paper is to obtain corresponding models to describe and understand an accelerated expansion of the Universe for the later stage of evolution. Phenomenology of the models describes by the phenomenological forms of $\Lambda(t)$ ($8 \pi G =c =1$). Concerning to the mathematical hardness we discuss results numerically and graphically. Obtained results give us hope that proposed models can work as good models for old Universe and in good agreement with observational data.
\end{abstract}

\section{\large{Introduction}}
Analysis of the observational data shows that our Universe for later stages of evolution indicates accelerated expansion. This conclusion is based on the observations of high redshift type SNIa supernovae [1]-[3]. The last problem is an interesting and important theoretical problem, and the best solutions of this problem based on an assumptions and carries phenomenological character. According to the data analysis we accept that in the Universe one of the main components is a Dark Energy and its negative pressure (positive energy density) has enough power to work against gravity and provide accelerated expansion of the Universe. To have a balance in Universe the second component known as Dark Matter is considered, which is responsible for the completely other phenomenon known as structure formation. According to different estimations Dark Energy occupies about $73\%$ of the energy of our universe, while dark matter, about $23\%$, and usual baryonic matter occupy about $4\%$. The surveys of clusters of galaxies show that the density of matter is very much less than critical density [4], observations of Cosmic Microwave Background (CMB) anisotropies indicate that the universe is flat and the total energy density is very close to the critical $\Omega_{tot} \simeq 1$ [5]. The simple model for the DE is the cosmological constant with two problems called fine-tuning and coincidence \cite{Weinberg}. These problems have opened ways for alternative models for the Dark energy including a dynamical form of dark energy, as a variable cosmological constant \cite{Sola}-\cite{Shapiro}, k-essence model \cite{Armendariz}-\cite{Armendariz1}, Chaplygin gas models \cite{Martiros}-\cite{Saadat4} to mention a few. In recent times were shown that certain type of interaction between DE and DM also could solve mentioned problems. On the other hand one can modify the left hand side of Einstein equation and obtain theories such as $f(R)$ theory of the gravity \cite{Sotiriou}-\cite{Martiros0}. Modifications of these types provide an origin of a fluid identified with dark energy. The origin of an accelerated expansion contributed from geometry were considered even before proposed modifications. But such theories with different forms of modifications still should pass experimental tests, because they contain ghosts, finite-time future singularities e.t.c, which is the base of other theoretical problems. One of the well studied Dark Energy models is a quintessence model \cite{Ratra}-\cite{Martiros8}, which is a scalar field model described by a field $\phi$ and $V(\phi)$ potential and it is the simplest scalar-field scenario without having theoretical problems such as the appearance of ghosts and Laplacian instabilities. Energy density and pressure of quintessence DE given as
\begin{equation}\label{eq:rhoQ}
\rho_{Q}=\frac{1}{2}\dot{\phi}^{2}+V(\phi),
\end{equation}   
and
\begin{equation}\label{eq:rhoP}
P_{Q}=\frac{1}{2}\dot{\phi}^{2}-V(\phi).
\end{equation}  
We consider the models of the Universe where an effective energy density and pressure assumed to be given as
\begin{equation}\label{eq:rhoeff}
\rho=\rho_{Q}+\rho_{b},
\end{equation}
and 
\begin{equation}\label{eq:Peff}
P=\rho_{Q}+\rho_{b},
\end{equation}
where $\rho_{b}$ and $P_{b}$ are energy density and pressure of a barotropic fluid which will model DM in Universe with $P_{b}=\omega_{b}\rho_{b}$ EoS equation. If we will model the background dynamics of the Universe within many component fluid, then we will have 
\begin{equation}
\rho=\sum_{i}{\rho_{i}},
\end{equation}
and
\begin{equation}
P=\sum_{i}{P_{i}},
\end{equation}
where $i$ represents the number of components. Last assumption is at the heart of the modern theoretical cosmology, and is a starting assumption for all articles in Cosmology. It can work particularly for old large scale Universe, where quantum and nonequilibrium effects are not considered. Would the last assumption work in early young Universe is an open question, because for early Universe with high energy/small scales quantum effects can have unexpected effects and how the situation should be modified is not clear yet. As well as we have other conceptual problems, for instance, we do not know how correctly we can model content of the early Universe (which is also open problem for old Universe). As in this work we will consider interaction between components, we would like to have a short discussion on that topic. Apart mathematical speculation concerning to the interaction between DE and DM, there is a question concerning to the physics of the interaction $Q$. Since there is not any reason in nature preventing or suppressing a nonminimal coupling between dark energy and dark matter, there may exist interactions between the two components. Moreover from observations, no piece of evidence has been so far presented against such interactions. Theoretically, several forms of the interaction discussed in recent years, give us possibility to alleviate the coincidence problem, therefore such approach gives a hope and it is one of the active questions considered in literature. Despite to the efforts the microscopic nature of the interaction is not clear (up to our knowledge) and considered models based on a phenomenological assumptions. The question how the interaction between two components arose is not answered yet. One of the assumptions concerning to the interaction between components, is probably the same origin of DE and DM which were from the very begining. However it is not clear why they should continue their connection, when they operate on different scales and responsible for different processes. Probably the final theory of Quantum Gravity can answer to this question, As the theory is missing and there is a posibility to make an assumption, we think a quantum effect like modified entanglement exist, informing DE and DM that they were the same for very early stages of evolution. Below, we would like to present several forms of interaction $Q$ between DE and DM, which are the results of mathematical speculations. One group of $Q$ with a general form described as
\begin{equation}\label{eq:Q}
Q=3H\sum_{i}{b_{i} \rho_{i}}+\sum_{i}{\gamma_{i}\dot{\rho}_{i}},
\end{equation}
where $b_{i}$ and $\gamma_{i}$ are positive constants. Typical value of them is about $0.01\div 0.03$. In the other group we can include interactions where on the base of classical forms of interactions some modifications are assumed, like $b_{i}$ or $\gamma_{i}$ to be a function of time $t$. In this paper the problem solving strategy and structure is the following we will assume that the form of the potential $V(\phi)$ is given
\begin{equation}\label{eq:potV}
V(\phi)=V_{0}e^{ \left [-\alpha\phi \right ]}.
\end{equation}
We already made several attempts to consider Cosmologies where $G(t)$ and $\Lambda(t)$ are varying functions instead of the constants. Consideration of varying $G$ and $\Lambda$ supposed modification of the field equations. In this work we will consider cosmological models defined in Lyra manifold with an effective varying $\Lambda$ term. Three different forms for the interaction between a barotropic fluid and a quintessence DE are considered. Each form of the interaction suppose an existance of the certain type of the model. In this paper we have 6 different models, because depends on the type of the interaction, we also considered a possibility that $\Lambda$ is a constant. Models ordered as Model 1, Model 2 and Model 3 described by the interaction terms $Q$
\begin{equation}\label{eq:Madel1}
Q=3Hb\rho_{Q}+\gamma (\rho_{b}-\rho_{Q})\frac{\dot{\phi}}{\phi},
\end{equation}
\begin{equation}\label{eq:Madel2}
Q=3Hb\rho+\gamma \dot{\rho},
\end{equation} 
and
\begin{equation}\label{eq:Madel3}
Q=bH^{1-2\gamma}\rho_{b}^{\gamma}\dot{\phi}^{2}.
\end{equation} 
The second group of the models (Model 4, Model 5 and Model 5) will be described by the same three forms of $Q$ with $\Lambda(t)$ given as
\begin{equation}
\Lambda(t)=H^{2}\phi^{-2}+\delta V(\phi),
\end{equation} 
where $\delta$ is a positive constant, $V(\phi)$ is the potential of the field. The first form of the interaction as well as the form of the varying $\Lambda(t)$ is considered by us in GR with varying $G(t)$ and $\Lambda(t)$ in Ref. \cite{Martiros9}. The second form of the interaction can be considered as one of the classical form of the interaction considered in literature over the years. The last form of the interaction were considered in Ref. \cite{Sar}\\\\
This paper is organized as the follows. In section 2 we review the field equations. In section 3 we analyse models corresponding to $\Lambda=const$ case. In section 4 we consider three models with varying $\Lambda(t)$. Finally, in section 5 we give conclusions.
\section{\large{The field equations}}
Field equations \cite{Shchigolev} that govern our model of consideration are 
\begin{equation}\label{eq:Einstein eq}
R_{\mu\nu}-\frac{1}{2}g_{\mu\nu}R-\Lambda g_{\mu \nu}+\frac{3}{2}\phi_{\mu}\phi_{\nu}-\frac{3}{4}g_{\mu \nu}\phi^{\alpha}\phi_{\alpha}=T_{\mu\nu}.
\end{equation}
Considering the content of the Universe to be a perfect fluid, we have
\begin{equation}\label{eq:T}
T_{\mu\nu}=(\rho+P)u_{\mu}u_{\nu}-Pg_{\mu \nu},
\end{equation}
where $u_{\mu}=(1,0,0,0)$ is a 4-velocity of the co-moving
observer, satisfying $u_{\mu}u^{\mu}=1$. Let $\phi_{\mu}$ be a time-like
vector field of displacement,
\begin{equation}
\phi_{\mu}=\left ( \frac{2}{\sqrt{3}}\beta,0,0,0 \right ),
\end{equation}
where $\beta=\beta(t)$ is a function of time alone, and the factor $\frac{2}{\sqrt{3}}$ is substituted in order to simplify the writing of all the following equations.
By using FRW metric for a flat Universe,
\begin{equation}\label{s2}
ds^2=-dt^2+a(t)^2\left(dr^{2}+r^{2}d\Omega^{2}\right),
\end{equation}
field equations can be reduced to the following Friedmann equations,
\begin{equation}\label{eq:f1}
3H^{2}-\beta^{2}=\rho+\Lambda,
\end{equation}
and
\begin{equation}\label{eq:Freidmann2}
2\dot{H}+3H^{2}+\beta^{2}=-P+\Lambda,
\end{equation}
where $H=\frac{\dot{a}}{a}$ is the Hubble parameter, and an overdot
stands for differentiation with respect to cosmic
time $t$, $d\Omega^{2}=d\theta^{2}+\sin^{2}\theta d\phi^{2}$, and $a(t)$
represents the scale factor. The $\theta$ and $\phi$ parameters are
the usual azimuthal and polar angles of spherical coordinates, with
$0\leq\theta\leq\pi$ and $0\leq\phi<2\pi$. The coordinates ($t, r,
\theta, \phi$) are called co-moving coordinates.\\ \\
The continuity equation reads as,
\begin{equation}\label{eq:coneq}
\dot{\rho}+\dot{\Lambda}+2\beta\dot{\beta}+3H(\rho+P+2\beta^{2})=0.
\end{equation}
With an assumption that 
\begin{equation}\label{eq:DEDM}
\dot{\rho}+3H(\rho+P)=0.
\end{equation}
Eq. (\ref{eq:coneq}) will give a link between $\Lambda$ and $\beta$ of the following form
\begin{equation}\label{eq:lbeta}
\dot{\Lambda}+2\beta\dot{\beta}+6H\beta^{2}=0.
\end{equation}
To introduce an interaction between DE and DM Eq. (\ref{eq:DEDM}) we should mathematically split it into two following equations
\begin{equation}\label{eq:inteqm}
\dot{\rho}_{DM}+3H(\rho_{DM}+P_{DM})=Q,
\end{equation}
and
\begin{equation}\label{eq:inteqG}
\dot{\rho}_{DE}+3H(\rho_{DE}+P_{DE})=-Q.
\end{equation}
Cosmological parameters of our interest are EoS parameters of each fluid components $\omega_{i}=P_{i}/\rho_{i}$, EoS parameter of composed fluid
\begin{equation}
\omega_{tot}=\frac{P_{m}+P_{\Lambda} }{\rho_{m}+\rho_{\Lambda}},
\end{equation}
deceleration parameter $q$, which can be writen as
\begin{equation}\label{eq:accchange}
q=\frac{1}{2}(1+3\frac{P}{\rho} ).
\end{equation}
\section{\large{Case of constant $\Lambda$}}
For a complete and full picture we will start our analyse from the models with constant $G$ and $\Lambda$. Without loss of generality we would like to describe equations allowing us to find dynamics of the models. According to the assumption with constant $\Lambda$ Eq.(\ref{eq:coneq}) will be modified
\begin{equation}
\dot{\rho}+2\beta\dot{\beta}+3H(\rho+P+2\beta^{2})=0.
\end{equation}
and with $\dot{\rho}+3H(\rho+P)=0$ we will obtain that
\begin{equation}
\dot{\beta}+3H\beta=0.
\end{equation} 
The last equation can be integrated very easily and the result is the following
\begin{equation}
\beta=\beta_{0}a^{-3},
\end{equation}
where $a(t)$ is the scale factor and $\beta_{0}$ is the integration constant. In our future calculations we use $\beta_{0}=1$ initial condition. Concerning to the form of the field equations, we need only to assume the form of $Q$ and we will obtain the cosmological solutions. Concerning to the mathematical hardness of the problem we will analyse models numerically and investigate graphical behavior of various important cosmological parameters. In the following subsections we consider our models with the particular forms of $Q$ considered in Introduction.  
\subsection{\large{Model 1: $Q=3Hb\rho_{Q}+\gamma (\rho_{b}-\rho_{Q})\frac{\dot{\phi}}{\phi}$}}
\begin{figure}[h!]
 \begin{center}$
 \begin{array}{cccc}
\includegraphics[width=50 mm]{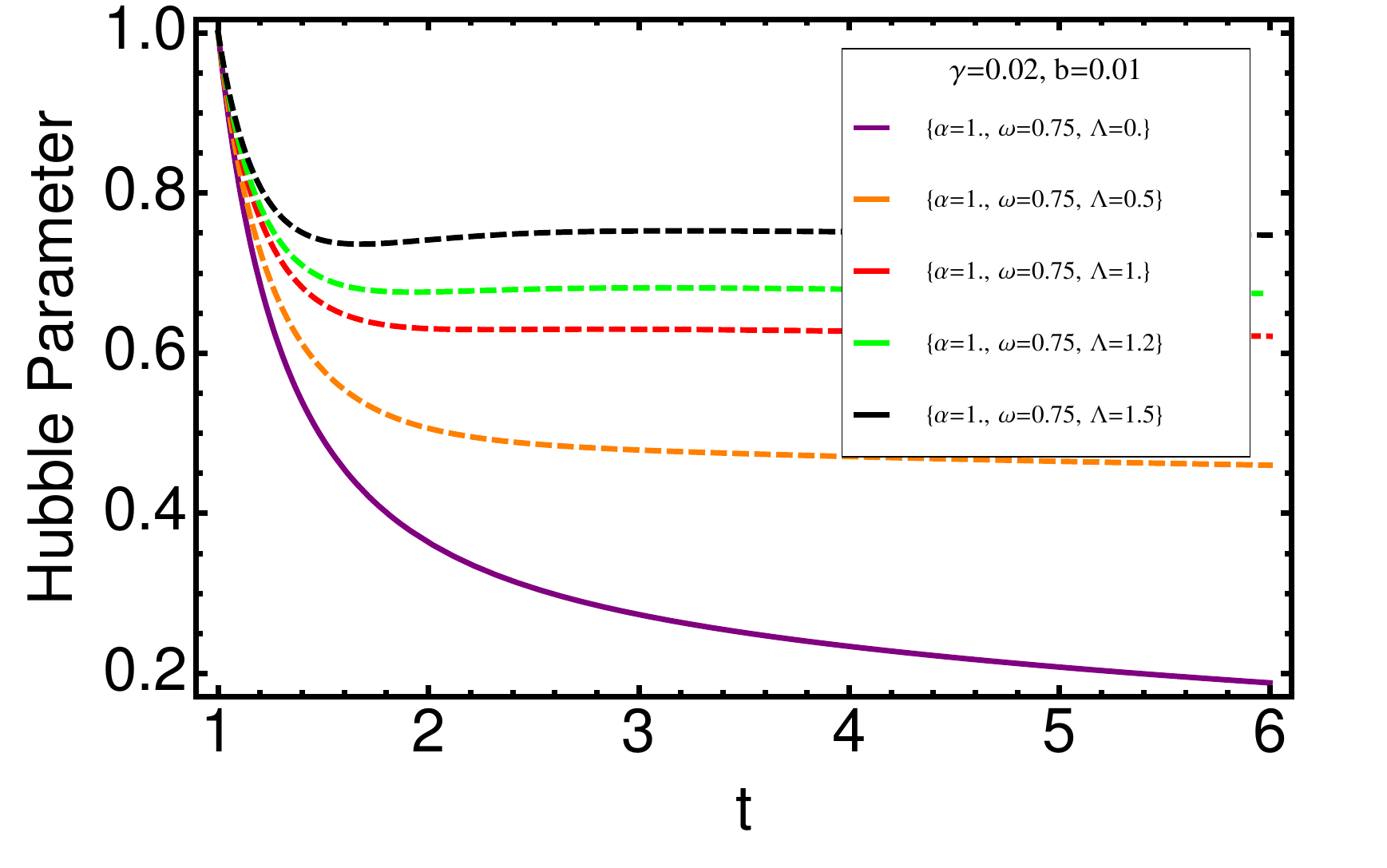} &
\includegraphics[width=50 mm]{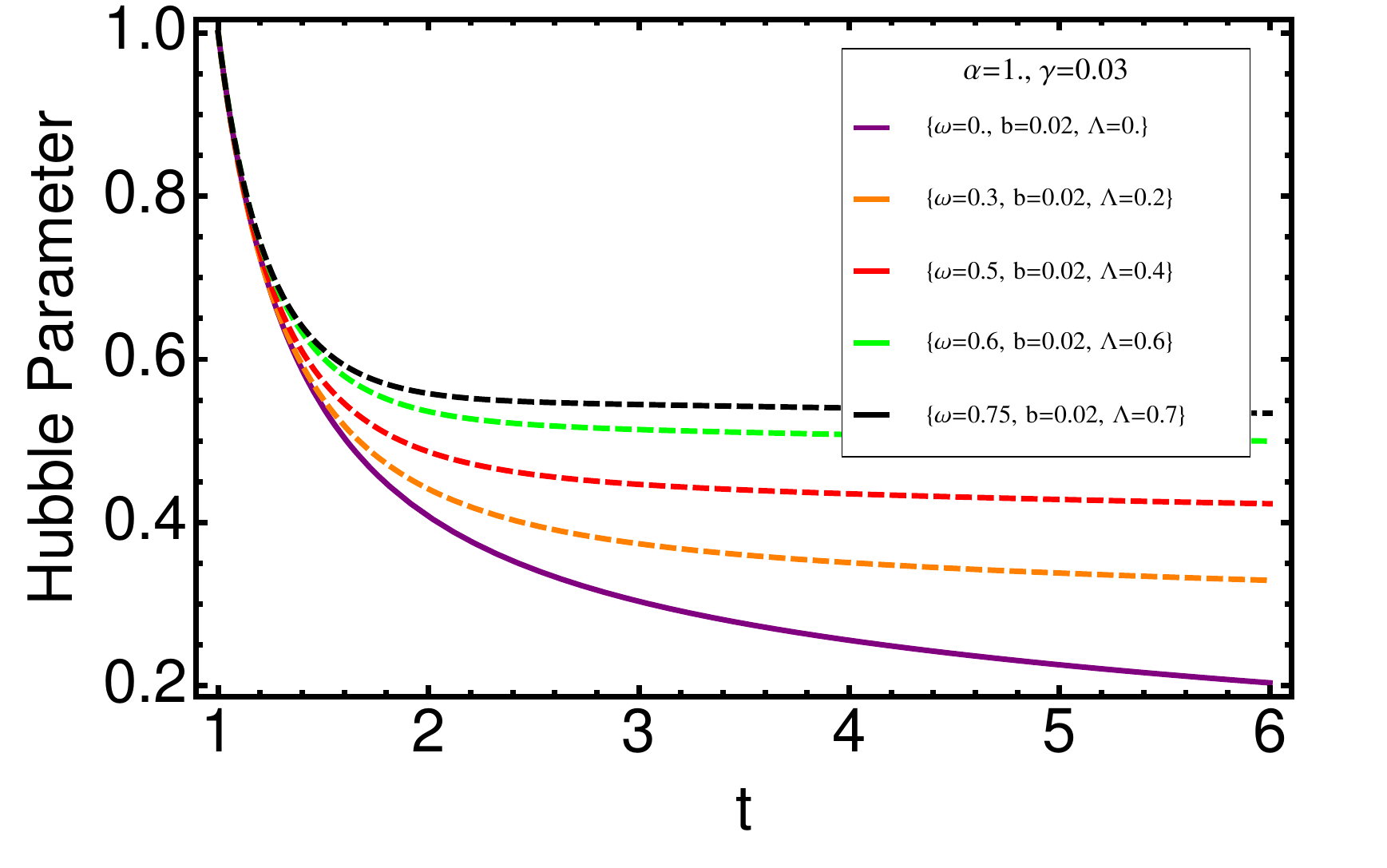}\\
\includegraphics[width=50 mm]{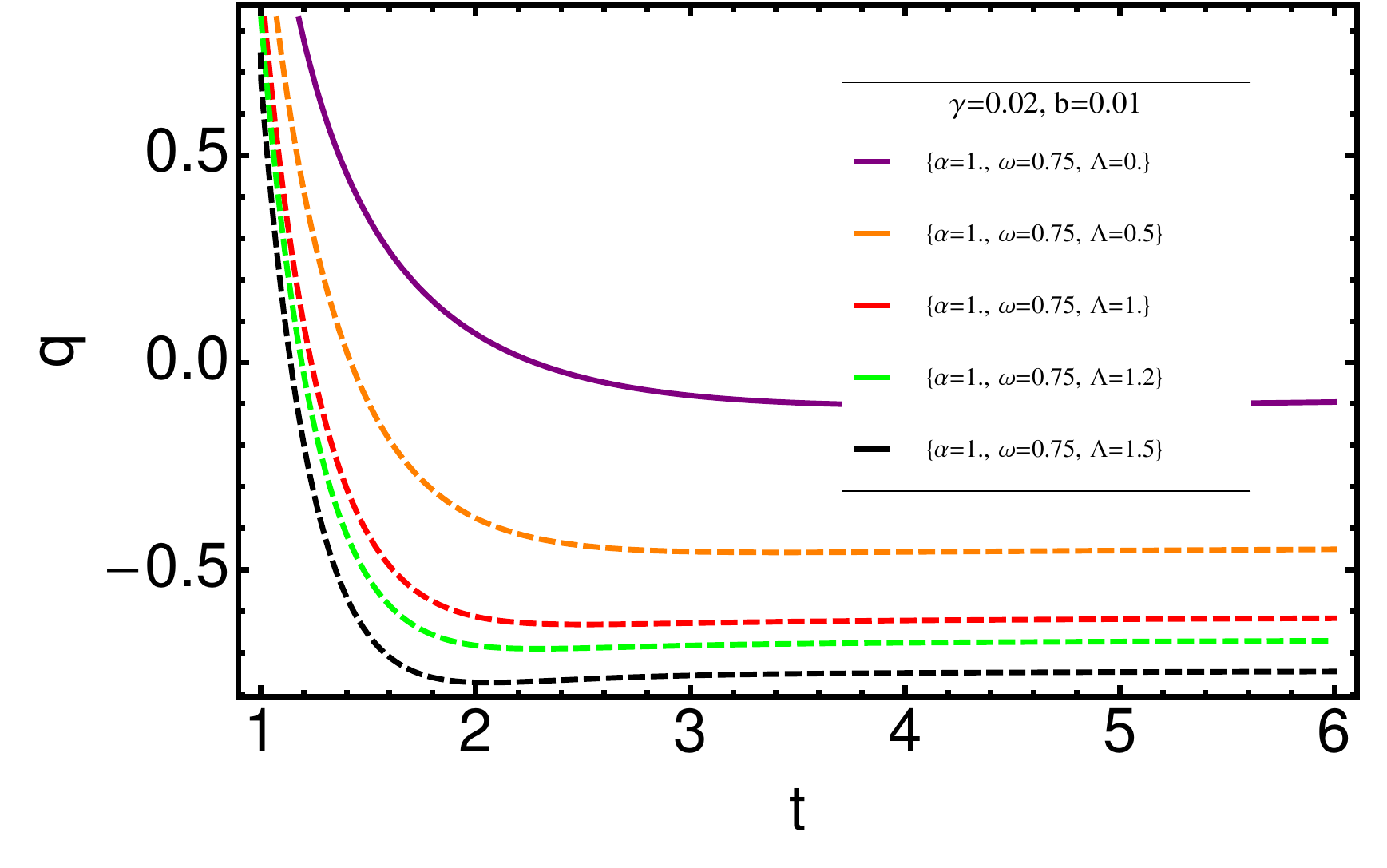} &
\includegraphics[width=50 mm]{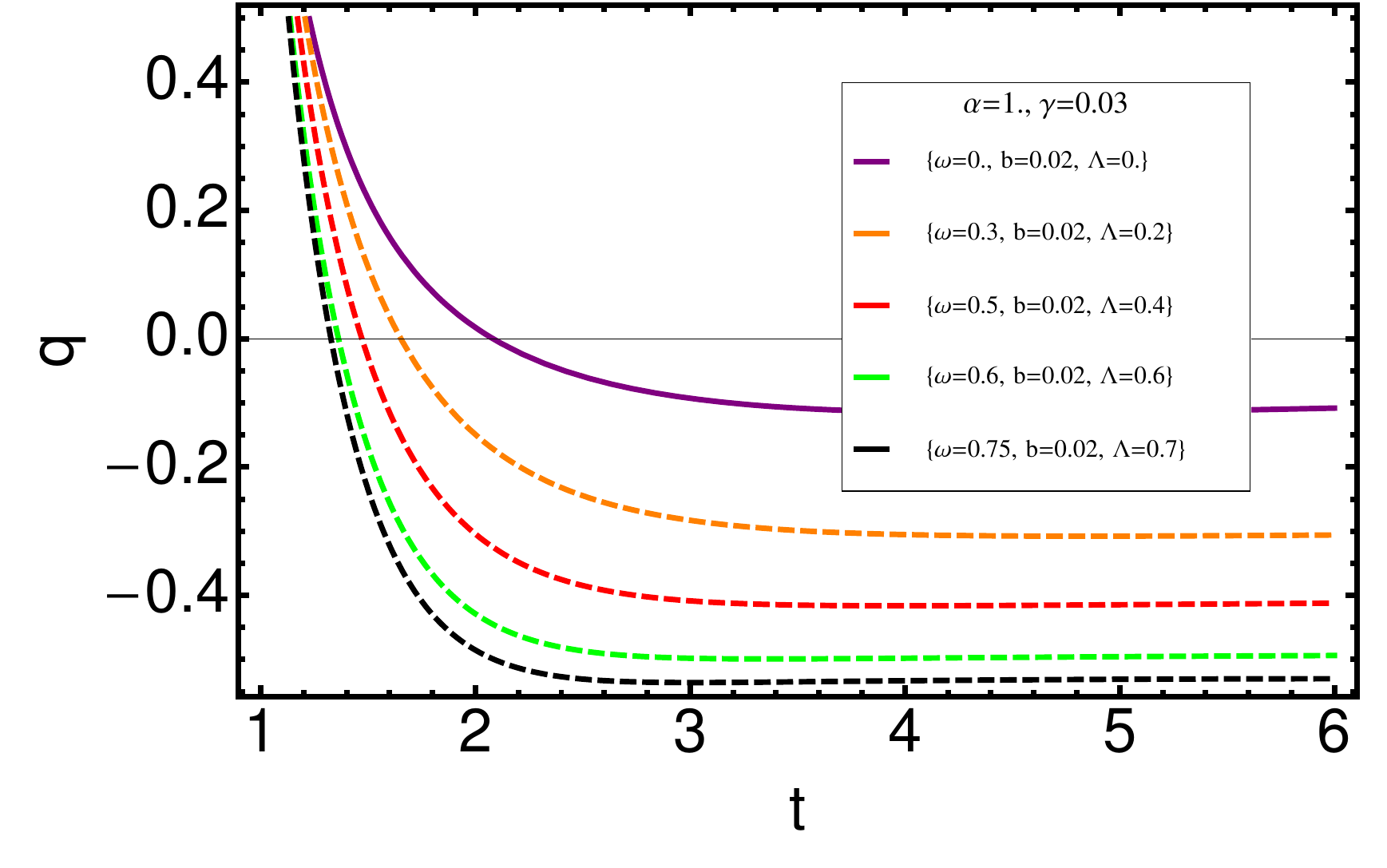}\\
 \end{array}$
 \end{center}
\caption{Behavior of Hubble parameter $H$ and  $q$ against $t$ for the constant $\Lambda$. Model 1}
 \label{fig:const1}
\end{figure}

\begin{figure}[h!]
 \begin{center}$
 \begin{array}{cccc}
\includegraphics[width=50 mm]{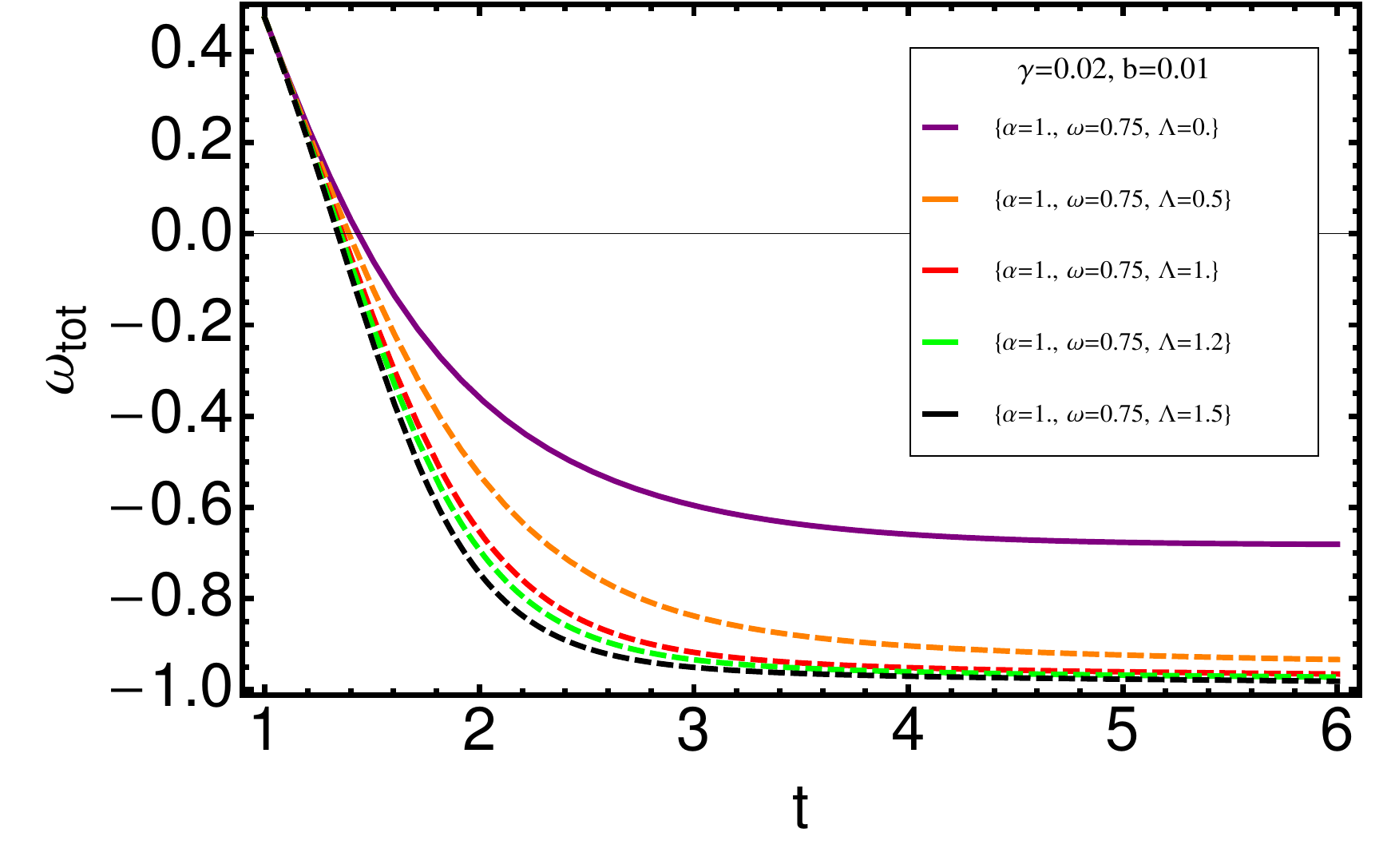} &
\includegraphics[width=50 mm]{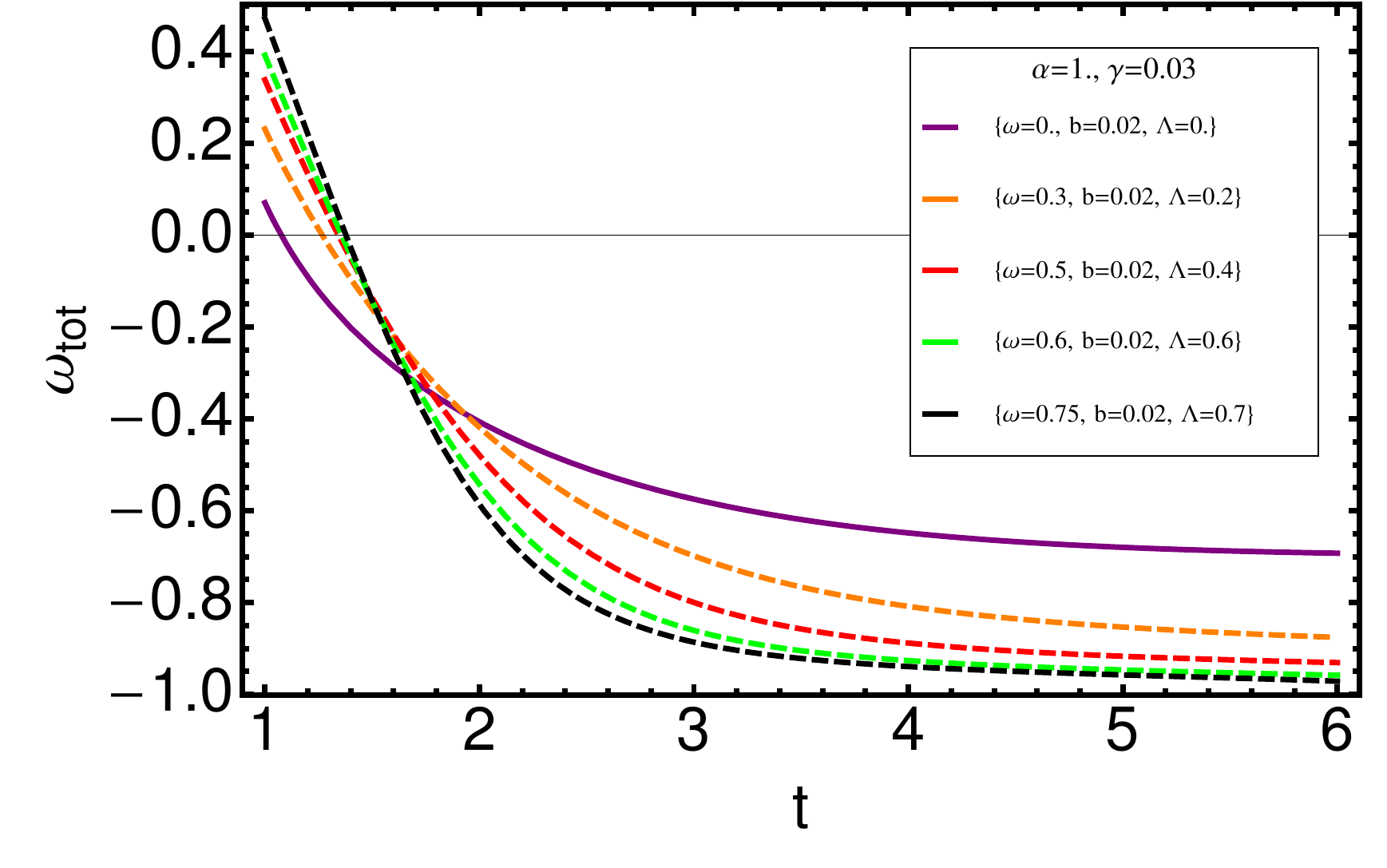}\\
\includegraphics[width=50 mm]{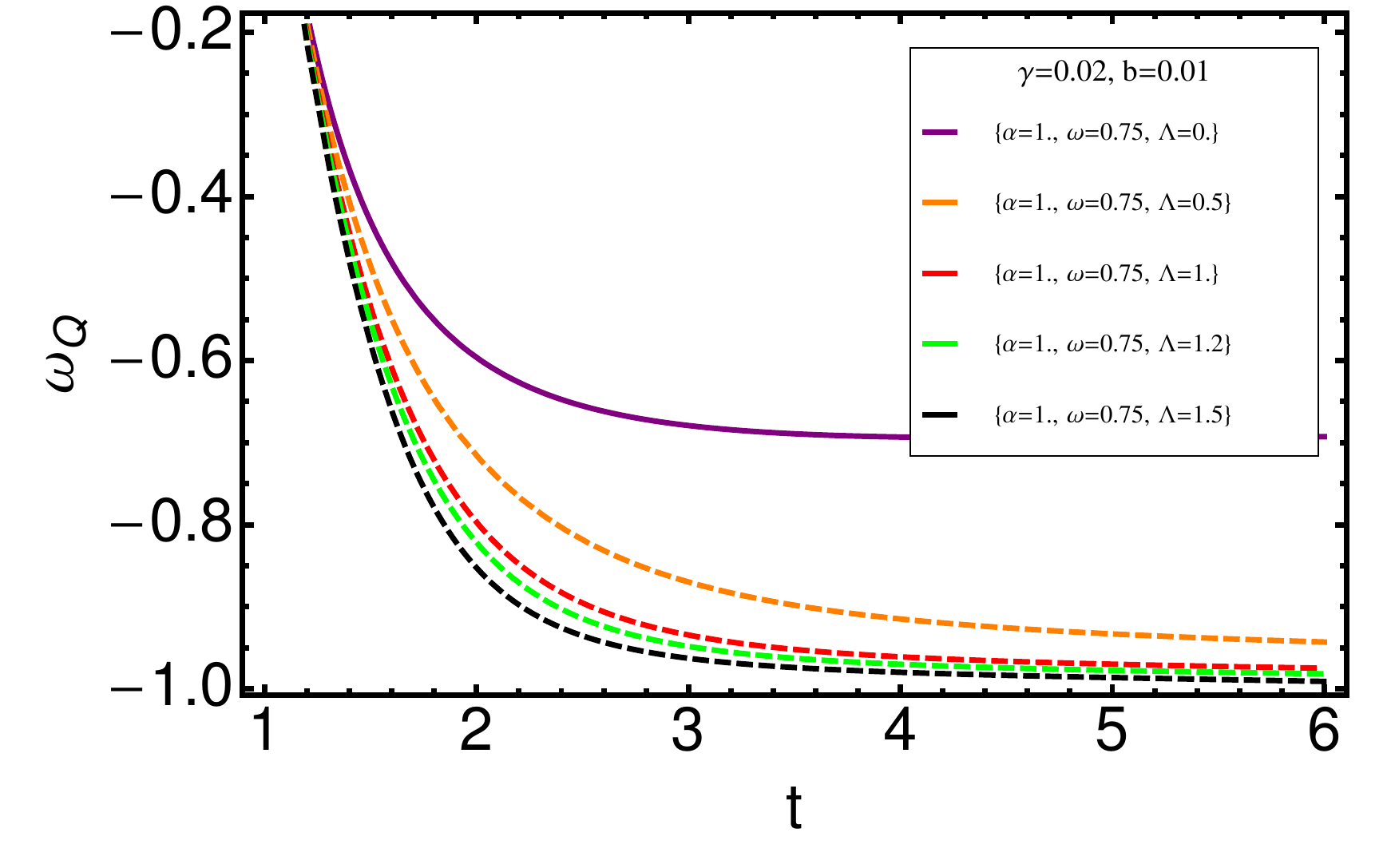} &
\includegraphics[width=50 mm]{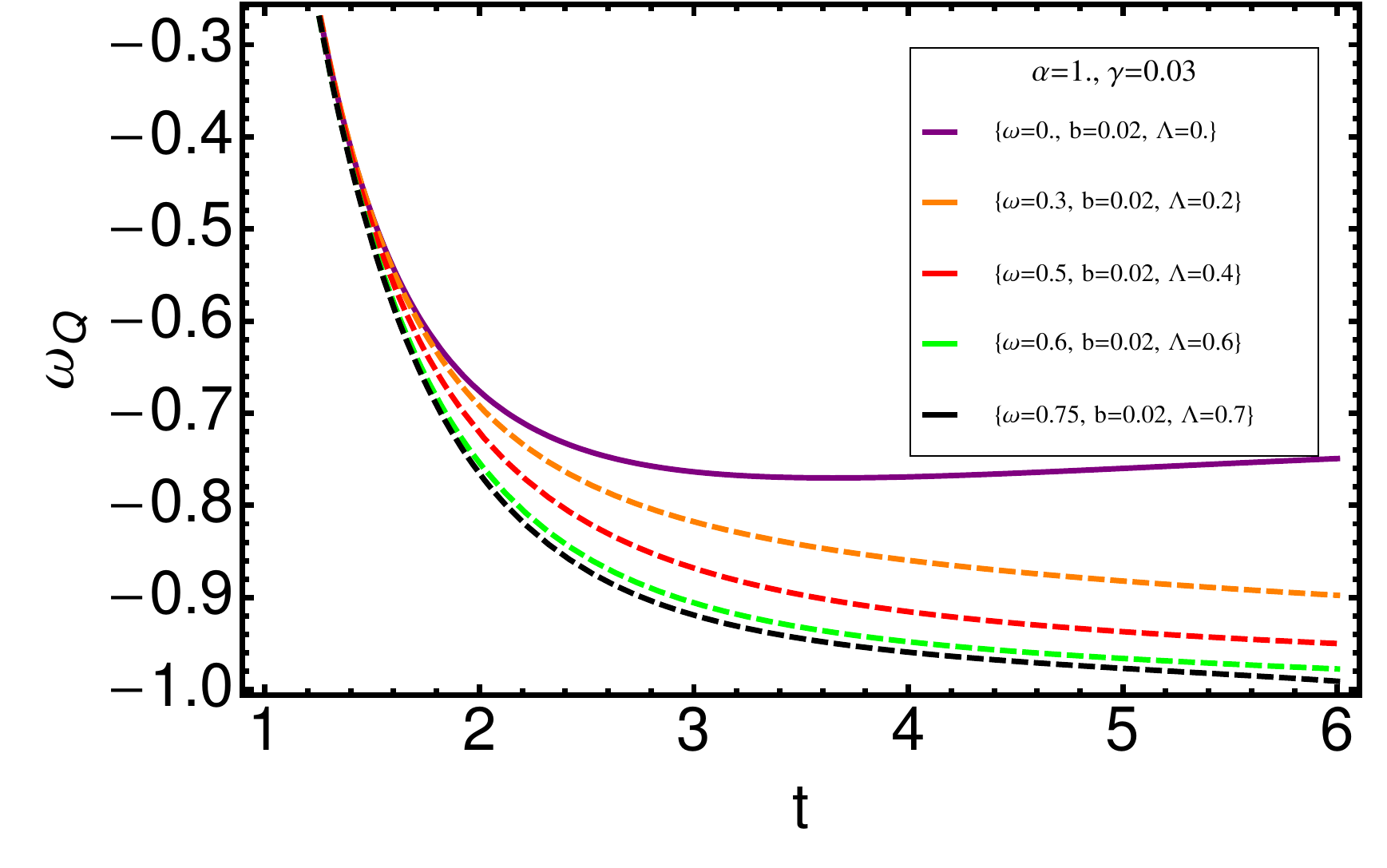}\\
 \end{array}$
 \end{center}
\caption{Behavior of EoS parameter $\omega_{tot}$ and $\omega_{Q}$ against $t$ for the constant $\Lambda$. Model 1}
 \label{fig:const2}
\end{figure}

\begin{figure}[h!]
 \begin{center}$
 \begin{array}{cccc}
\includegraphics[width=50 mm]{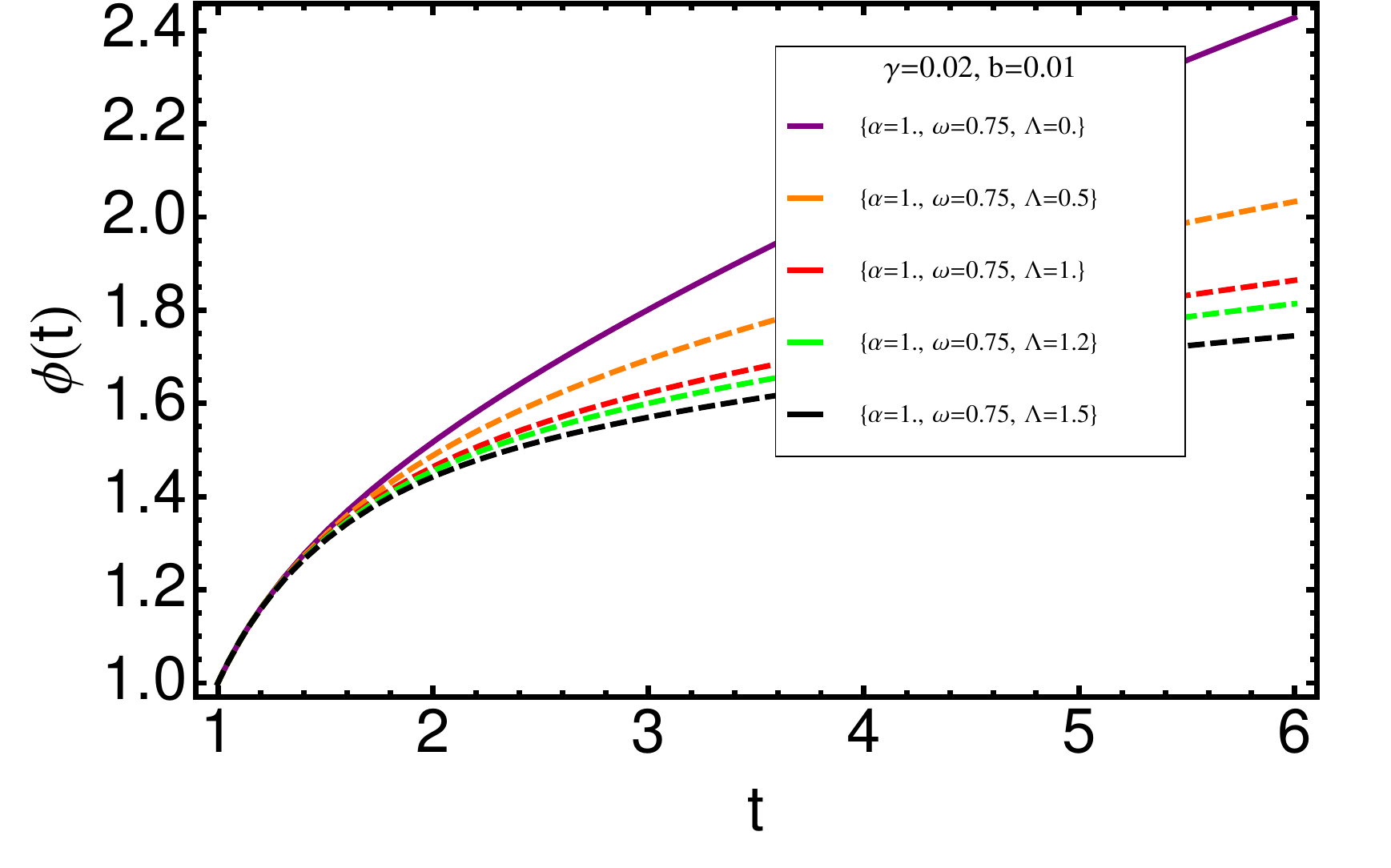} &
\includegraphics[width=50 mm]{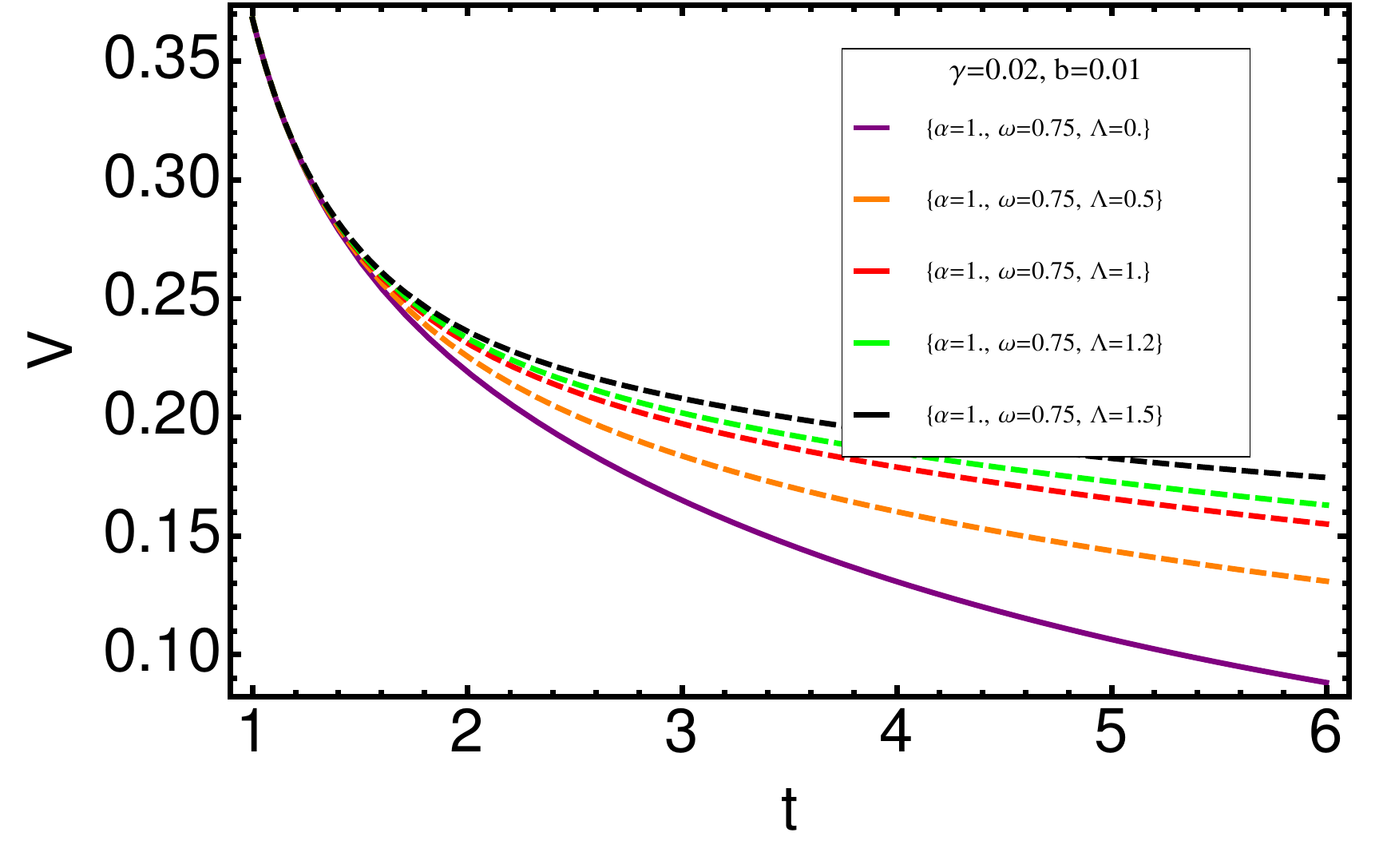}\\
 \end{array}$
 \end{center}
\caption{Behavior of filed $\phi$ and potential $V$ against $t$ for the constant $\Lambda$. Model 1}
 \label{fig:const3}
\end{figure}

In this section we will pay our attention to the first toy model. Within this and other models of this work we would like to examine the behavior of the Universe, to analyse and see if within our assumptions accelerated expansion of the Universe can be observed as well as analyse behavior of important cosmological parameters. In this model an interaction between DE and DM are assumed to be 
\begin{equation}
Q=3Hb\rho_{Q}+\gamma (\rho_{b}-\rho_{Q})\frac{\dot{\phi}}{\phi},
\end{equation} 
where $b$ is positive constant, $\phi$ is a field, $H$ is the Hubble parameter and $\rho_{b}$ and $\rho_{Q}$ represents energy densities of DM and DE. Therefore the dynamics of the energy density of the barotropic fluid can be found from
\begin{equation}
\dot{\rho}_{b}+3H \left (1+\omega_{b}-\frac{\gamma}{3H}\frac{\dot{\phi}}{\phi} \right )\rho_{b}=3H \left ( b-\frac{\gamma}{3H}\frac{\dot{\phi}}{\phi} \right )\rho_{Q}.
\end{equation}
Using the same mathematics we can obtain dynamics of DE
\begin{equation}
\dot{\rho}_{Q}+3H \left ( 1+b +\omega_{Q} -\frac{\gamma}{3H} \frac{\dot{\phi}}{\phi}\right  ) \rho_{Q}=-\gamma\frac{\dot{\phi}}{\phi}\rho_{b}.
\end{equation}
From the graphical analysis of the Hubble parameter and deceleration parameter $q$ we conclude 
that in the case of a constant $\Lambda$ the Hubble parameter is a decreasing function, which for later stages of evolution becomes a constant. Also we observe that with increasing numerical value of the $\Lambda$ we increase the numerical value of the Hubble parameter. For the deceleration parameter $q$ we see that the transition from the decelerated phase to the accelerated expansion phase in the history of Universe can be seen. Moreover, we see that for later stages of evolution $q$ behaves as a constant and its numerical value is in well agreement with observational facts. Therefore we can declare that proposed model is in good agreement with the observations. Discussed behaviors for the Hubble parameter and deceleration parameter $q$ can be seen in Fig. \ref{fig:const1}. We also analyse behavior of $\omega_{tot}$ and $\omega_{Q}$ and results can be found in Fig. \ref{fig:const2}. We see that that both parameters are a decreasing functions. $\omega_{tot}$ is a positive for early stages of evolution, then it is a negative and for later stages of evolution it is a constant. With an increasing the numerical value of the $\Lambda$ we can satisfy $\omega_{tot}=-1$ condition i.e for later stages of evolution in the dynamics of the Universe a cosmological constant has an important place. Behavior of $\omega_{Q}$ shows quintessence behavior of DE. The parameters of the Model 1 were fixed to obtain the well know fact that $V \rightarrow 0$ when $t \rightarrow \infty$. The field $\phi$ appears to be an increasing function of time (Fig. \ref{fig:const3}). 

\subsection{\large{Model 2: $Q=3Hb\rho+\gamma \dot{\rho}$}}
\begin{figure}[h!]
 \begin{center}$
 \begin{array}{cccc}
\includegraphics[width=50 mm]{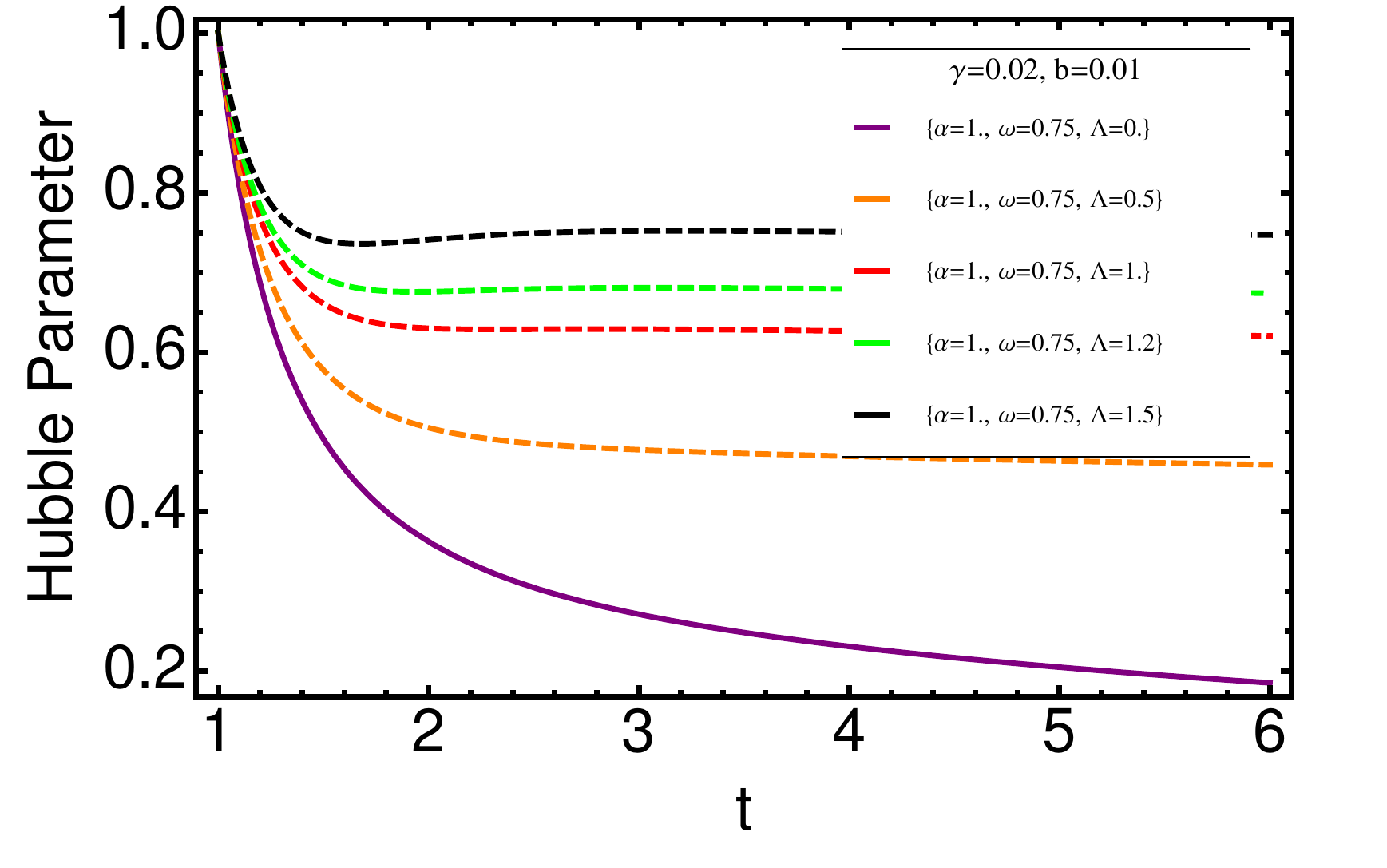} &
\includegraphics[width=50 mm]{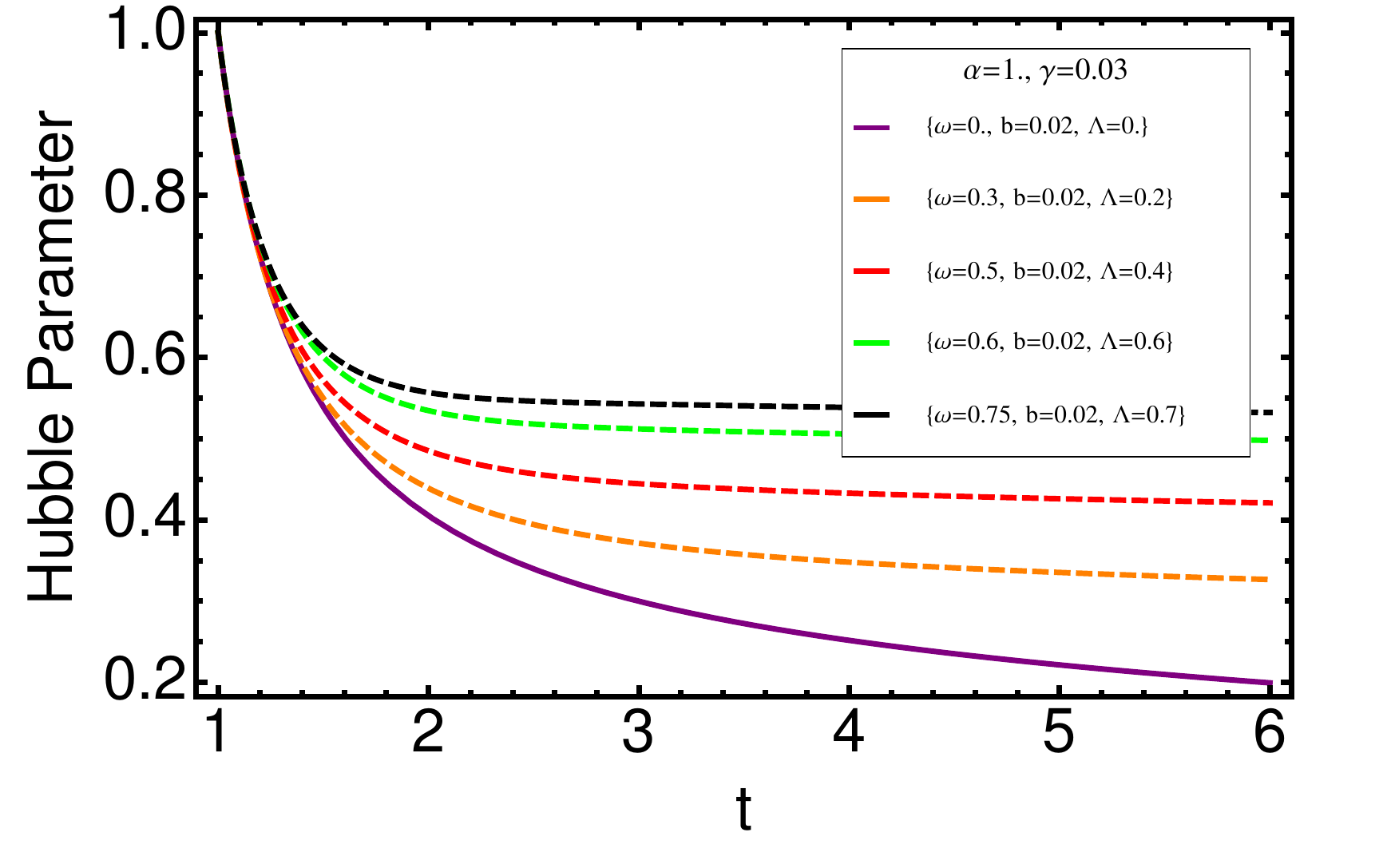}\\
\includegraphics[width=50 mm]{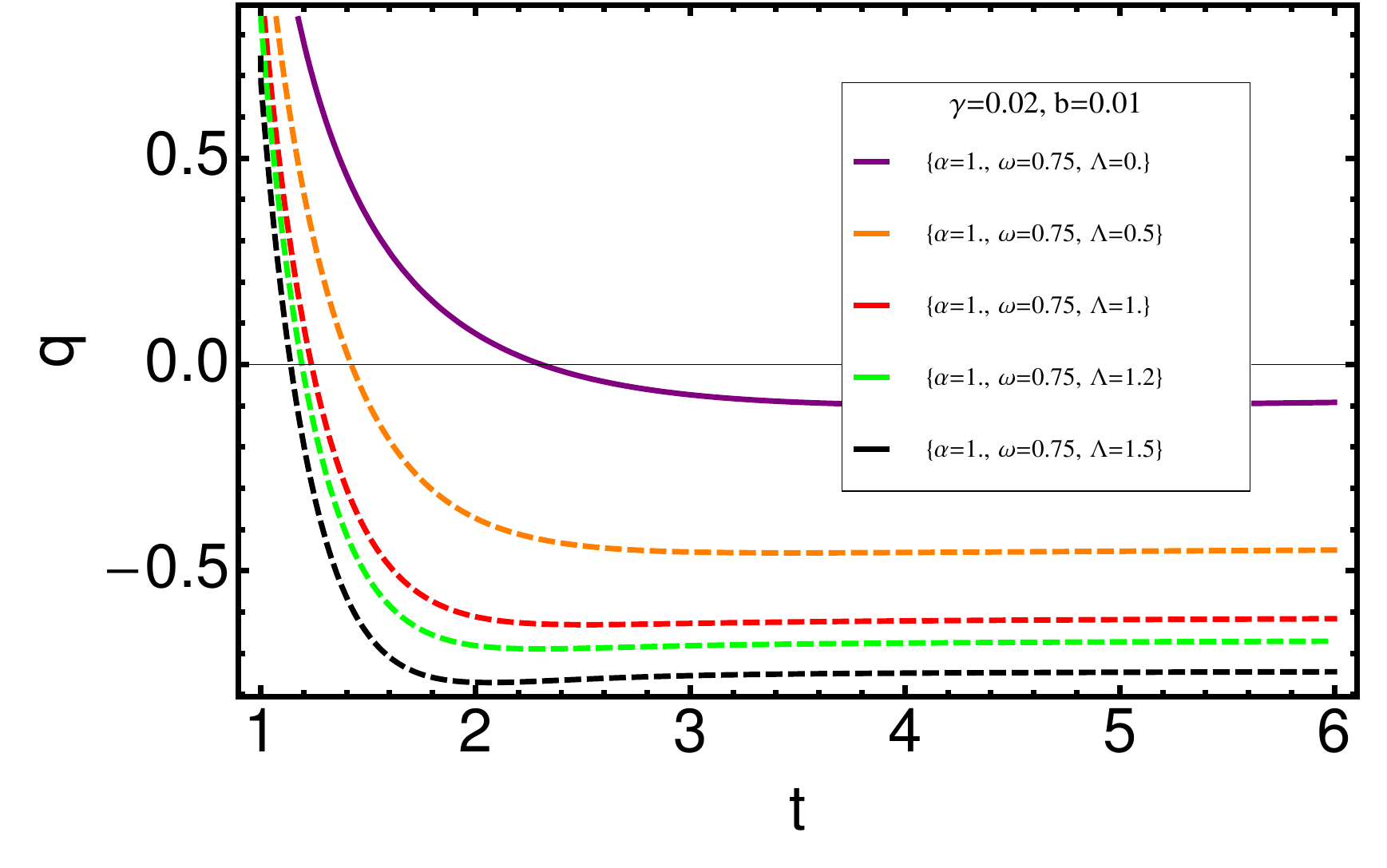} &
\includegraphics[width=50 mm]{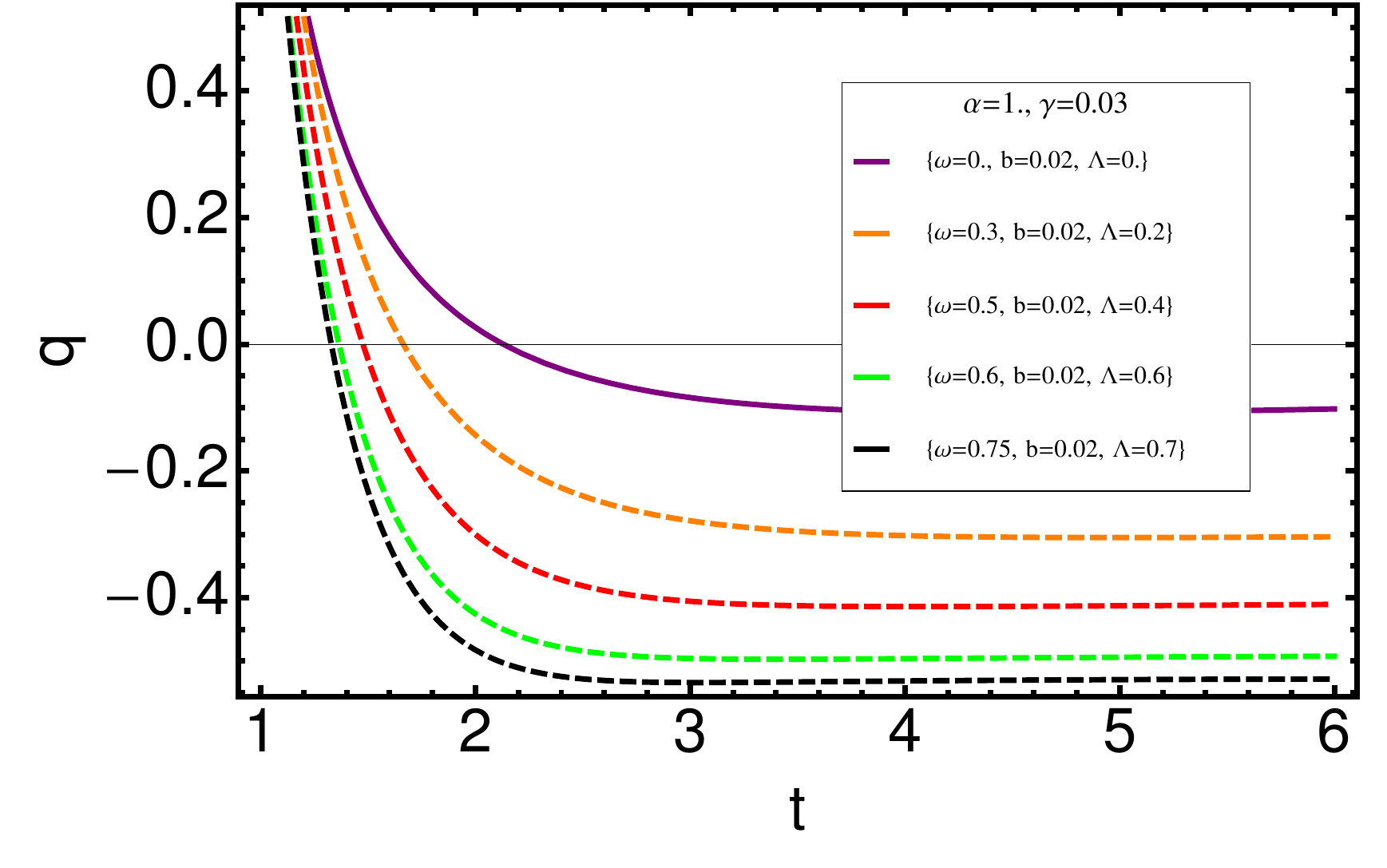}\\
 \end{array}$
 \end{center}
\caption{Behavior of Hubble parameter $H$ and  $q$ against $t$ for the constant $\Lambda$. Model 2}
 \label{fig:const4}
\end{figure}

\begin{figure}[h!]
 \begin{center}$
 \begin{array}{cccc}
\includegraphics[width=50 mm]{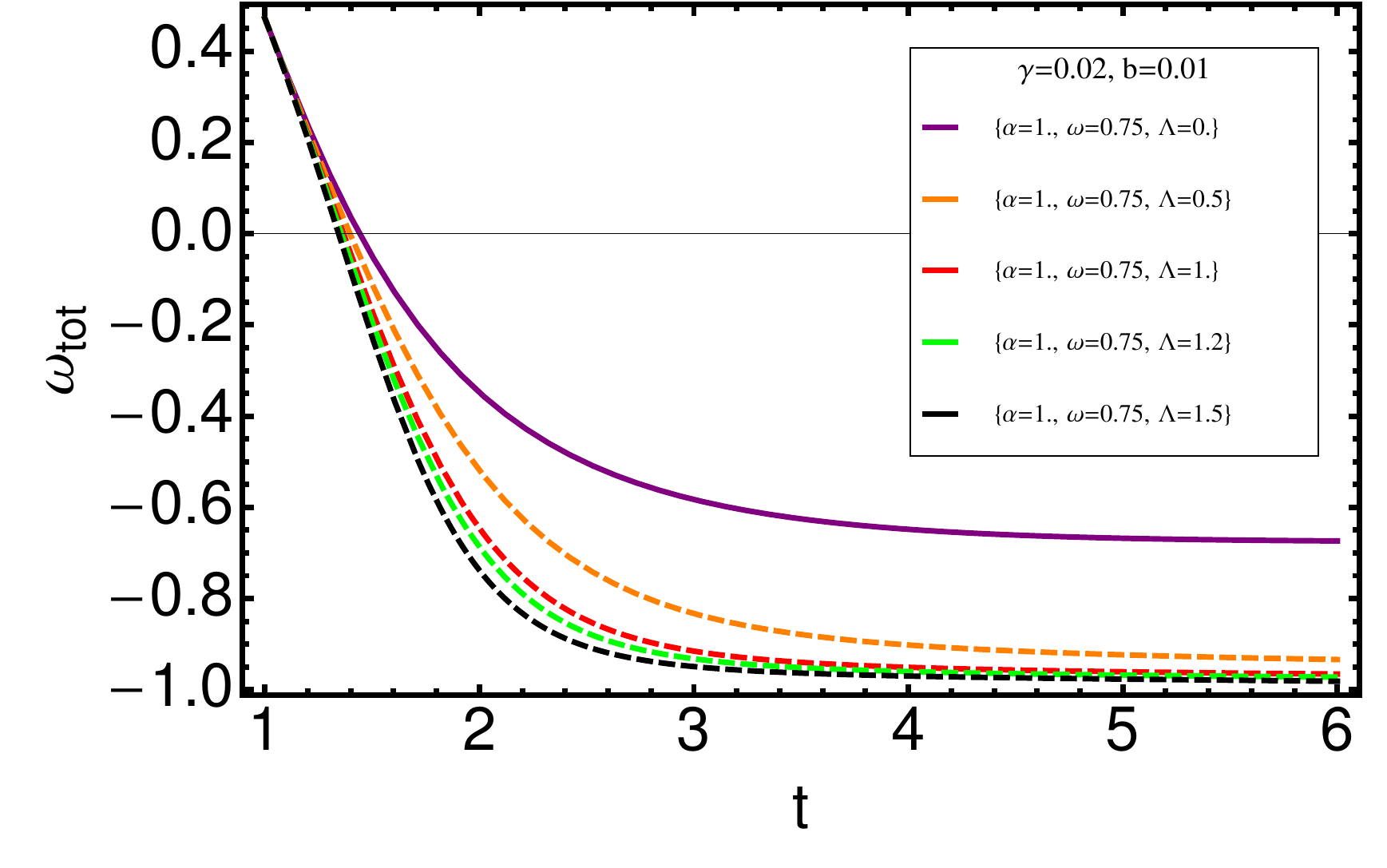} &
\includegraphics[width=50 mm]{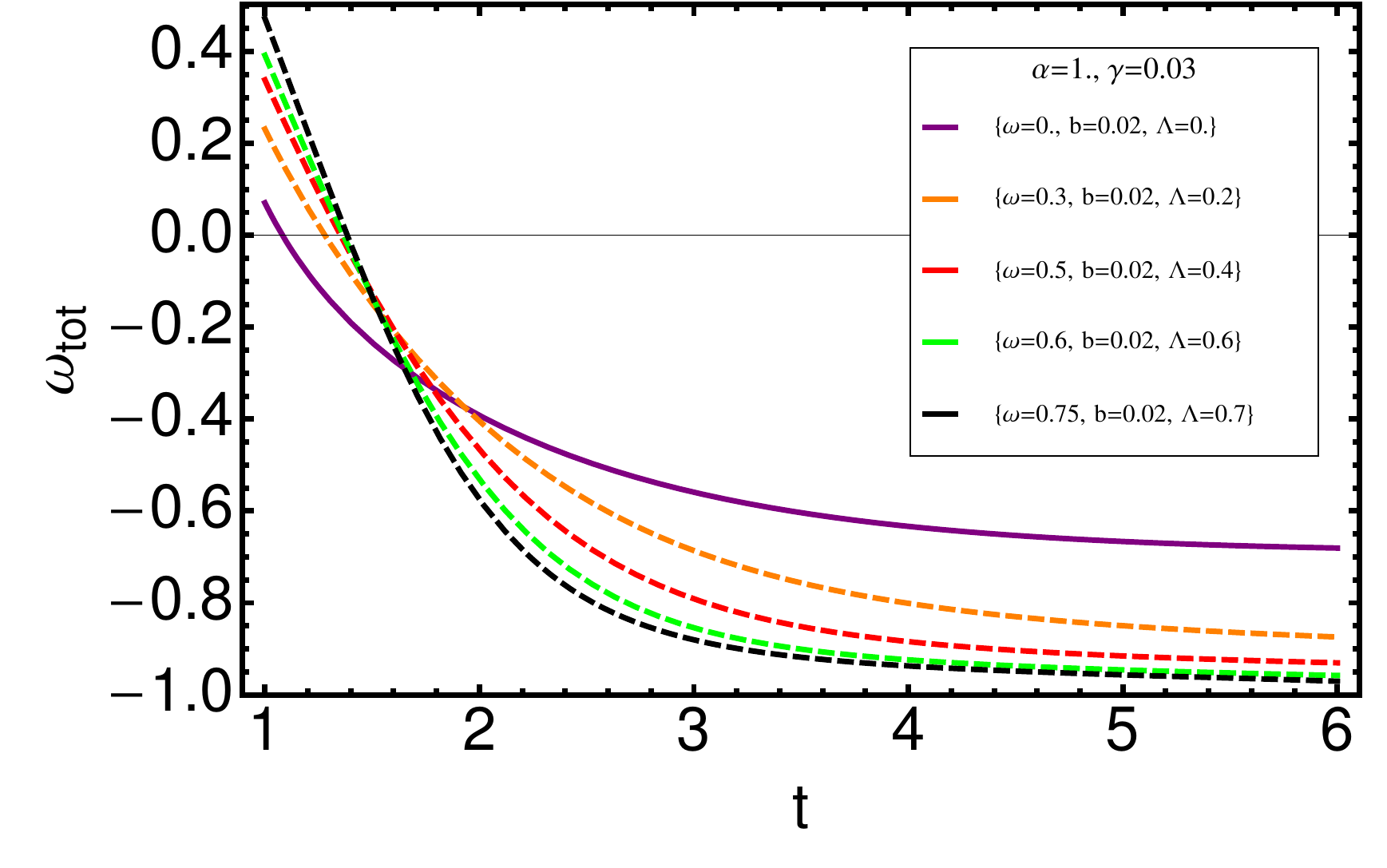}\\
\includegraphics[width=50 mm]{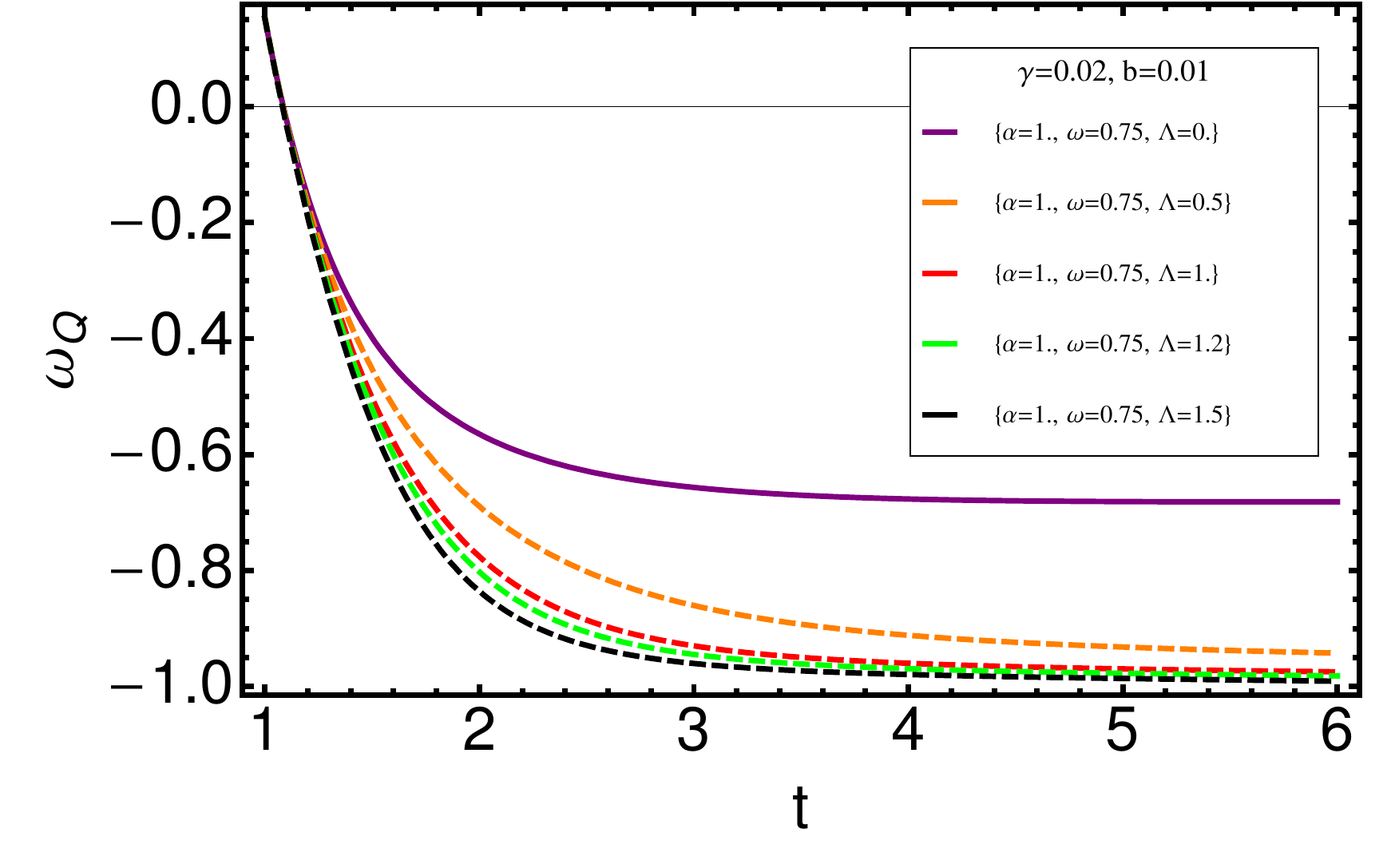} &
\includegraphics[width=50 mm]{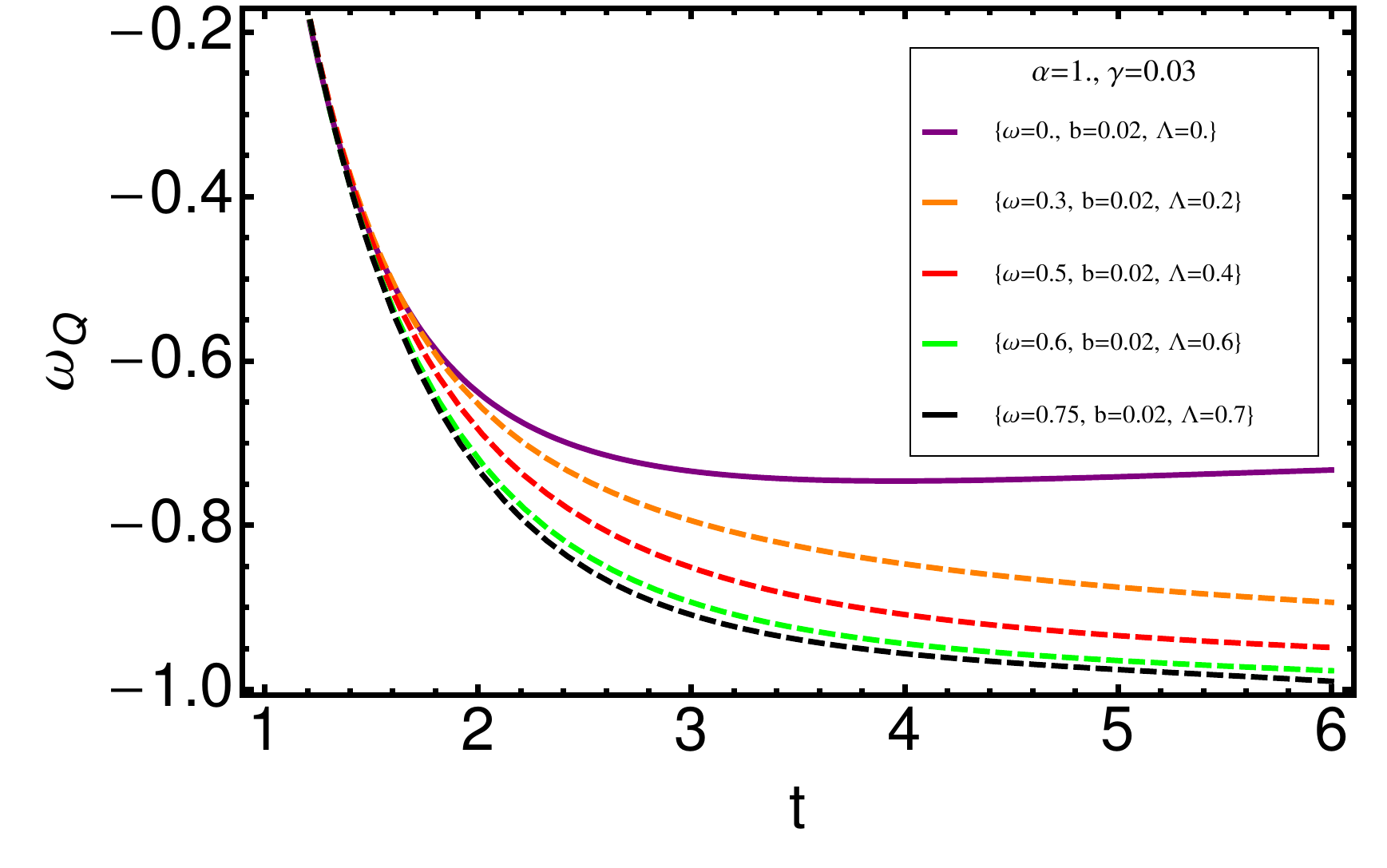}\\
 \end{array}$
 \end{center}
\caption{Behavior of EoS parameter $\omega_{tot}$ and $\omega_{Q}$ against $t$ for the constant $\Lambda$. Model 2}
 \label{fig:const5}
\end{figure}

\begin{figure}[h!]
 \begin{center}$
 \begin{array}{cccc}
\includegraphics[width=50 mm]{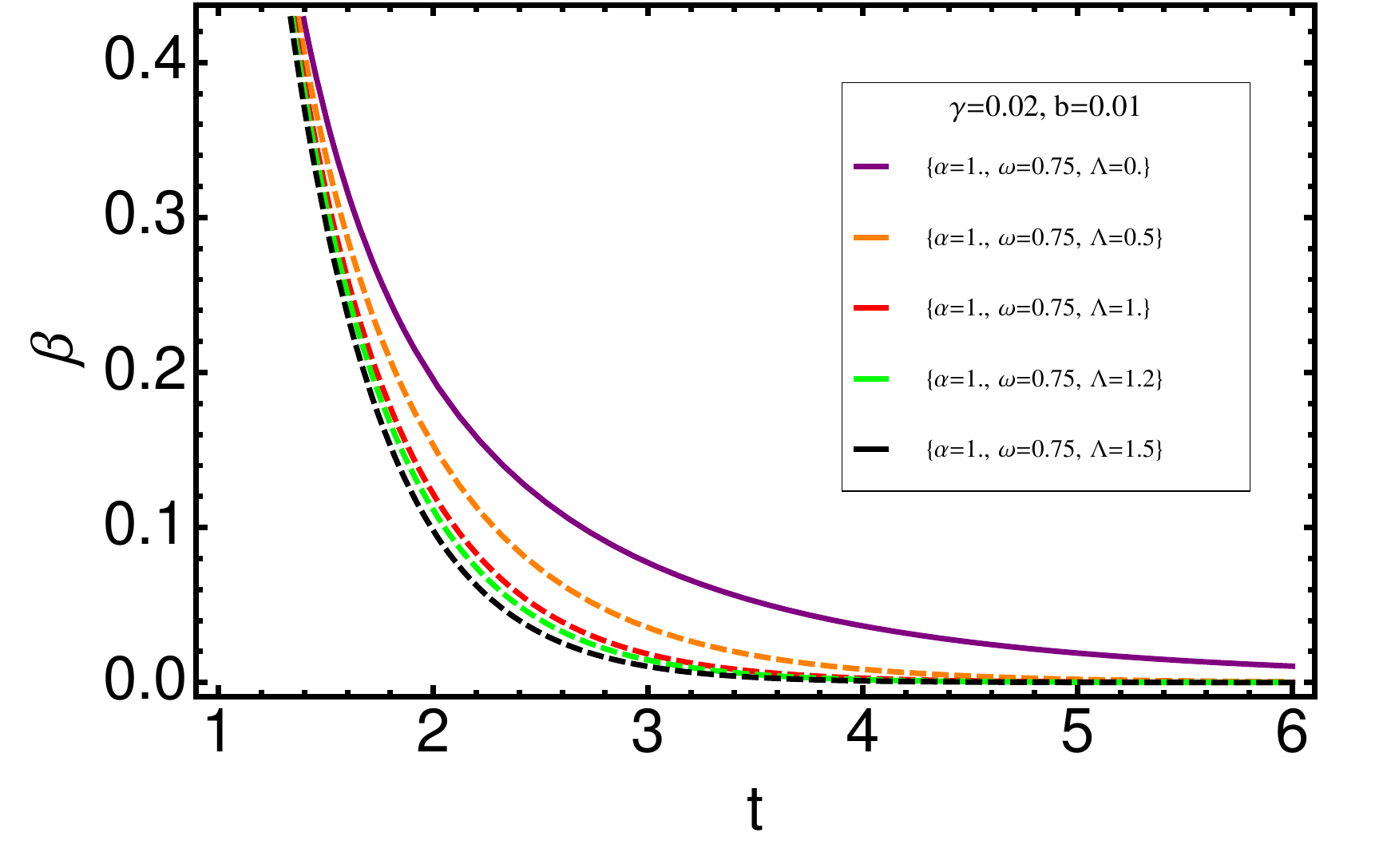} &
\includegraphics[width=50 mm]{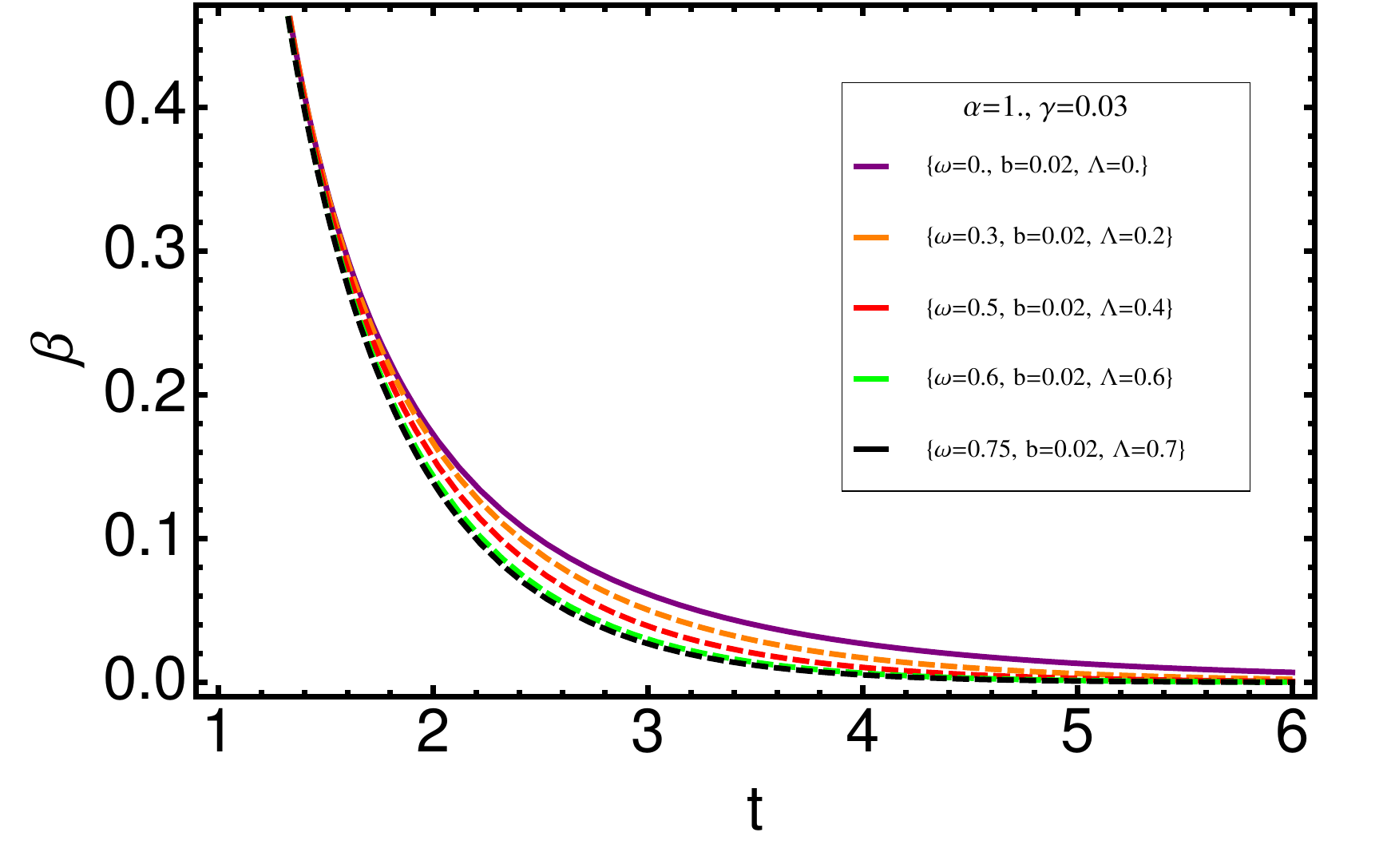}\\
 \end{array}$
 \end{center}
\caption{Behavior of $\beta$ against $t$ for the constant $\Lambda$. Model 2}
 \label{fig:const6}
\end{figure}

In this section we will analyse the second model (Model 2), with the interaction term $Q$, which has the following form
\begin{equation}
Q=3Hb\rho+\gamma \dot{\rho},
\end{equation}
where $\rho$ is the total energy density. The last form can be considered as a classical form of the general form for the interaction, because such interaction were considered intensively in literature. In this case our interaction is a function also from time derivative of energy density. The dynamics of DM and DE can be found after some mathematical transformations and we have the following forms
\begin{equation}
(1-\gamma)\dot{\rho}_{b}+3H(1+\omega_{b}-b)\rho_{b}=3Hb\rho_{Q}+\gamma \dot{\rho}_{Q},
\end{equation}
and
\begin{equation}
(1+\gamma)\dot{\rho}_{Q}+3H(1+\omega_{Q}+b)\rho_{Q}=-3Hb\rho_{b}-\gamma \dot{\rho}_{b}.
\end{equation}
The graphical behavior of the Hubble parameter shows us that it is a decreasing function within the evolution of the Universe. Also we see that it is a constant for later stages of evolution, moreover we see that with an appropriate choose of the numerical values of the model parameters we can find an agreement with observational data. We also would like to add, that with increasing numerical value of $\Lambda$ we increase numerical value of the Hubble parameter. An investigation of the deceleration parameter $q$, we found that this model is also in good agreement with observations. Behavior of the $q$ also can explain the well known fact that our Universe has transition from $q>0$ phase to the accelerated expansion phase with $q<0$. The combination of several observational data it is concluded that for the our Universe deceleration parameter should be greater than $-1$ which is illustrated for this model. In conclusion we would like to indicate that this model is also a good model. This analysis is based on the plots of Fig. \ref{fig:const4}. For the behavior of $\omega_{tot}$ which is the EoS parameter for our interacting two component fluid reveals quintessence-like behavior of the Universe for intermediate phases, while it is a positive for early stages of evolution, and for the old Universe it is a cosmological constant for high values of $\Lambda$ (Fig. \ref{fig:const5}).  The decreasing behavior of $\beta(t)$ is illustrated in Fig. \ref{fig:const6} 

\subsection{\large{Model 3: $Q=bH^{1-2\gamma}\rho_{b}^{\gamma}\dot{\phi}^{2}$}}
For the third model we assume that interaction between DI and DM can be modeled within
\begin{equation}
Q=bH^{1-2\gamma}\rho_{b}^{\gamma}\dot{\phi}^{2}.
\end{equation}
This type for the interaction is already considered in literature, and we wonder about it role in Lyra manifold. Next, we gave differential equations for the dynamics for energy densities. For barotropic DM it can be writen as
\begin{equation}
\dot{\rho}_{b}+3H(1+\omega_{b}-\frac{b}{3}H^{2(1-\gamma)}\rho_{b}^{\gamma-1}\dot{\phi}^{2}),
\end{equation}
and for DE we will have
\begin{equation}
\dot{\rho}_{Q}+3H(1+\omega_{Q})\rho_{Q}+bH^{1-2\gamma}\rho_{b}^{\gamma}\dot{\phi}^{2}.
\end{equation}
This form of interaction is a nonlinear function from the Hubble parameter, energy density of the barotropic fluid. It is a function also from derivative of the field $\phi$. For illustration we analyse behavior of the Hubble parameter, deceleration parameter $q$, $\omega_{tot}$ and $omega_{Q}$ as a function of $b$, $\gamma$ with increasing $\Lambda$. Our analysis shows that this model is in good agreement with observations, therefore in conclusion of this section we would like to mention that considered three interacting models in Lyra manifold can serve as a good models. As the starting models they could be generalized and investigated from different corners in order to understand the viability of them. In the next section we will consider the varying $\Lambda(t)$ case. The form which we consider in this work is already constructed by us and considered in usual GR for the general case.
\begin{figure}[h!]
 \begin{center}$
 \begin{array}{cccc}
\includegraphics[width=50 mm]{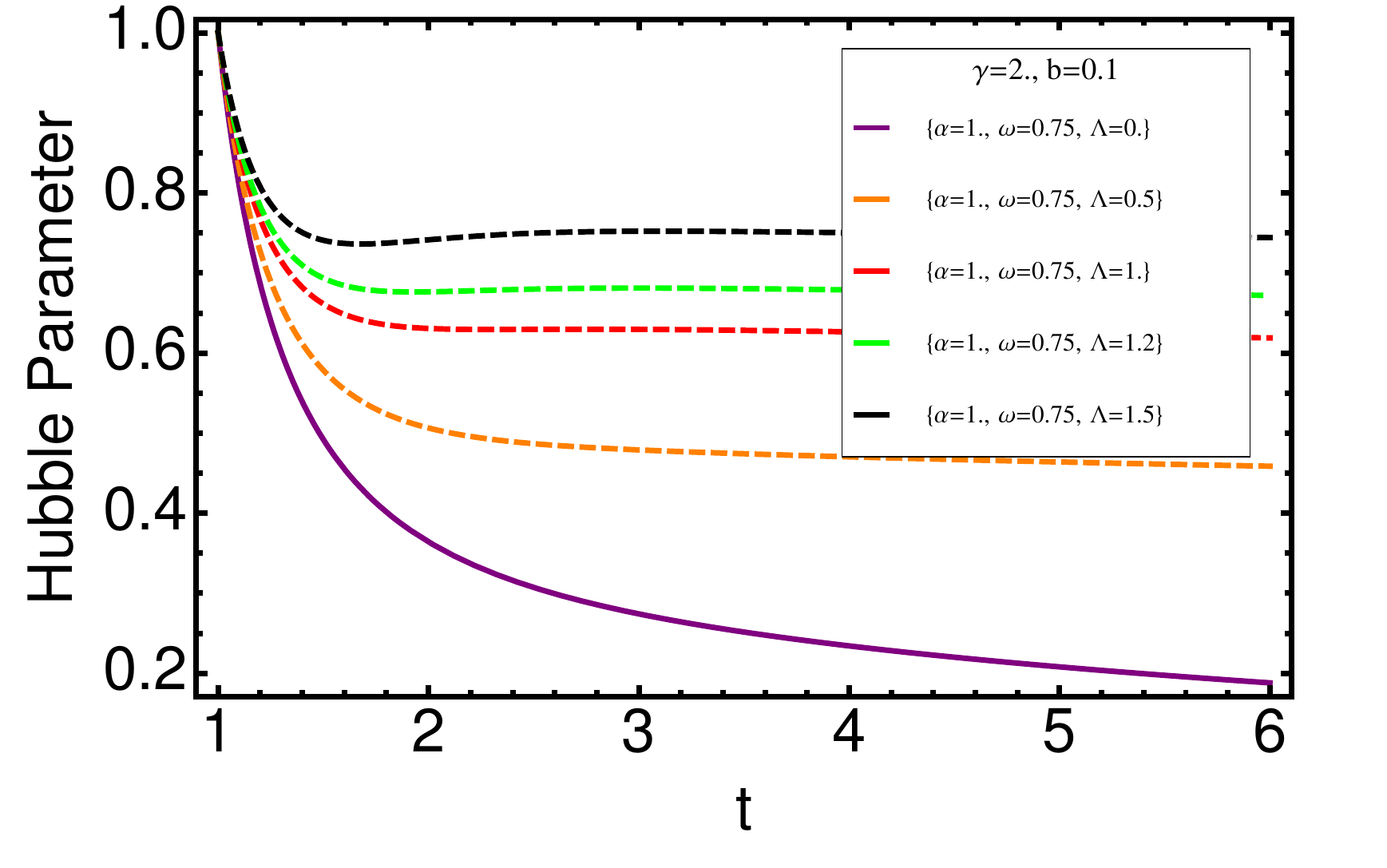} &
\includegraphics[width=50 mm]{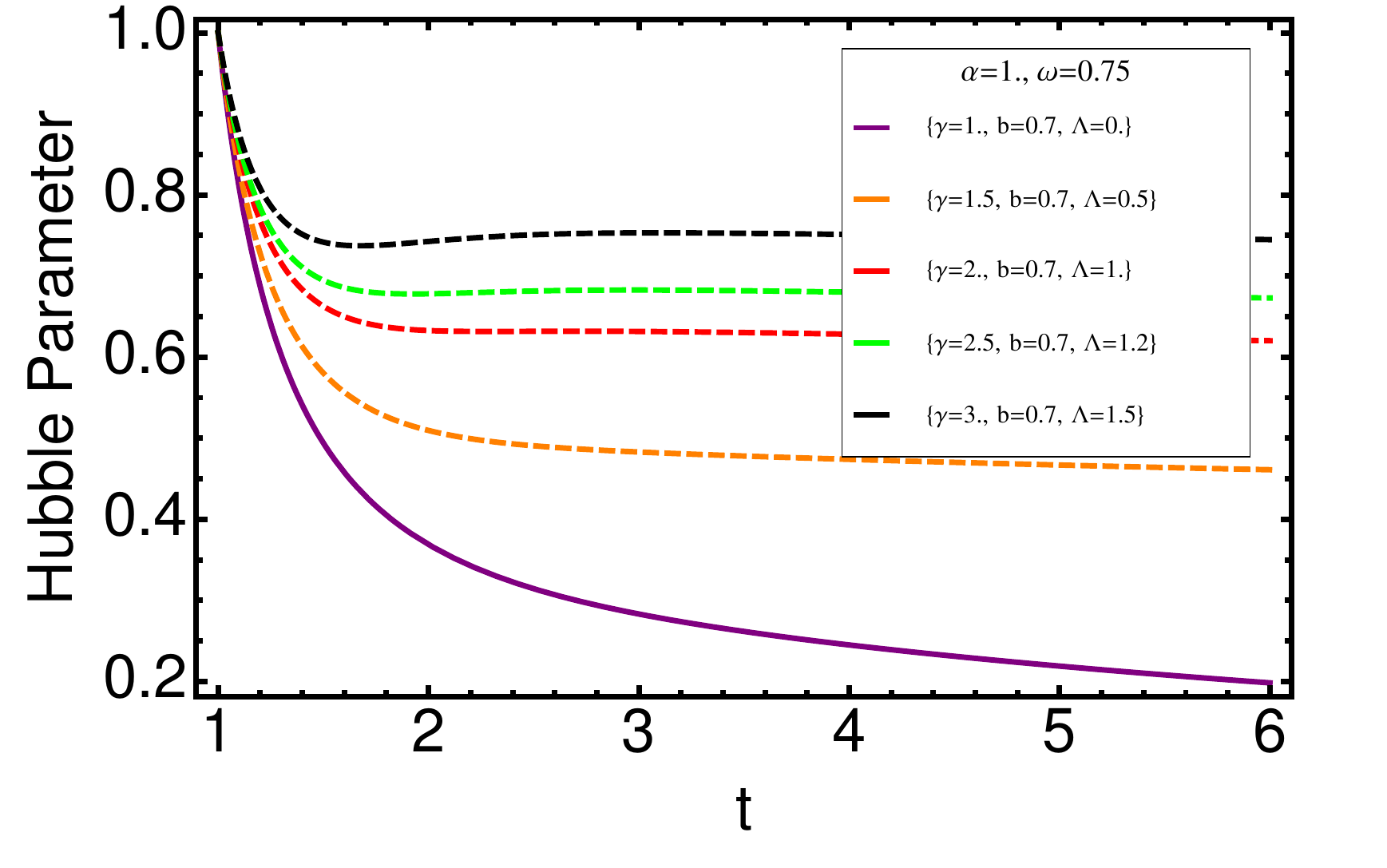}\\
\includegraphics[width=50 mm]{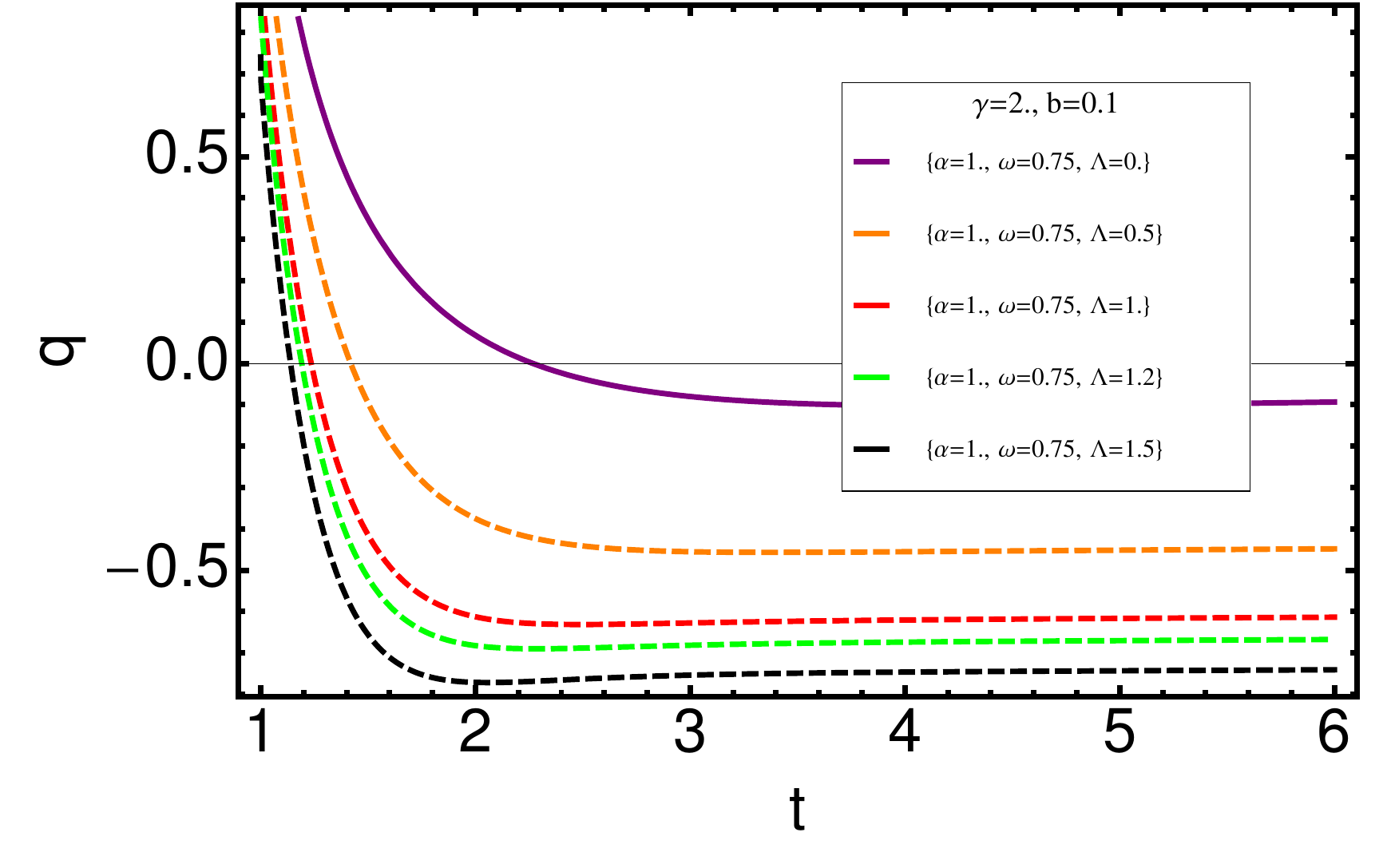} &
\includegraphics[width=50 mm]{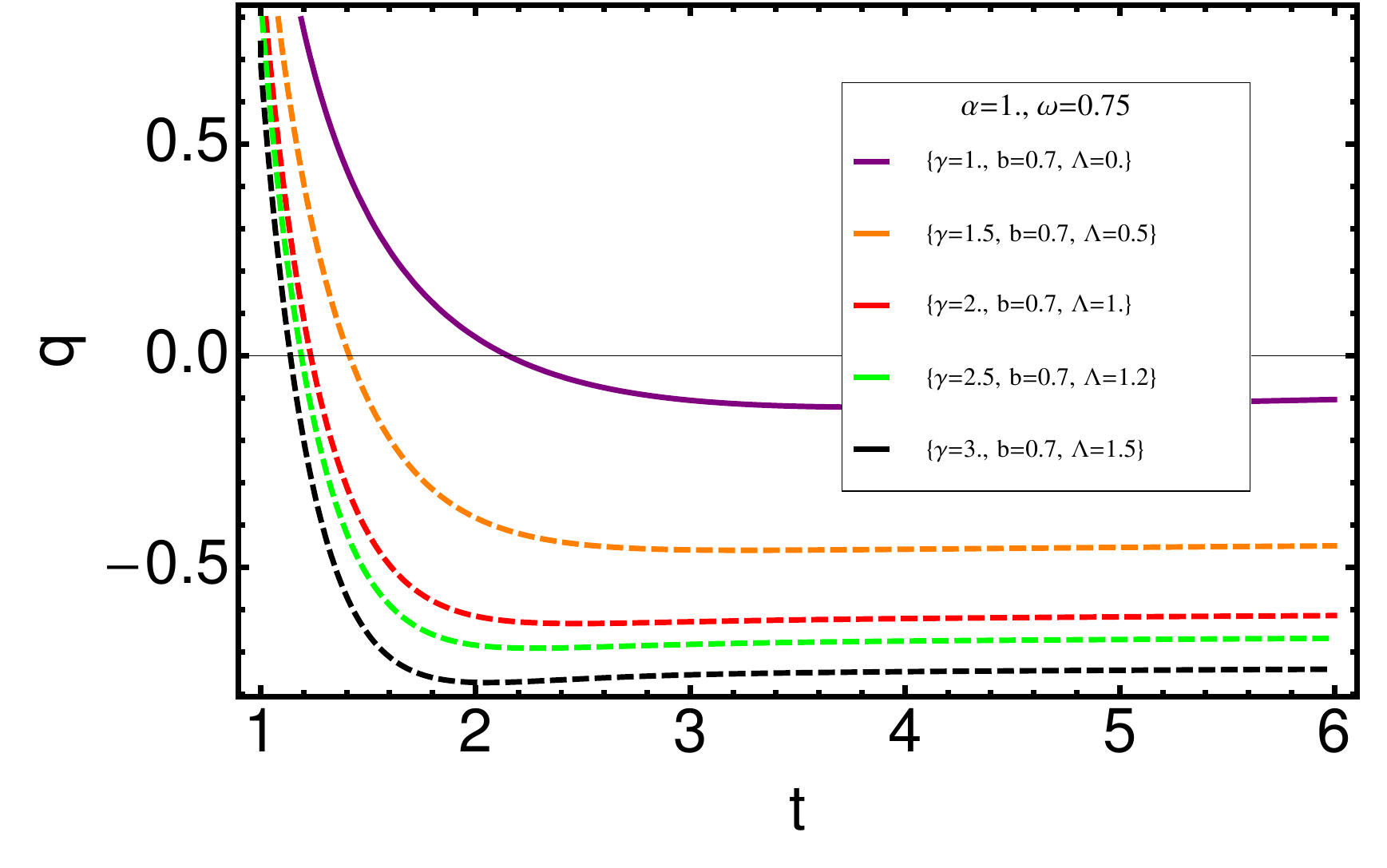}\\
 \end{array}$
 \end{center}
\caption{Behavior of Hubble parameter $H$ and  $q$ against $t$ for the constant $\Lambda$. Model 3}
 \label{fig:const7}
\end{figure}

\begin{figure}[h!]
 \begin{center}$
 \begin{array}{cccc}
\includegraphics[width=50 mm]{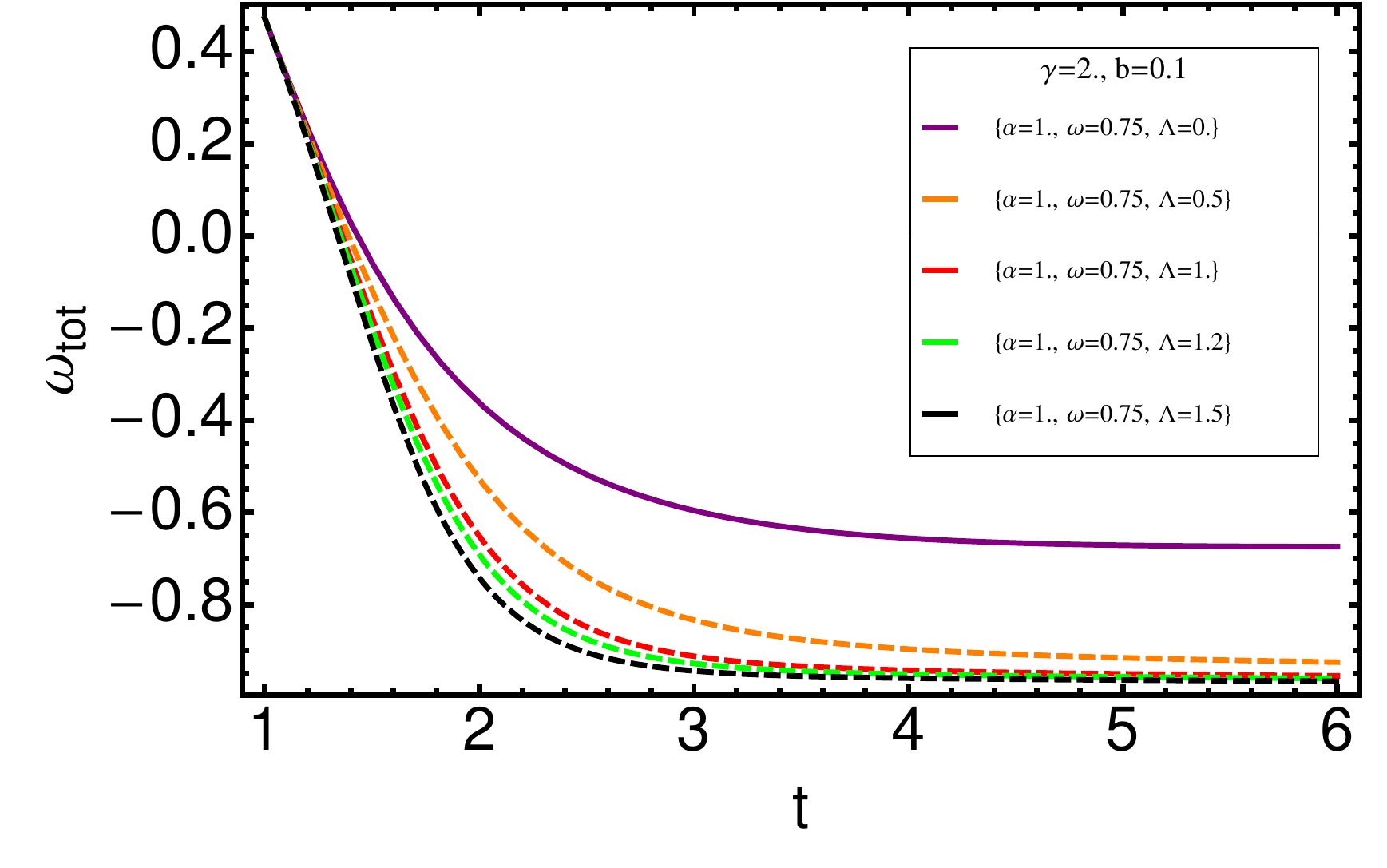} &
\includegraphics[width=50 mm]{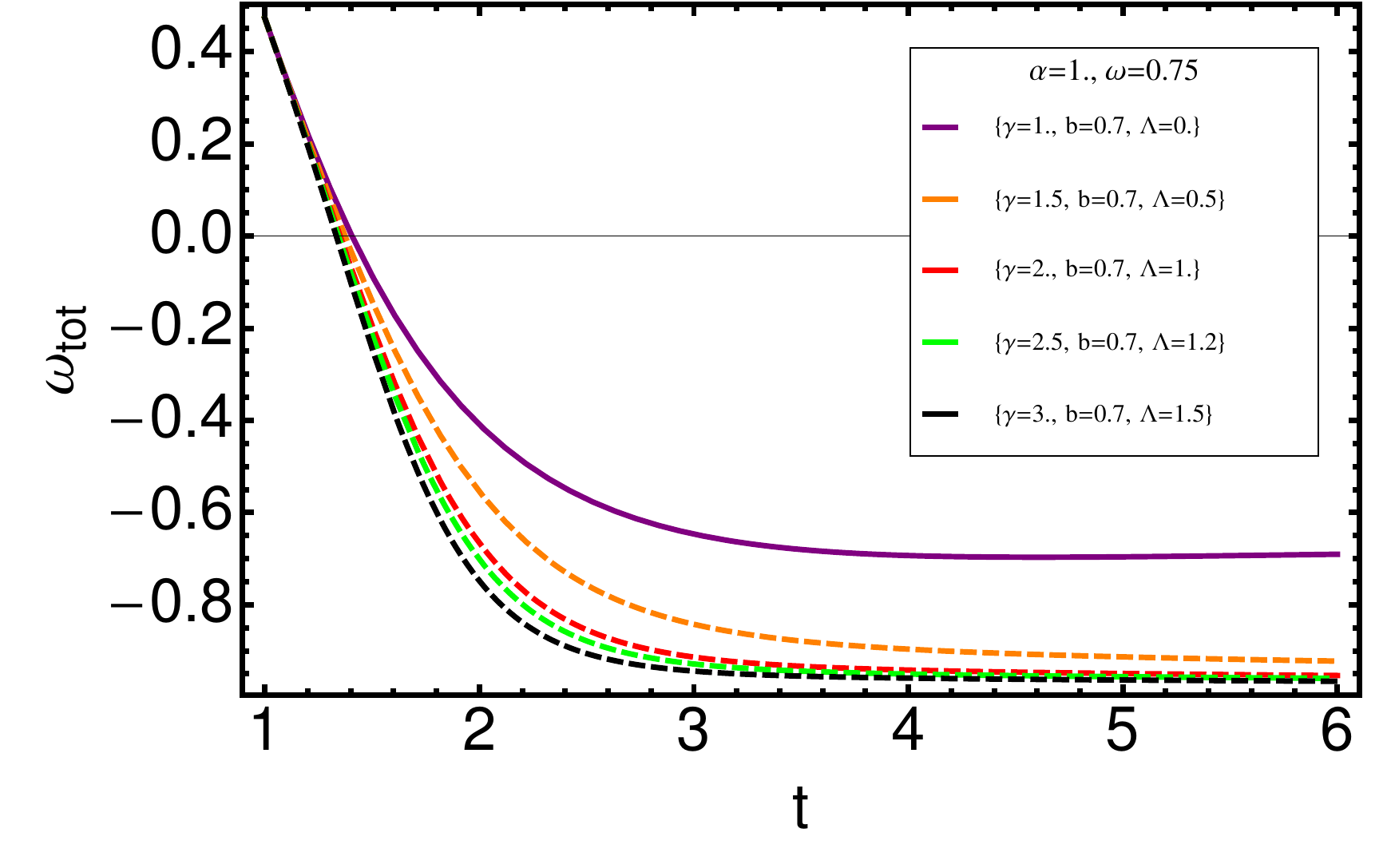}\\
\includegraphics[width=50 mm]{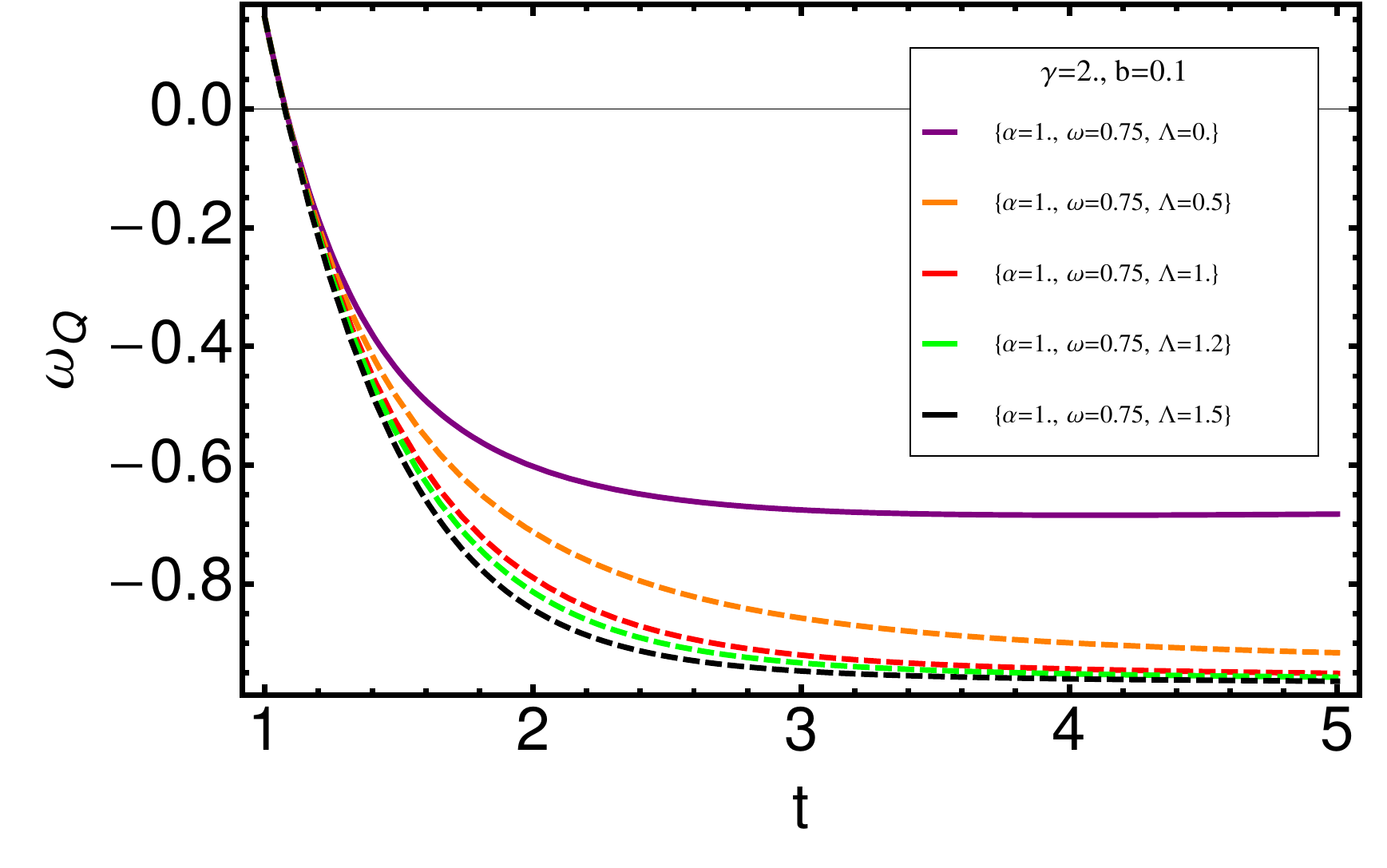} &
\includegraphics[width=50 mm]{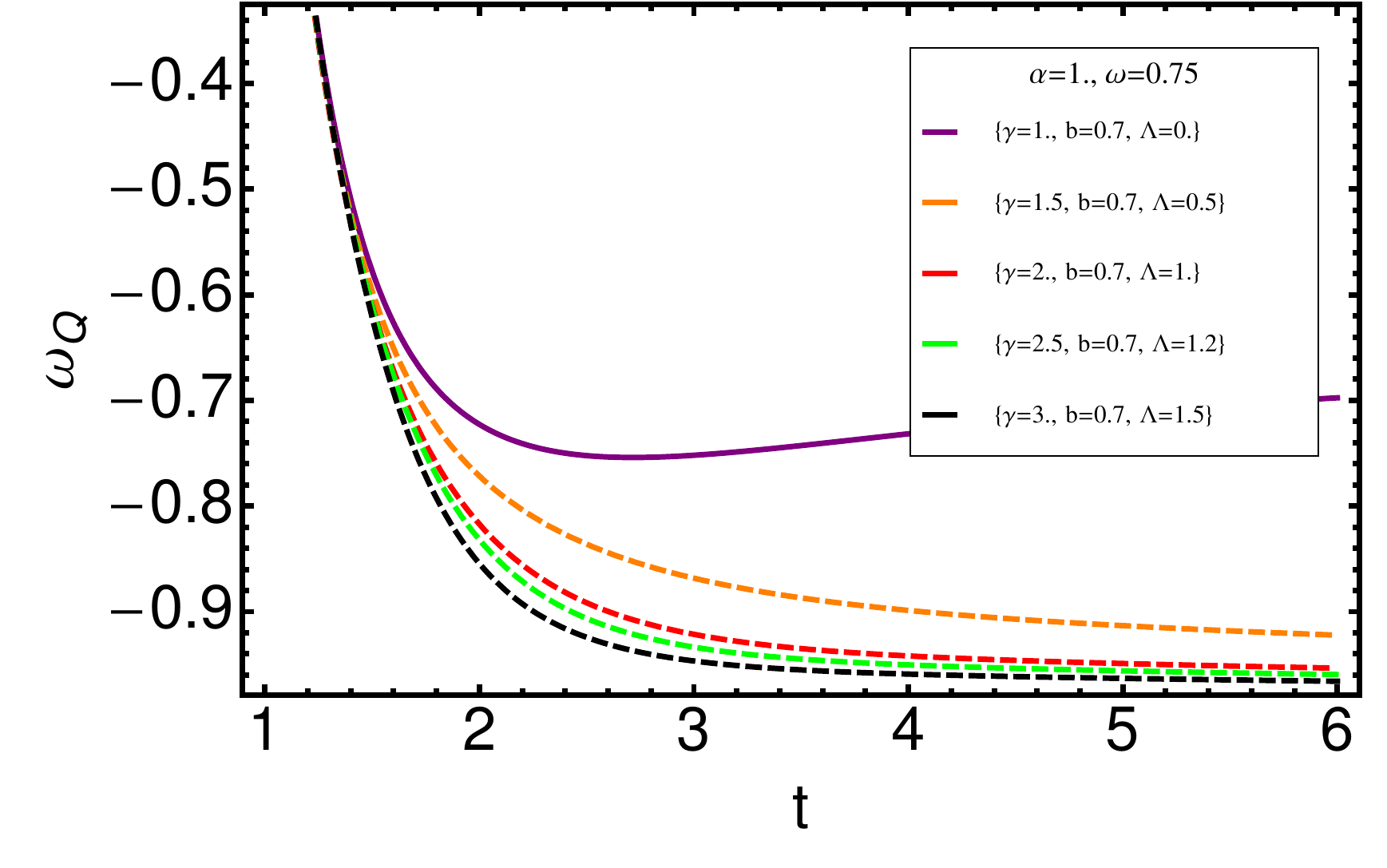}\\
 \end{array}$
 \end{center}
\caption{Behavior of EoS parameter $\omega_{tot}$ and $\omega_{Q}$ against $t$ for the constant $\Lambda$. Model 3}
 \label{fig:const8}
\end{figure}

\section{\large{Varyin $\Lambda(t)=\Lambda(H,\phi,V(\phi))$}}
In this section we will consider three interacting fluid models and will investigate cosmological parameters like the Hubble parameter $H$, deceleration parameter $q$, EoS parameters of the total fluid and DE $\omega_{Q}$. Based on numerical solutions, we will discuss graphical behaviors of the cosmological parameters. For the $\Lambda(t)$ we take a phenomenological form which was considered by us recently. The formula of $\Lambda(t)$ is
\begin{equation}
\Lambda(t) = H^{2} \phi^{-2}+\delta V(\phi),
\end{equation}
which is a function of the Hubble parameter, potential of the scalar field, and time derivative of the scalar field. For the potential we take a simple form $V(\phi)=e^{[-\alpha \phi]}$, therefore the form of $\Lambda(t)$ can be writen also in the following way as only a function of the filed $\phi$
\begin{equation}
\Lambda(t)=H^{2}\phi^{-2}+\delta e^{[-\alpha \phi]}.
\end{equation}
Therefore, the dynamics of $\beta$ can be obtained from the following differential equation
\begin{equation}
2\beta \dot{\beta}+6H\beta^{2}+2H\dot{H}\phi^{-2}-2H^{2}\phi^{-3}\dot{\phi}-\delta \alpha e^{[-\alpha \phi]} \dot{\phi}=0.
\end{equation}
In forthcoming subsections within three different forms of $Q$ we will investigate the dynamics of the Universe. The question of the dynamics for the energy densities of the DE and DM is already discussed in previous section, therefore we will not consider them here and we will start with the comments on the graphical behaviors of the cosmological parameters of the models. We will start with the model where $Q=3Hb\rho_{Q}+\gamma (\rho_{b}-\rho_{Q})\frac{\dot{\phi}}{\phi}$.

\subsection{\large{Model 4: $Q=3Hb\rho_{Q}+\gamma (\rho_{b}-\rho_{Q})\frac{\dot{\phi}}{\phi}$}}
\begin{figure}[h!]
 \begin{center}$
 \begin{array}{cccc}
\includegraphics[width=50 mm]{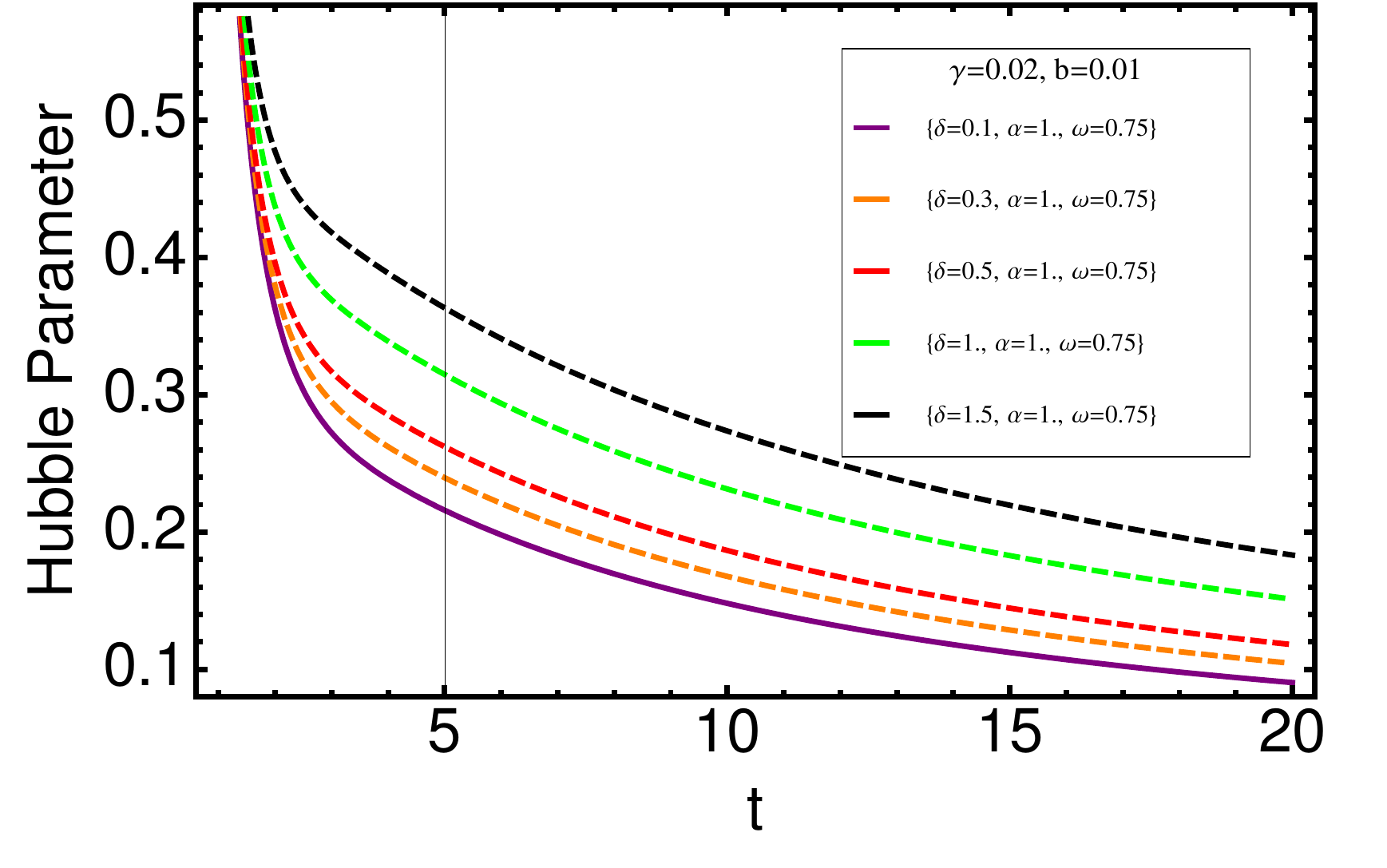} &
\includegraphics[width=50 mm]{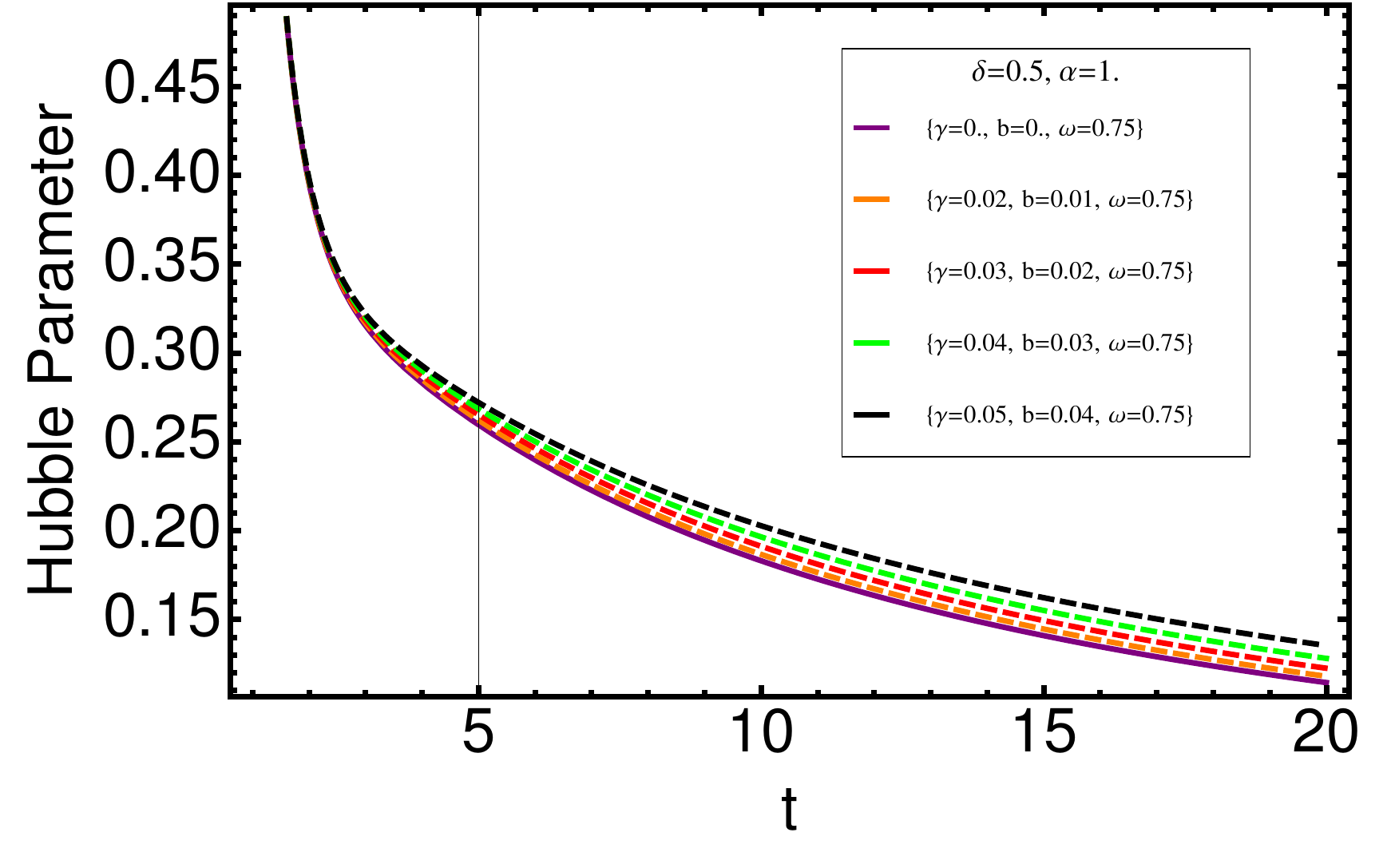}&
\includegraphics[width=50 mm]{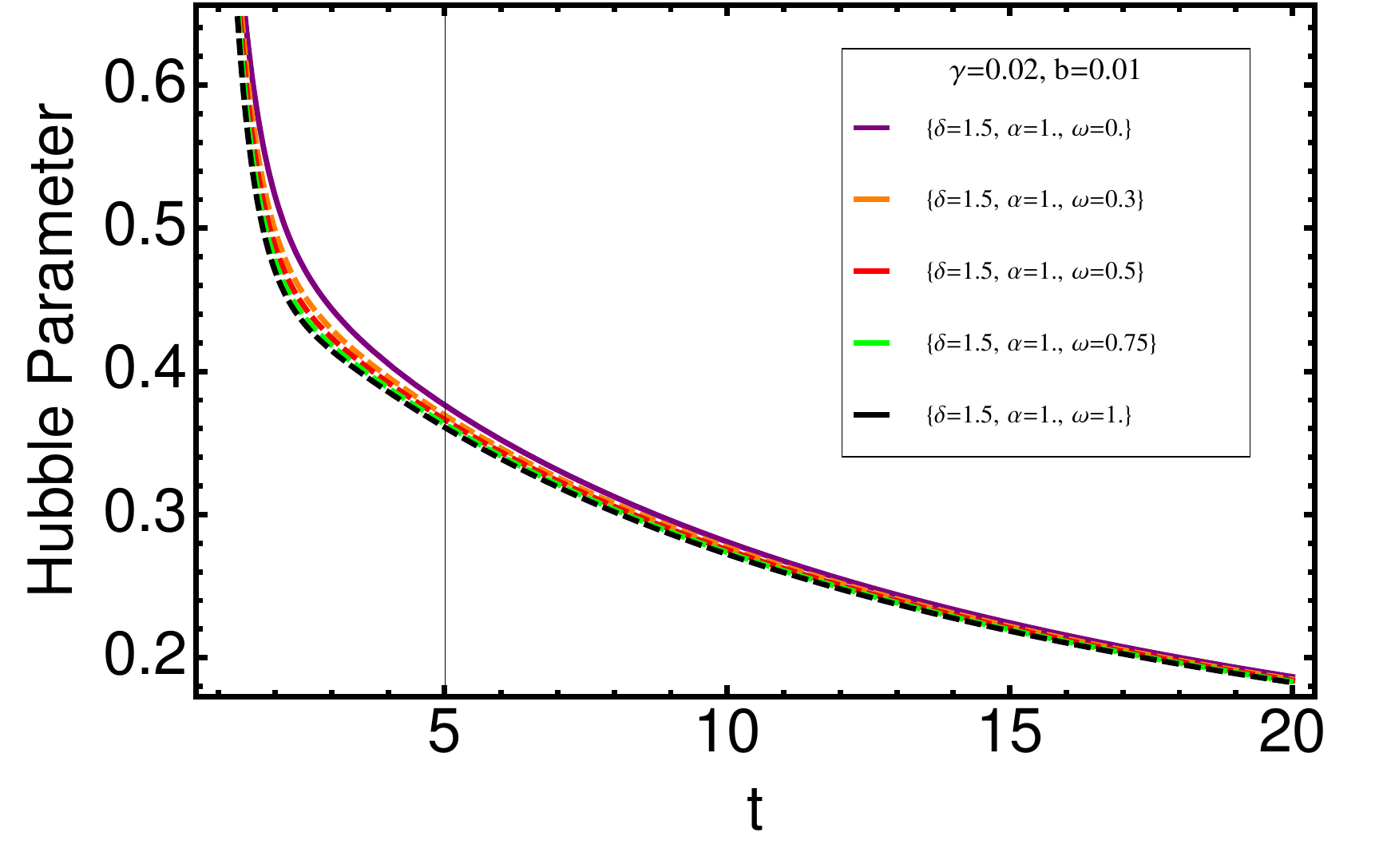}\\
\includegraphics[width=50 mm]{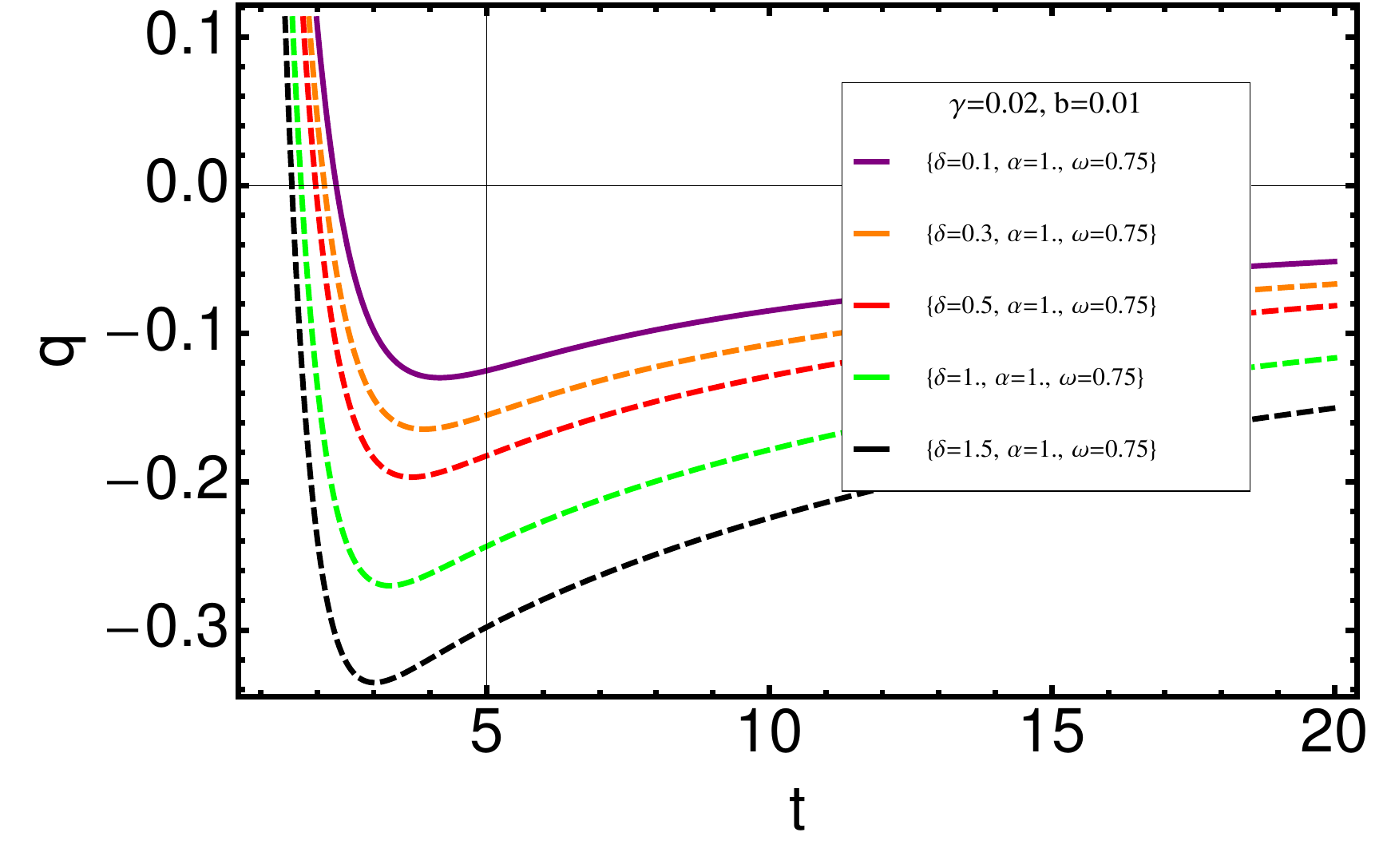} &
\includegraphics[width=50 mm]{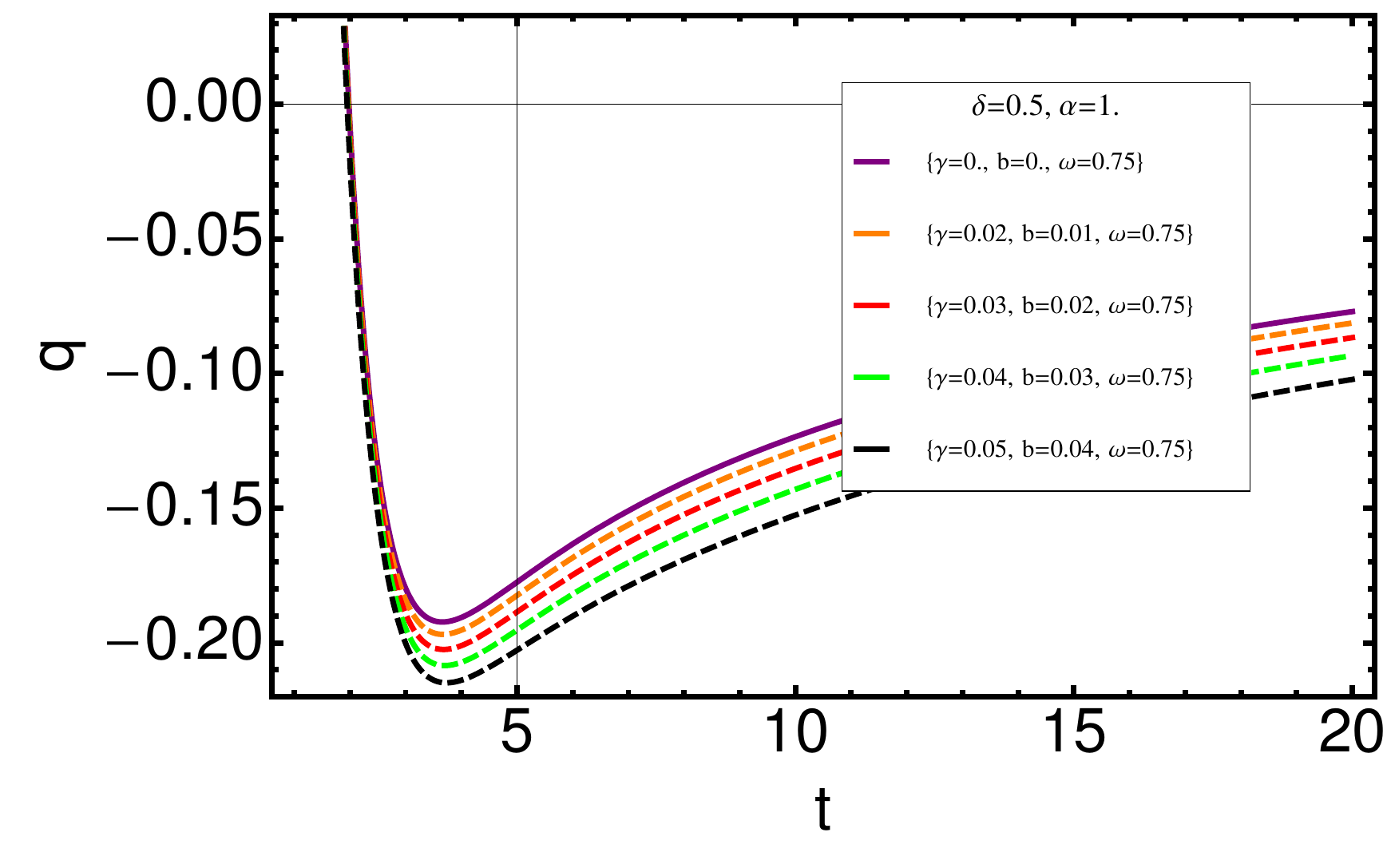}&
\includegraphics[width=50 mm]{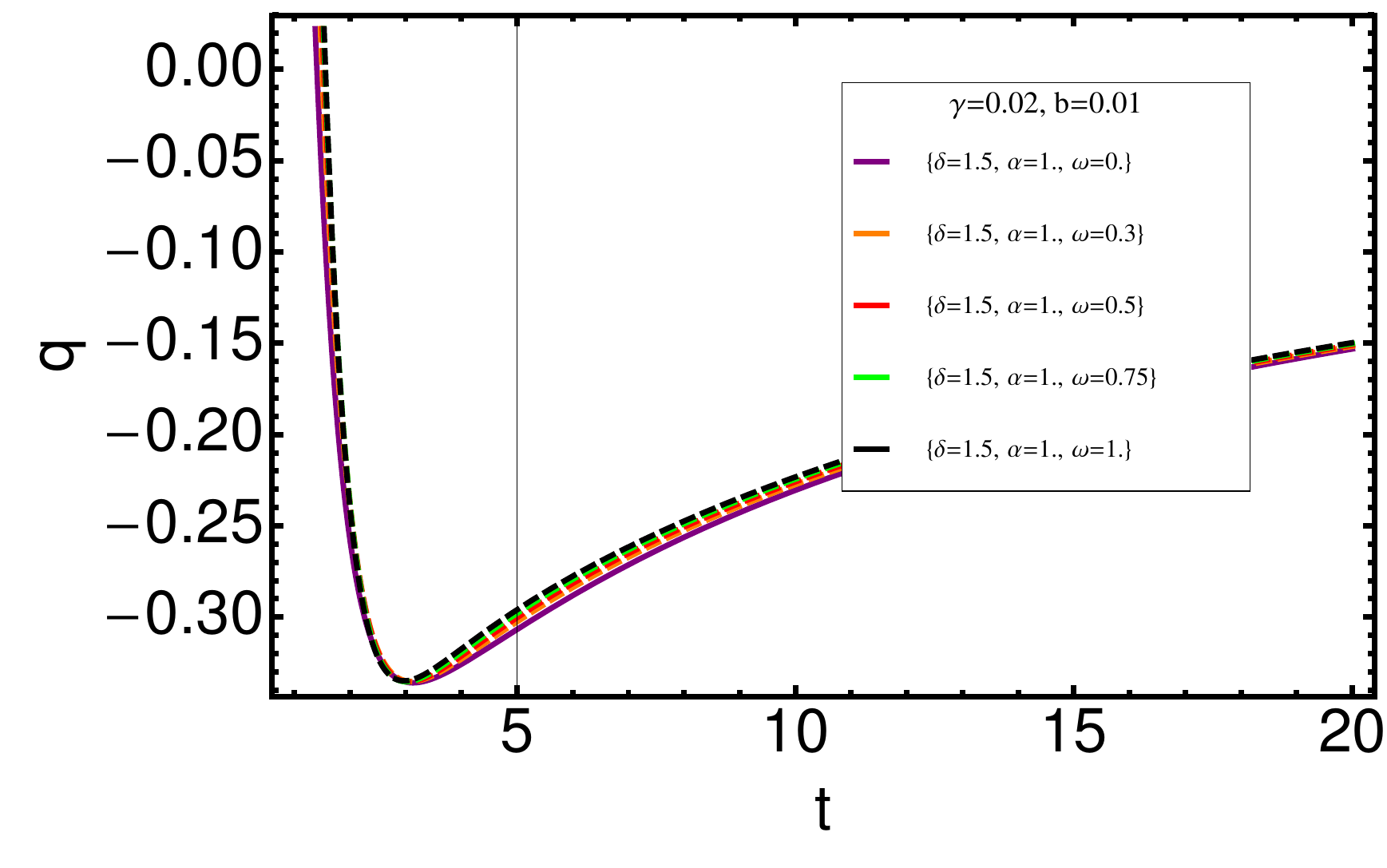}
 \end{array}$
 \end{center}
\caption{Behavior of Hubble parameter $H$ and deceleration parameter $q$ against $t$ for Model 4.}
 \label{fig:1}
\end{figure}

\begin{figure}[h!]
 \begin{center}$
 \begin{array}{cccc}
\includegraphics[width=50 mm]{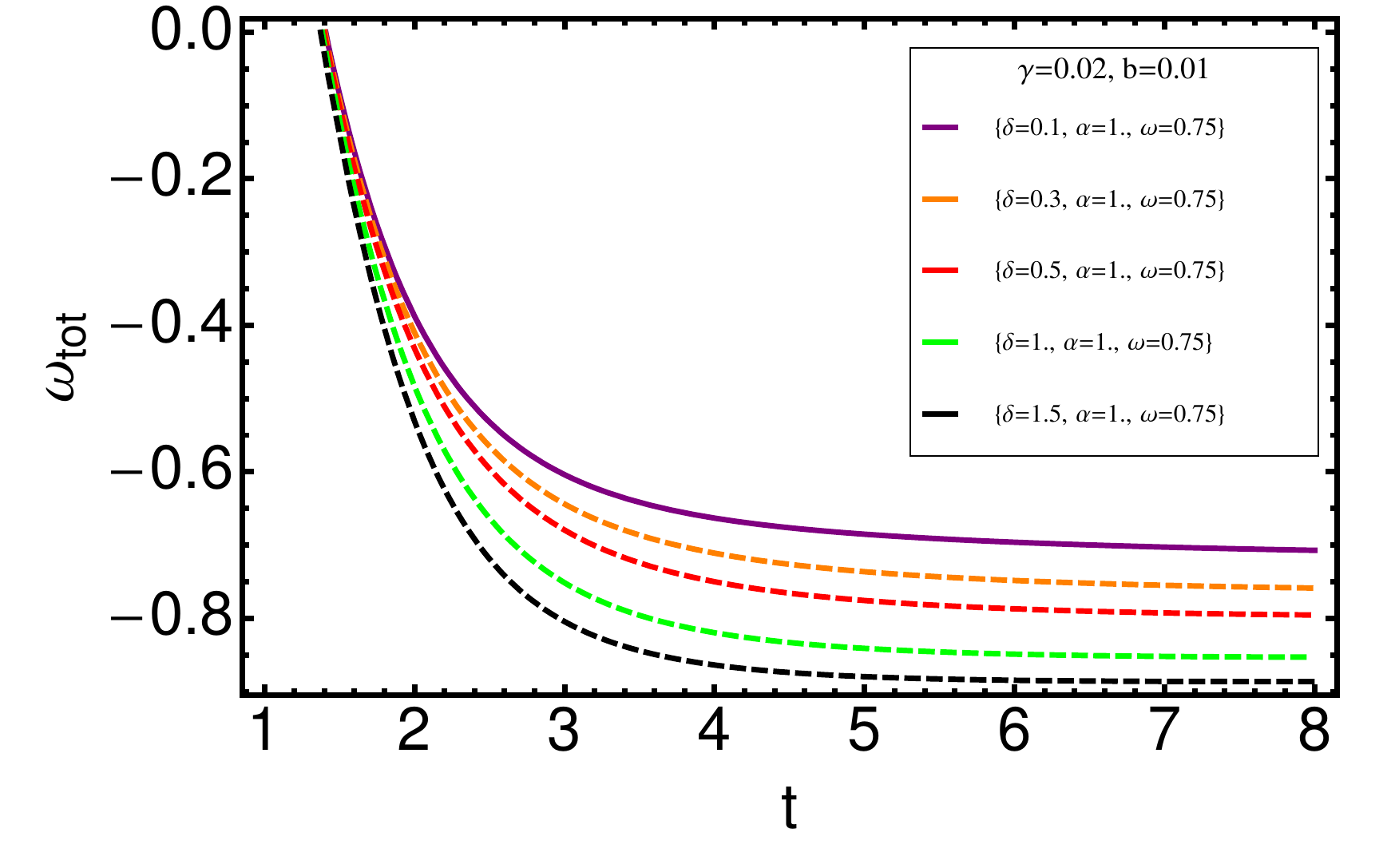} &
\includegraphics[width=50 mm]{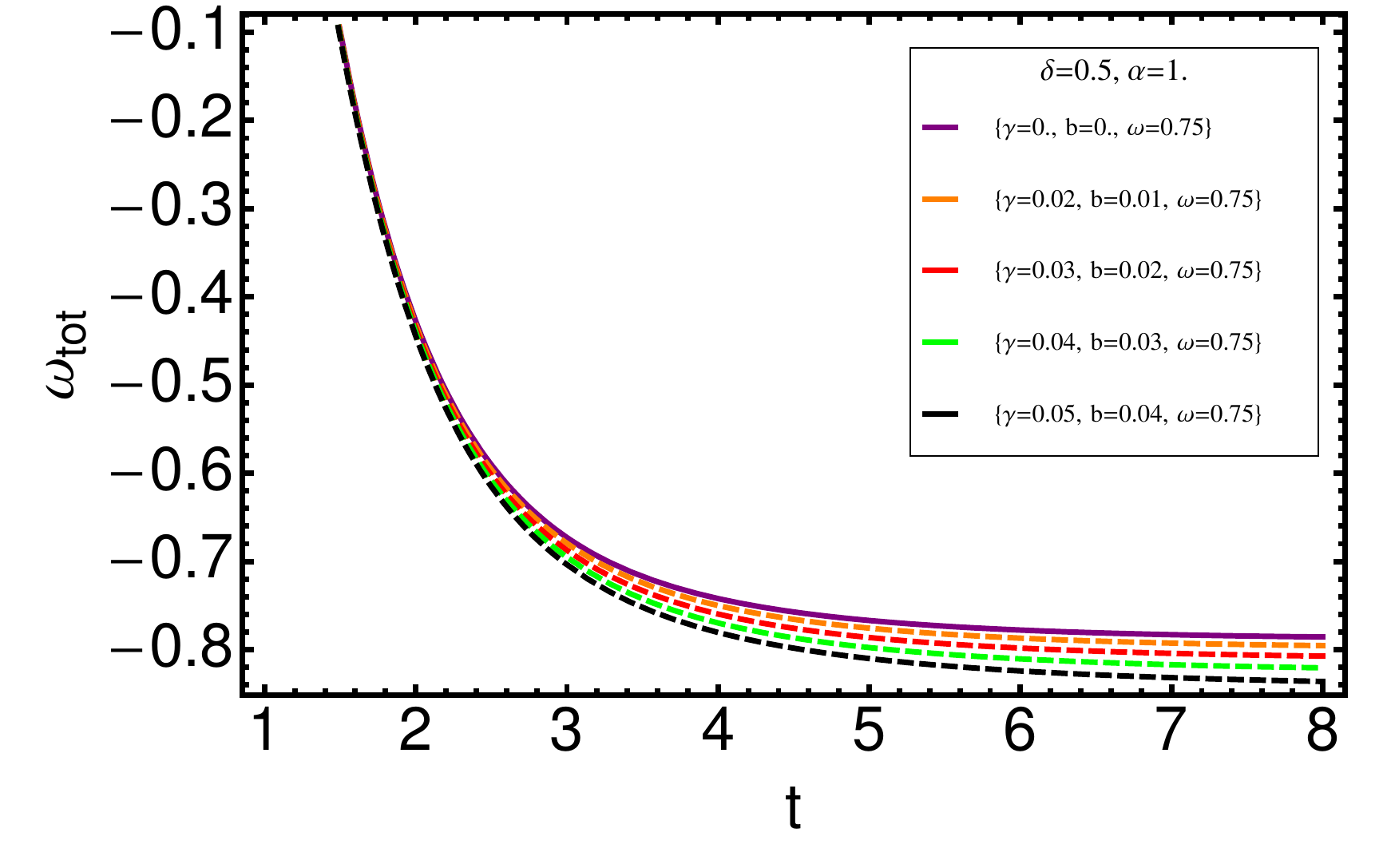}&
\includegraphics[width=50 mm]{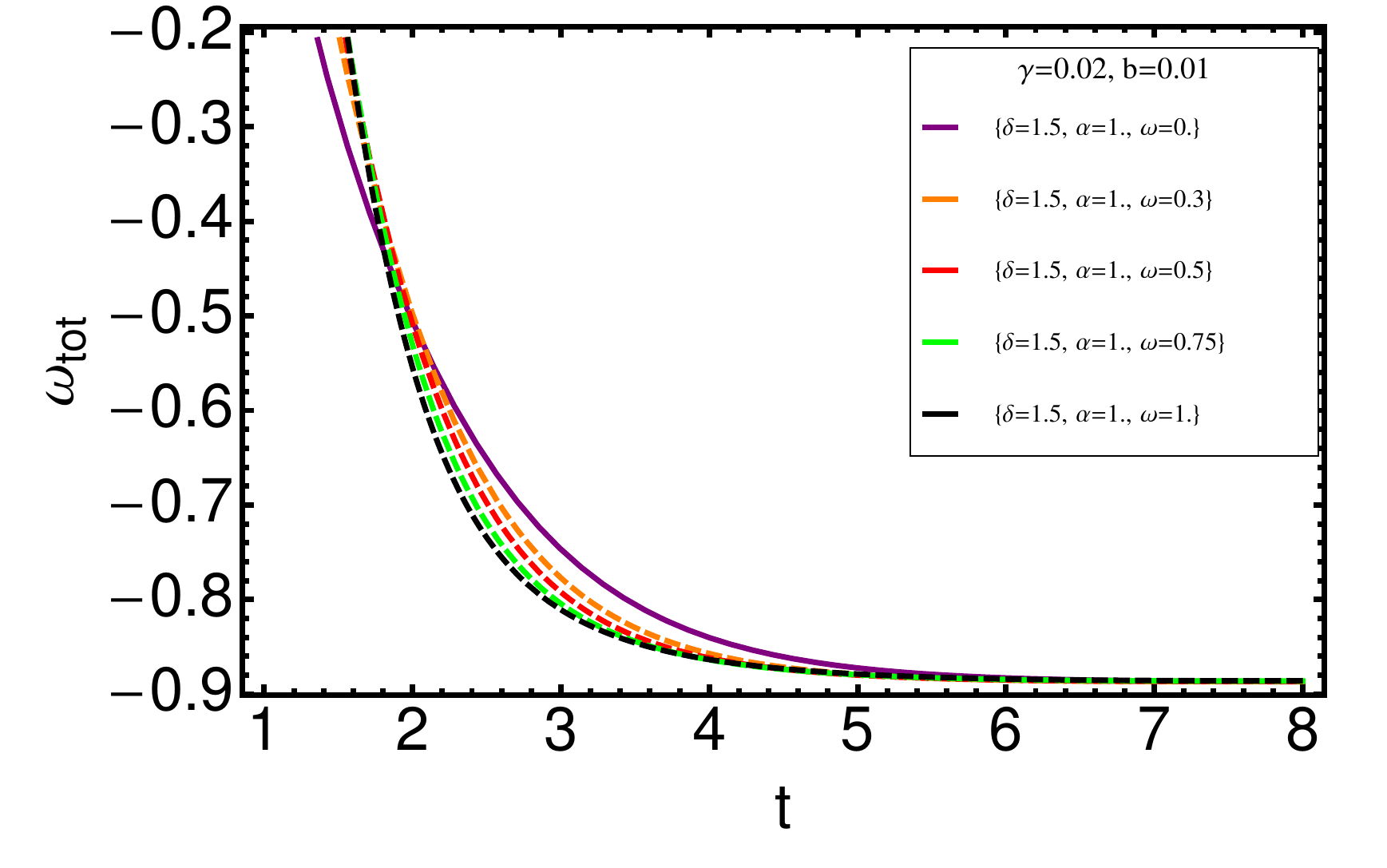}\\
\includegraphics[width=50 mm]{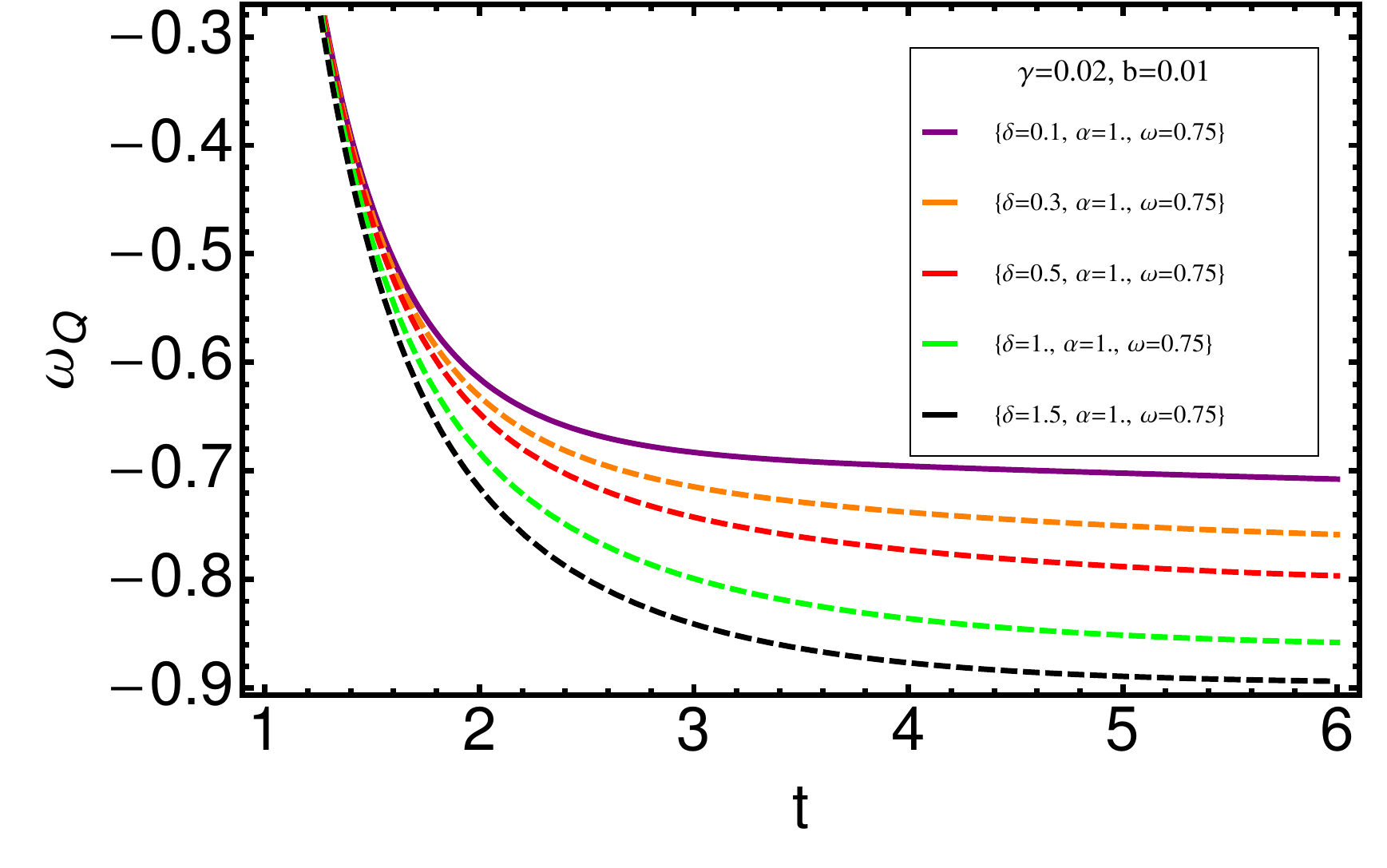} &
\includegraphics[width=50 mm]{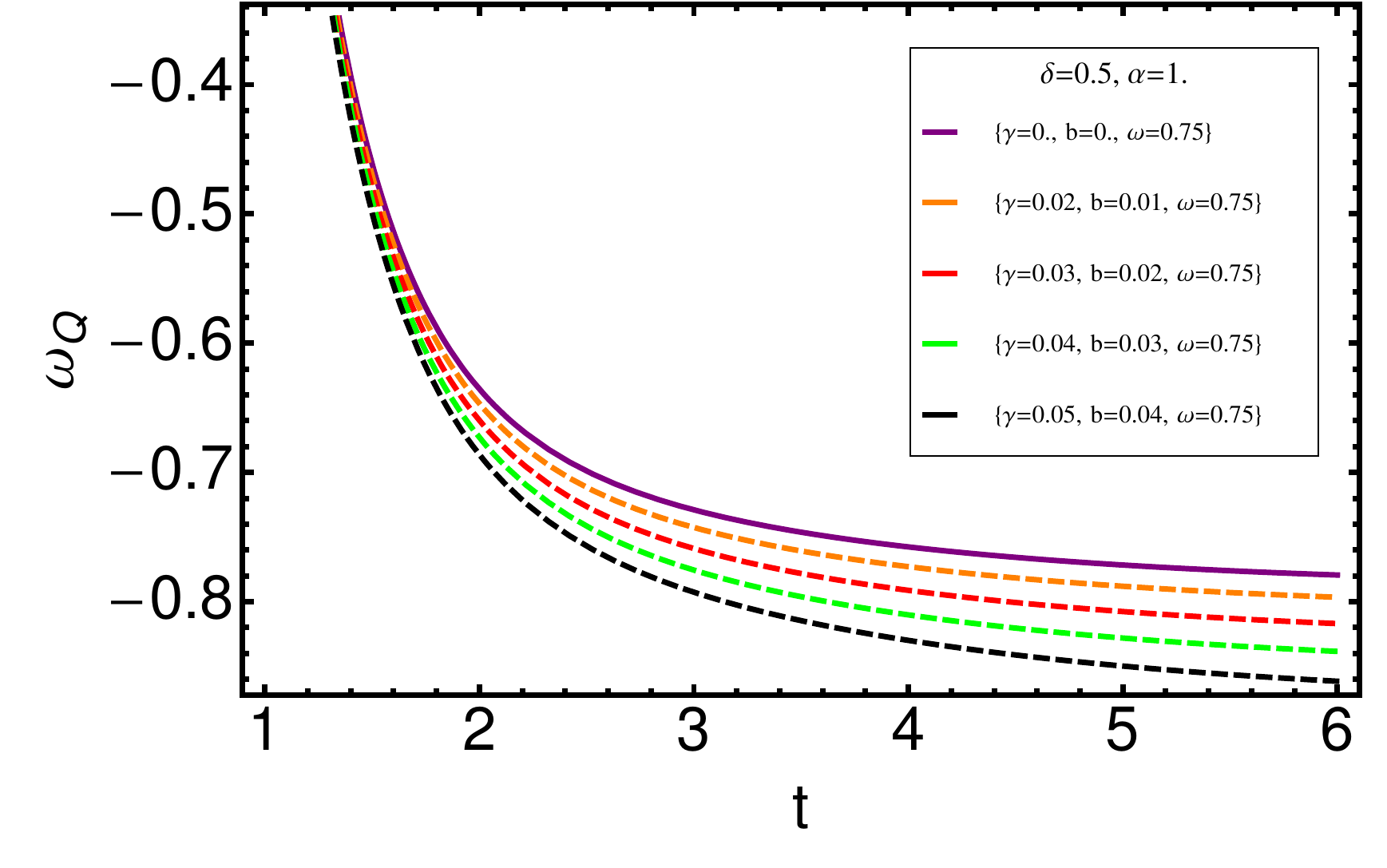}&
\includegraphics[width=50 mm]{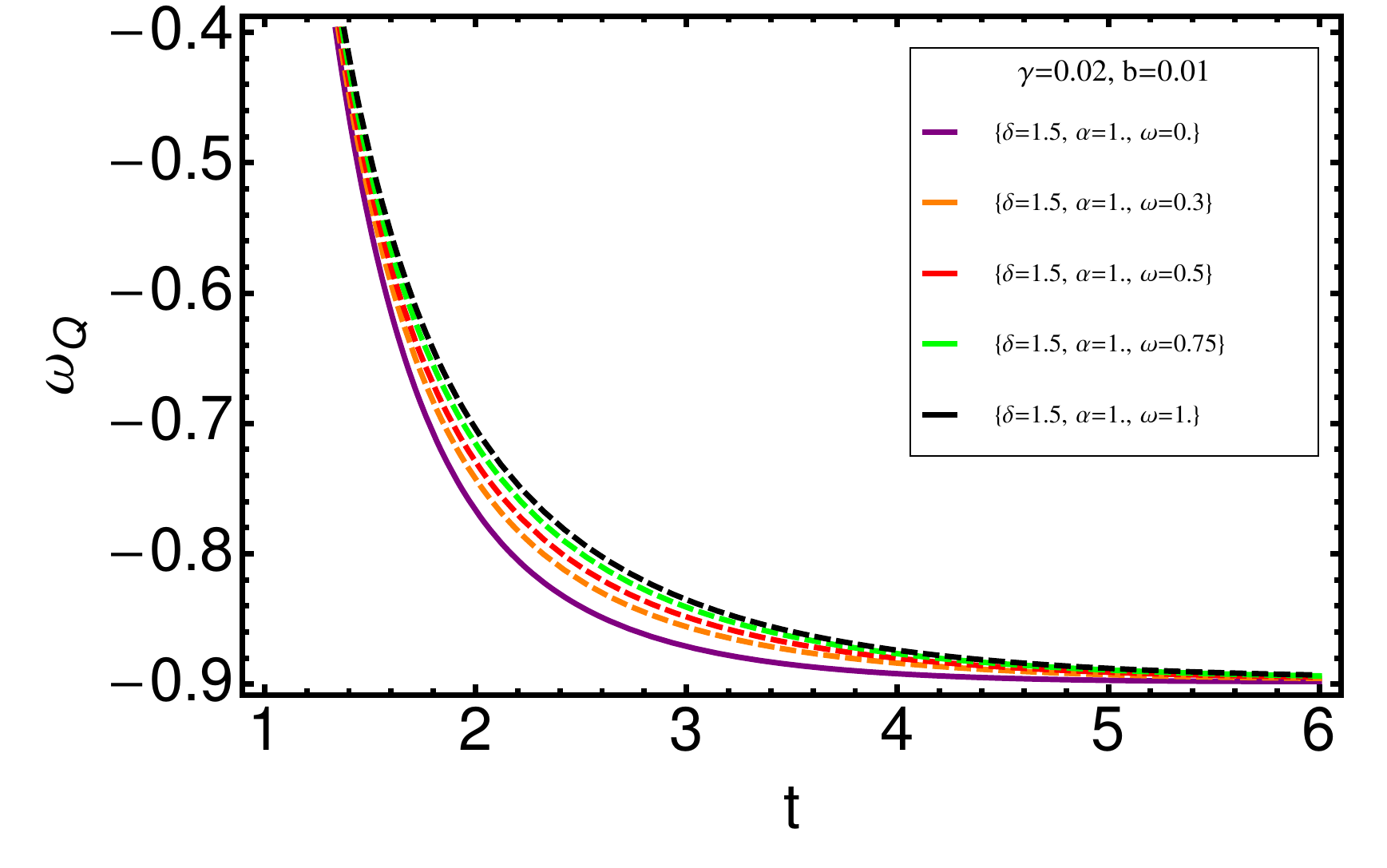}\\
 \end{array}$
 \end{center}
\caption{Behavior of EoS parameter $\omega_{tot}$ and $\omega_{Q}$ against $t$ for Model 4.}
 \label{fig:2}
\end{figure}

We start the analysis of the model 5 from discussions about graphical behavior of the Hubble parameter and deceleration parameter presented in Fig. \ref{fig:1}. The Hubble parameter is a decreasing function and gets constant value at relatively far future. At the top panel we consider three cases corresponding to the behavior of the Hubble parameter. From the first plot we see that when $\gamma=0.02$, $b=0.01$ and $\omega_{b}=0.75$ with increasing $\delta$ we will increase the value of the Hubble parameter. The middle plot represents behavior of Hubble parameter as a function of interaction parameters $b$ and $\gamma$. We see that with an increasing numerical values of the parameters we increase numerical value of the Hubble parameter, when other parameters $\delta$ and $\omega_{b}$ are fixed. The third plot presents behavior of the Hubble parameter from the $\omega_{b}$. We see that for later stages of evolution the Hubble parameter is practically does not depend from the $\omega_{b}$. At the bottom panel we represent graphical behavior of the deceleration parameter. We can declare that at early stages of the evolution transition from $q>0$ to $q<0$ can be realised. The first plot indicates the strong dependence of the $q$ from the $\delta$ for $\gamma=0.02$, $\beta=0.01$ and $\omega_{b}=0.75$. An increasing the numerical value of the $\delta$ decreases the numerical value of the $q$. We also see that for the later stages of evolution $q$ increases and becomes a constant. From the middle plot we can obtain information about behavior of the $q$ as a function of $\gamma$ and $b$. For early stages of the evolution we do not observe any dependence, which becomes apparent only for the later stages of evolution. With an increasing both $\gamma$ and $b$ we decrease the value of $q$. Finally, the last plot shows $q$ dependence from $\omega_{b}$. We see almost independent behavior (Fig. \ref{fig:1}) from $\omega_{b}$. We also investigate the behaviors of $\omega_{tot}$ and $\omega_{Q}$ and results are presented in Fig. \ref{fig:2}. The behavior of $\omega_{tot}$ predicts quintessence-like behavior for the Universe. Comparision of the results of this model with the Model 1, where constant $\Lambda$ were assumed, showed that in this model $\omega_{tot}$ remains strictly above $-1$. Also the value of the $q$ for the later stages of the evolution is higher then in Model 1.
  
\subsection{\large{Model 5: $Q=3Hb\rho+\gamma \dot{\rho}$}}
\begin{figure}[h!]
 \begin{center}$
 \begin{array}{cccc}
\includegraphics[width=50 mm]{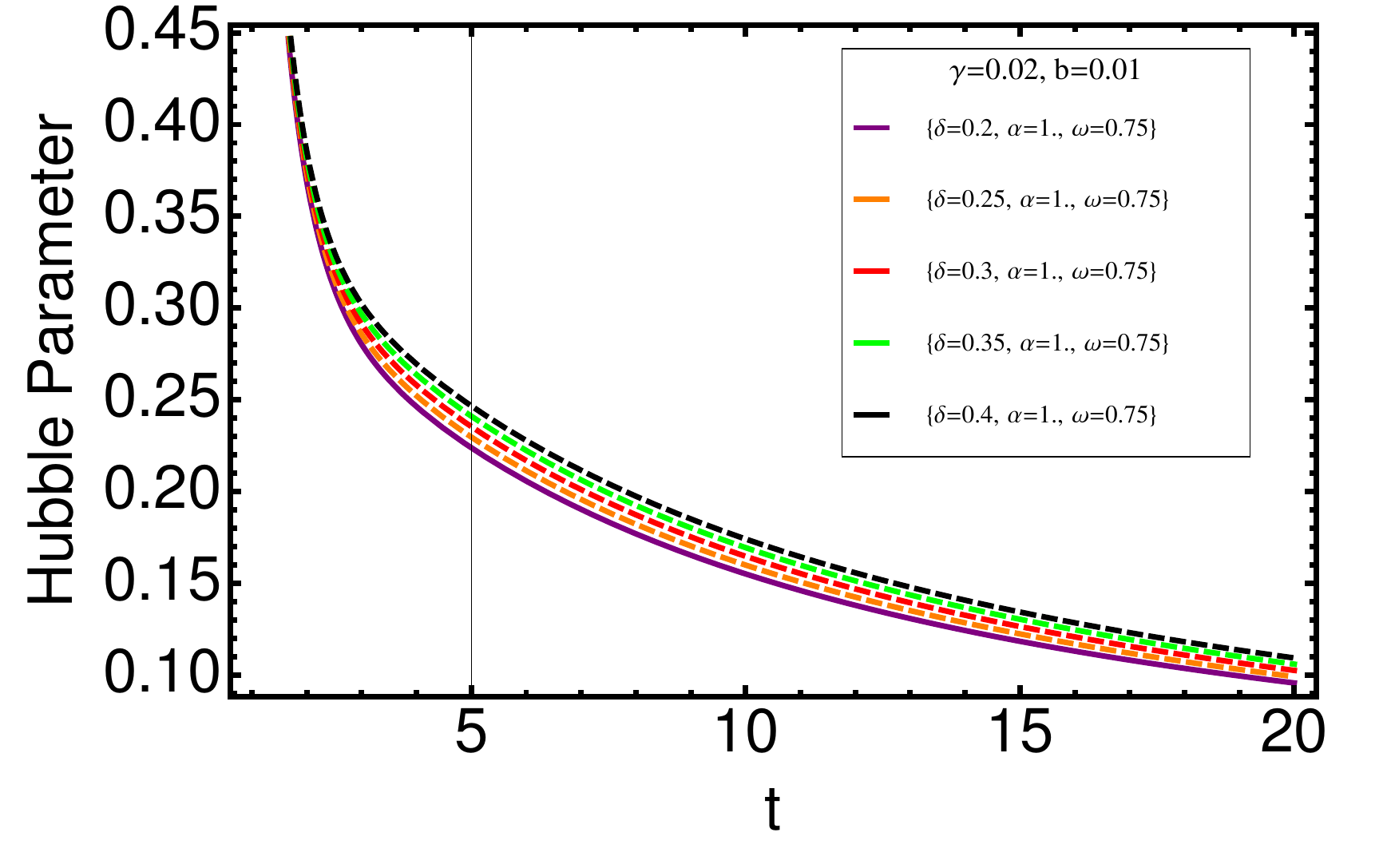} &
\includegraphics[width=50 mm]{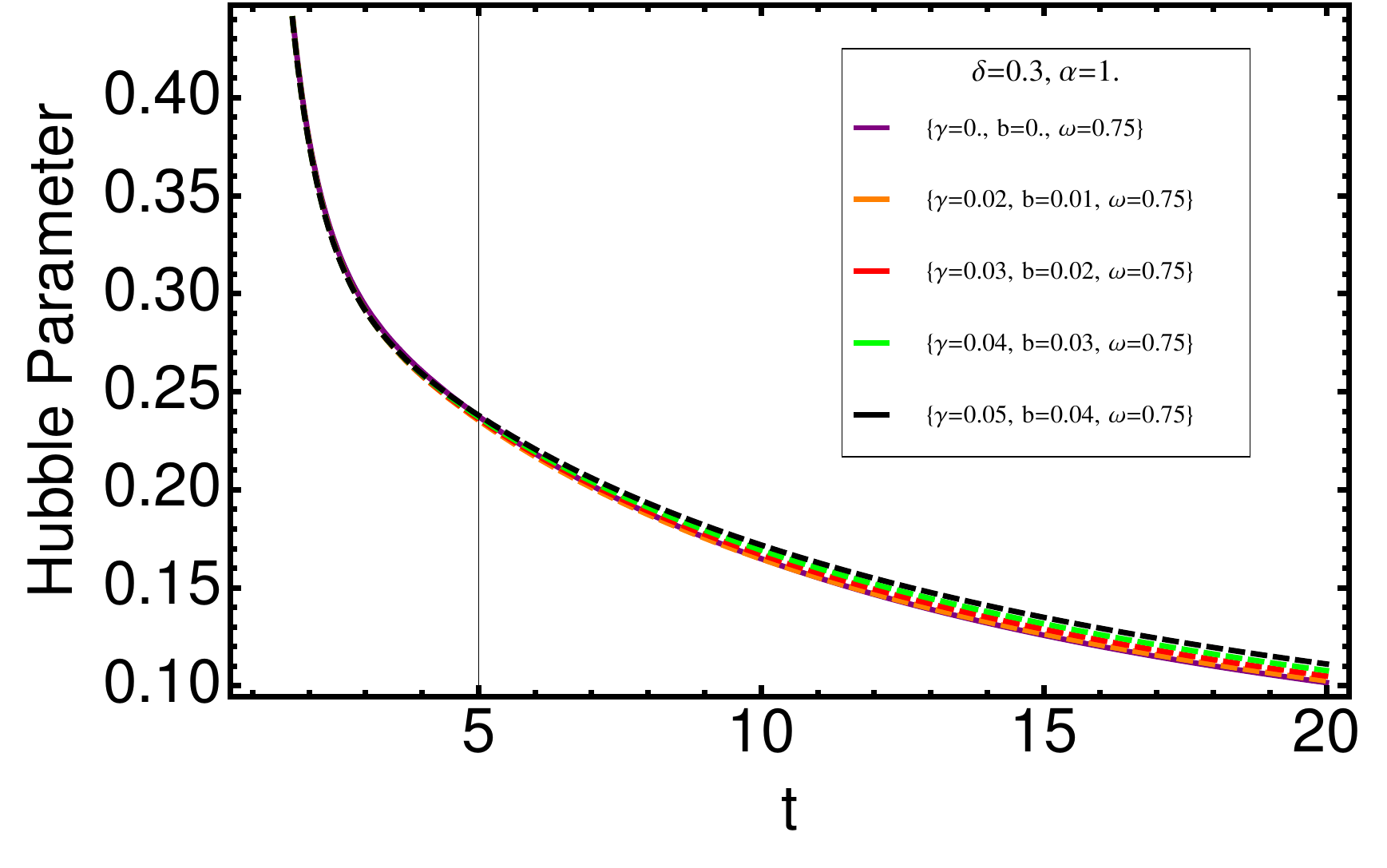}&
\includegraphics[width=50 mm]{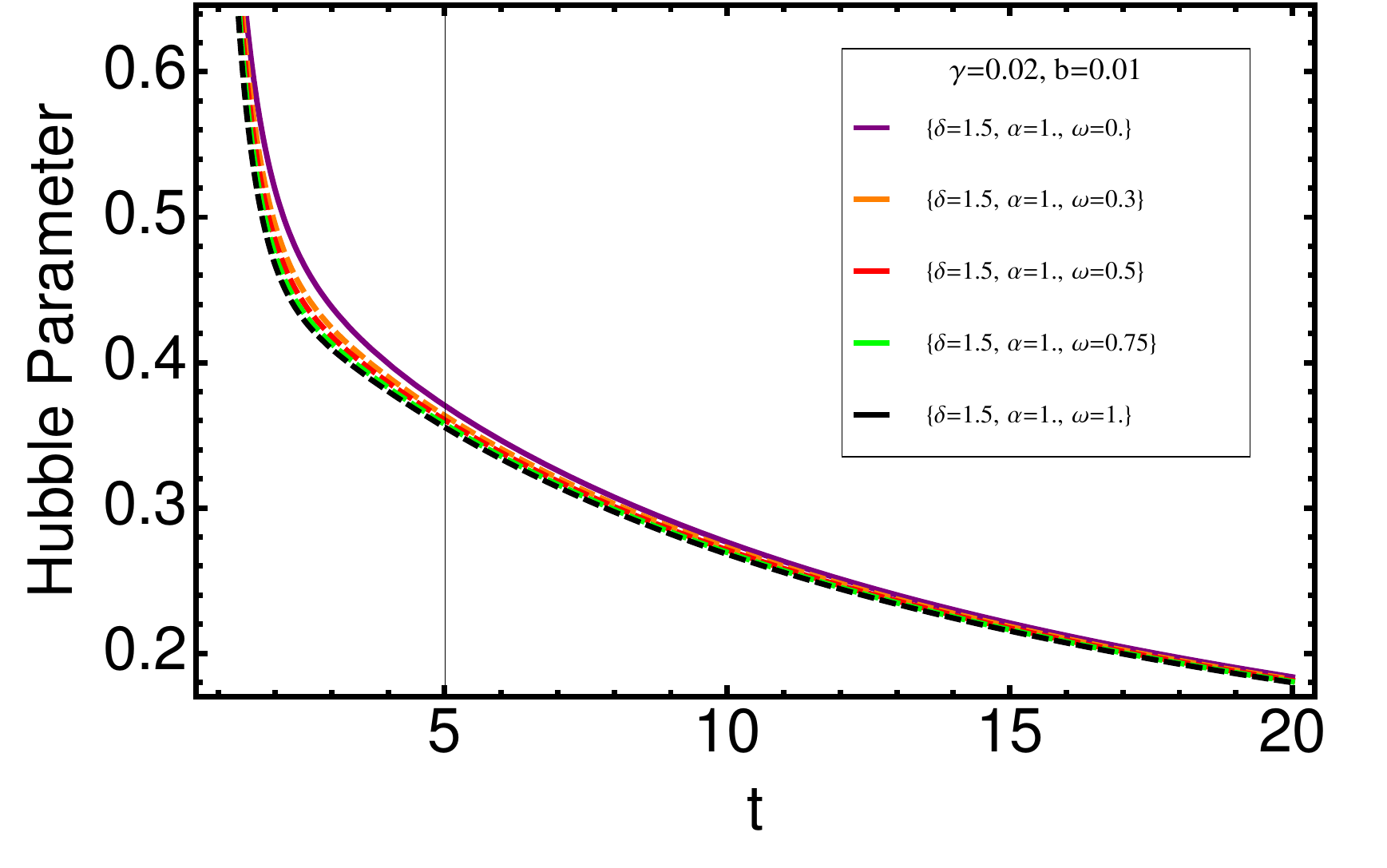}\\
\includegraphics[width=50 mm]{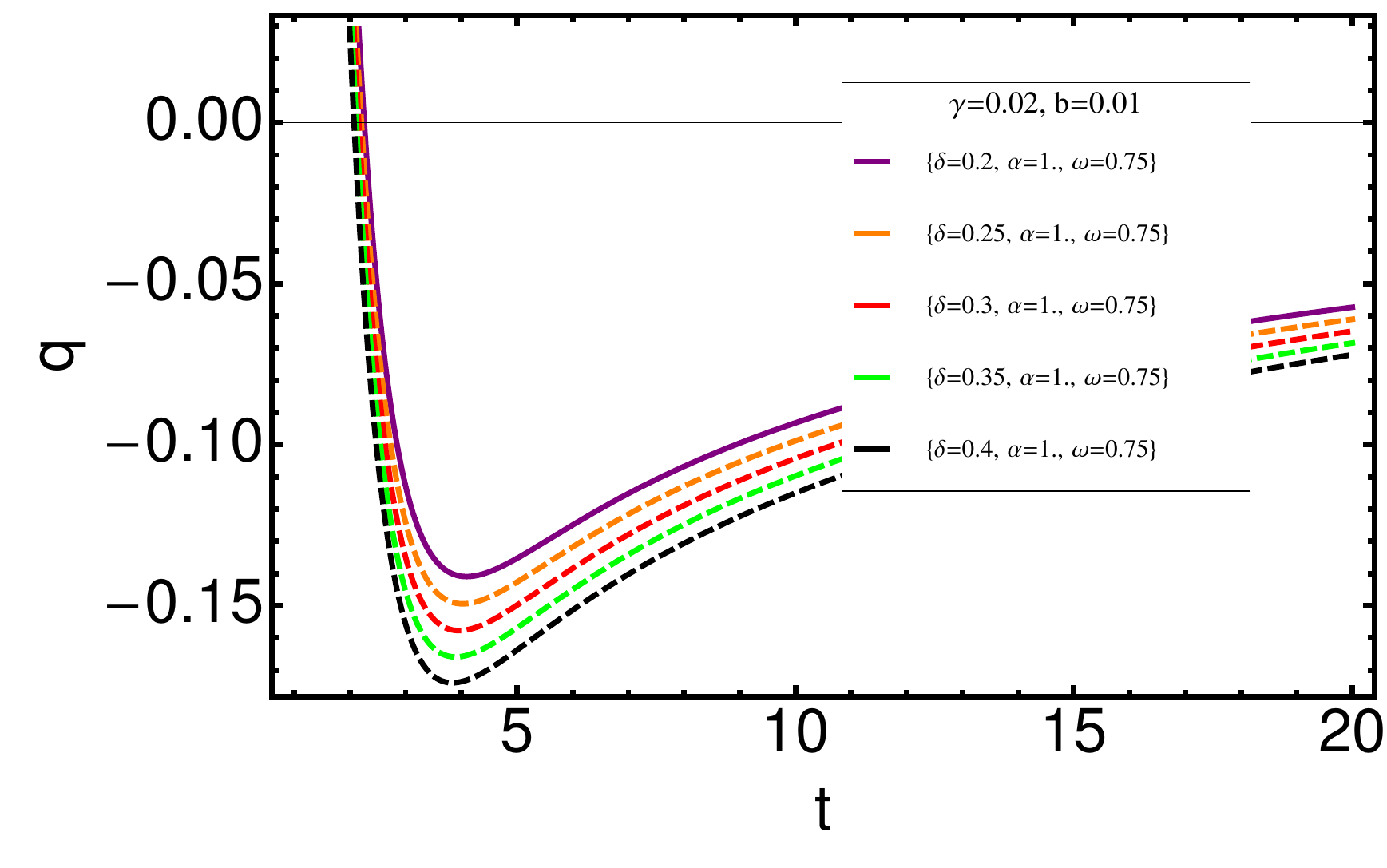} &
\includegraphics[width=50 mm]{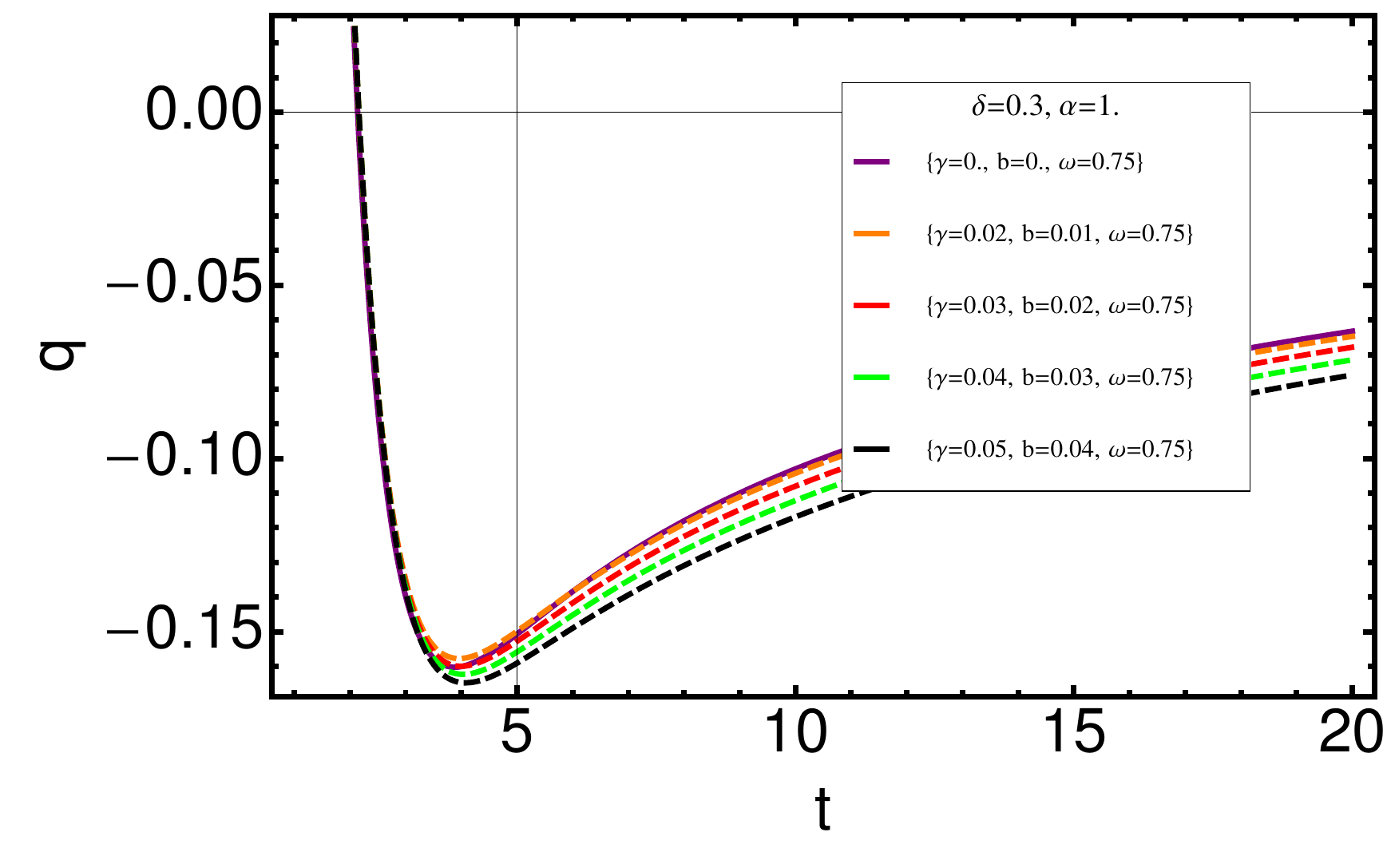}&
\includegraphics[width=50 mm]{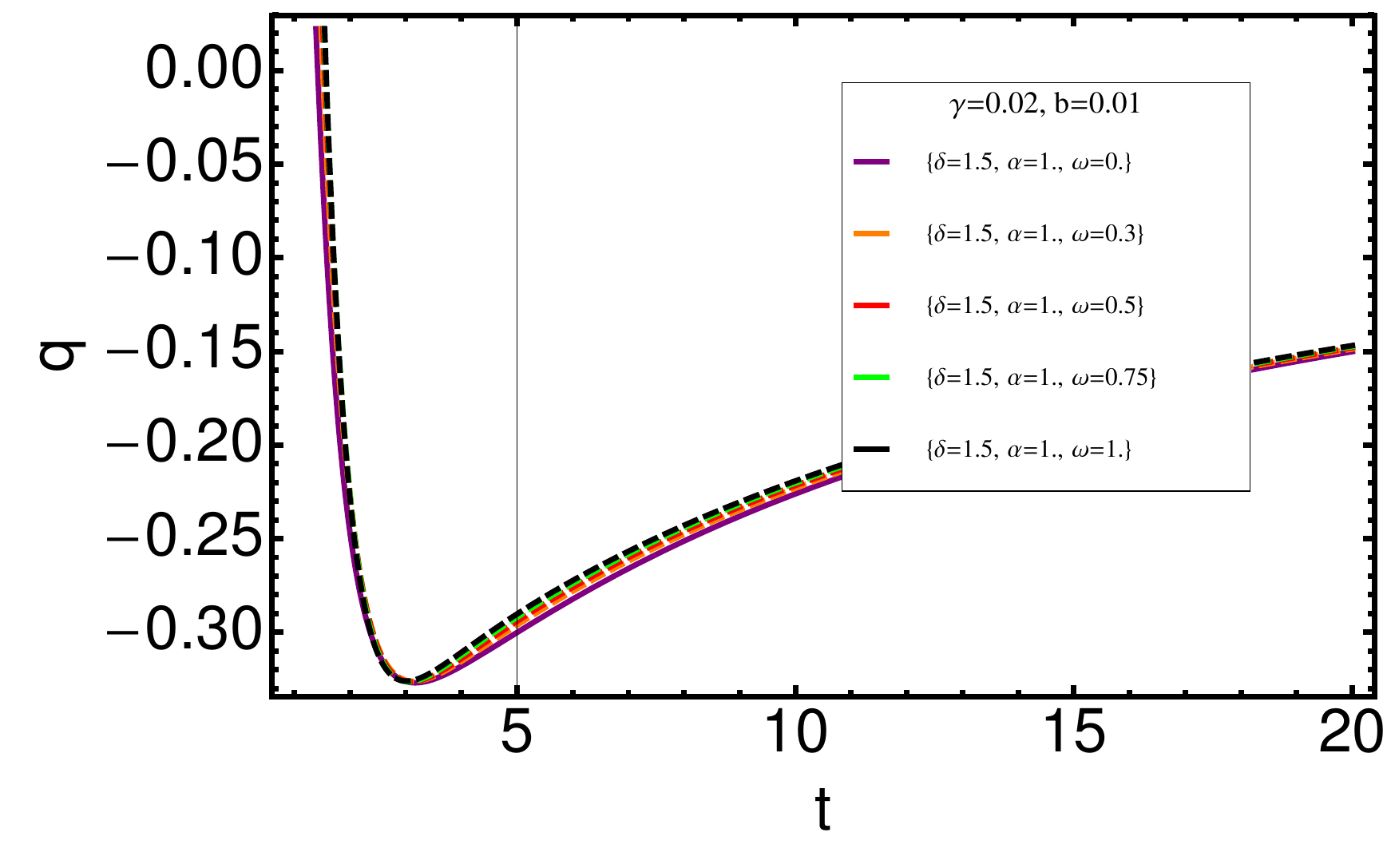}
 \end{array}$
 \end{center}
\caption{Behavior of Hubble parameter $H$ and deceleration parameter $q$ against $t$ for Model 5.}
 \label{fig:3}
\end{figure}

Here we will analyse the model where the interaction between DE and DM is the form $Q=3Hb\rho+\gamma \dot{\rho}$. The results of the Model 5 should be compared with Model 2, where instead of the varying $\Lambda(t)$ the constant $\Lambda$ were considered. The Hubble parameter is a decreasing function and it becomes a constant for the later stages of the evolution. Three plots of the top panel of Fig. \ref{fig:3} give a general idea about the behavior of the Hubble parameter as a function of $\delta$ (first plot) for $\gamma=0.02$, $\beta=0.01$ and $\omega_{b}=0.75$. From the  middle plot we have information about the Hubble parameter describing dependence of it from the interaction parameters, when the numerical values of the $\delta$ and $\omega_{b}$ are taken in advance. The last plot gives time evolution of the Hubble parameter as a function from $\omega_{b}$. The bottom of the same Figure dedicated to the deceleration parameter $q$. Like to the other models, in this case as well, we have the Universe where transition to $q<0$ is possible and we see that it can realised for early stages of evolution. To understand behavior of $q$ from the model parameters we considered 3 cases allowing a variation one of the model parameters. We see that for later stages of the evolution $q$ from decreasing function becomes an increasing function and for very late stages becomes a constant. Moreover, compared with observational facts known about $q$, we can conclude that this model is also can be considered as a good model.  Comparision between Model 2 and Model 5, allows us to see that when $\Lambda$ is a constant, then $\omega_{tot}$ for later stages is $-1$, while for the varying model with the same form of the interaction term $\omega_{tot}>-1$ indicating quintessence-like Universe. For both models $\omega_{tot}$ and $\omega_{Q}$ are decreasing functions over the time and a constant for later stages of the evolution. For the Model 5 numerical value of the $q$ when it is a constant is higher than for the Model 2. In the next section we will examine our last model. Comparision between it and Model 3 will take a place to see differences between varying $\Lambda(t)$ and constant $\Lambda$ cases. For all cases to make a real comparison of the models the parameters describing the models assumed to be the same.
 
\begin{figure}[h!]
 \begin{center}$
 \begin{array}{cccc}
\includegraphics[width=50 mm]{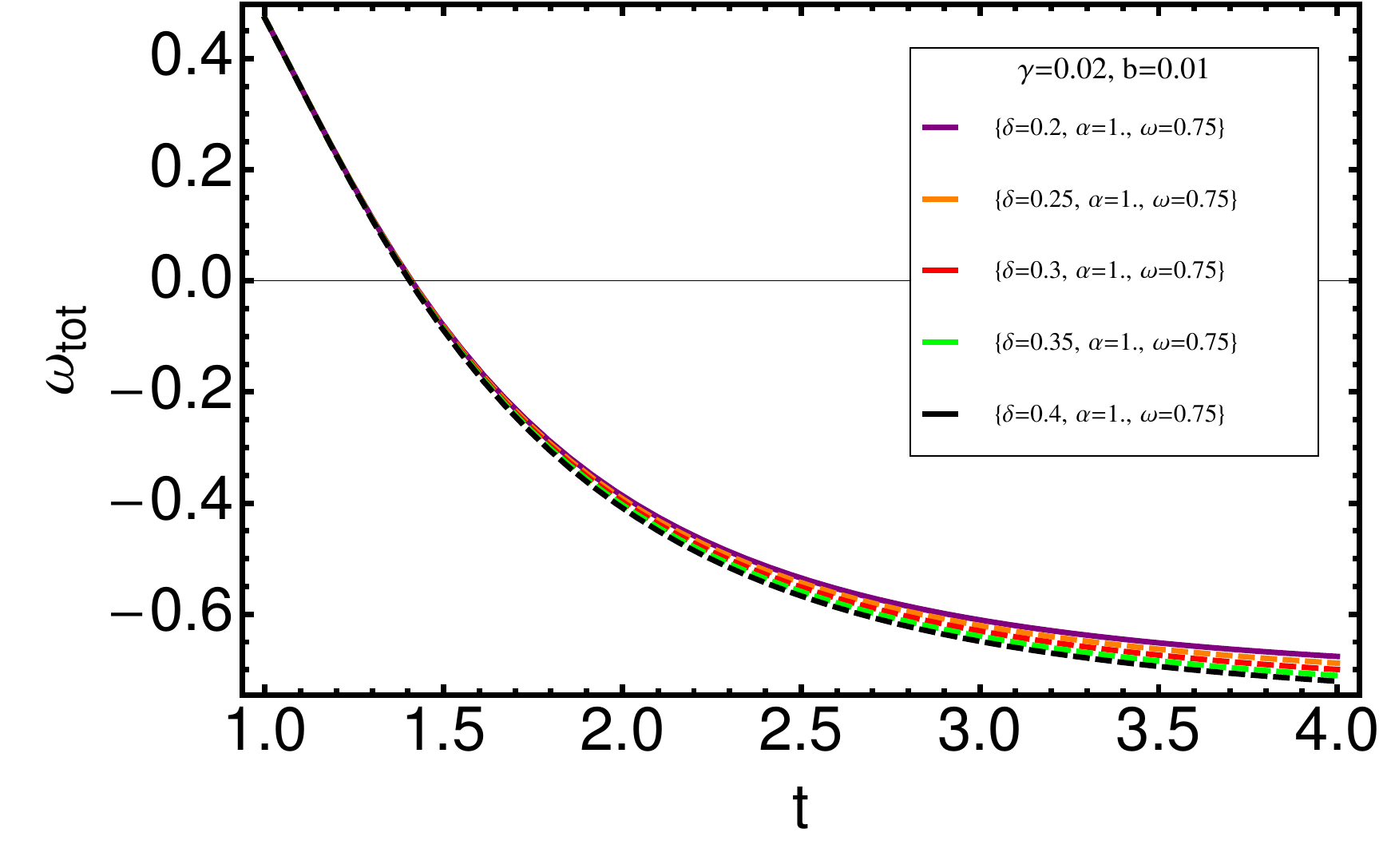} &
\includegraphics[width=50 mm]{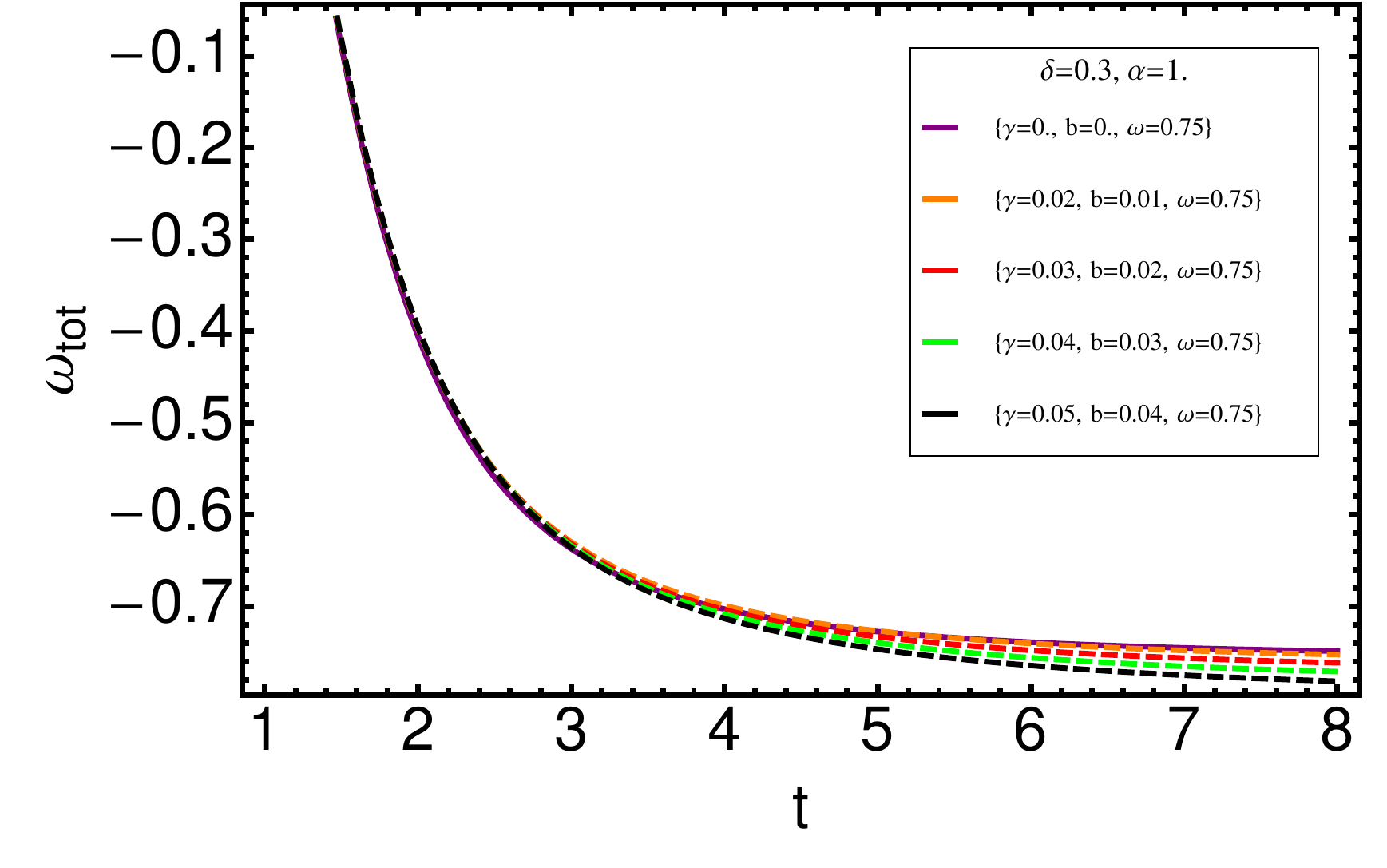}&
\includegraphics[width=50 mm]{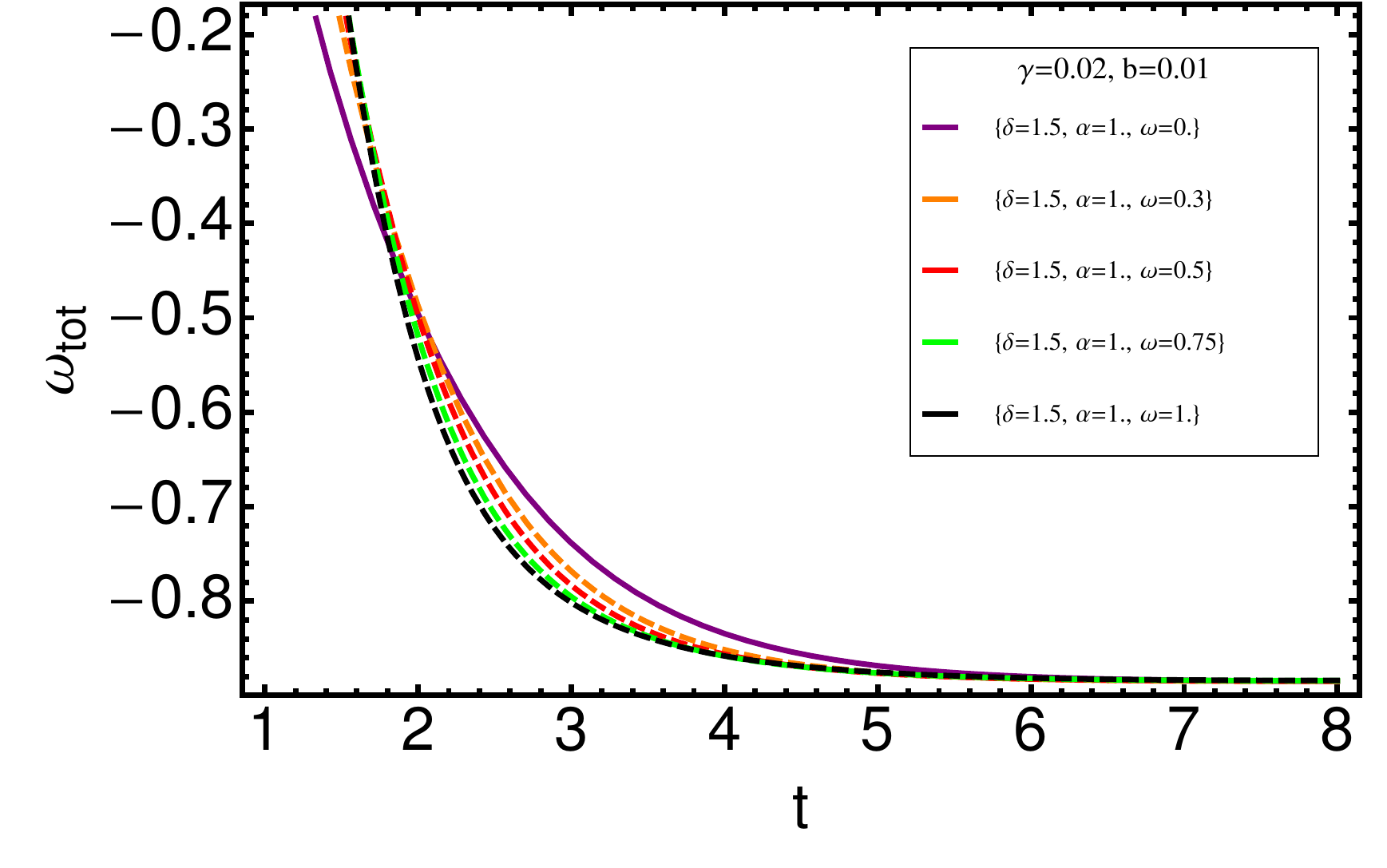}\\
\includegraphics[width=50 mm]{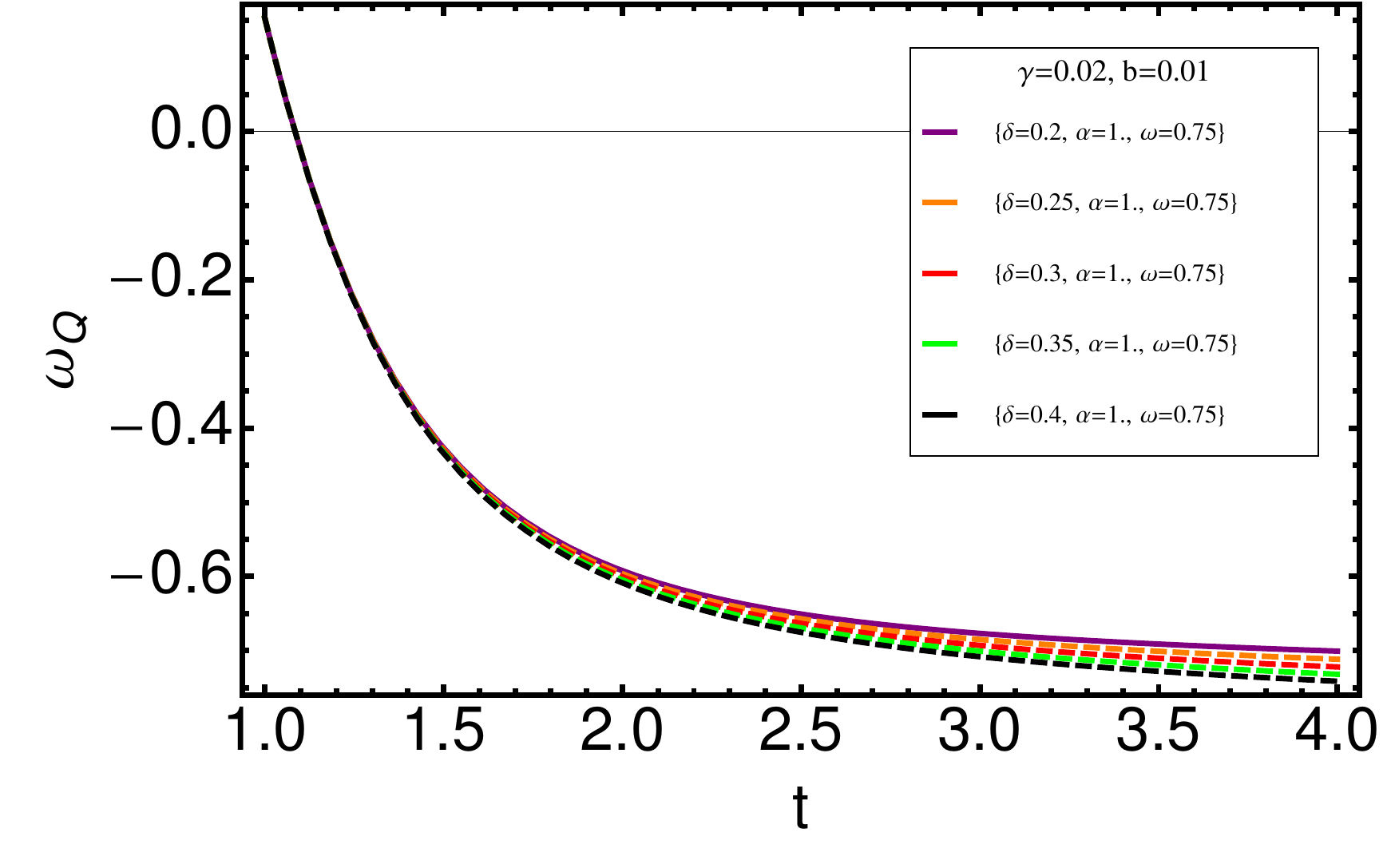} &
\includegraphics[width=50 mm]{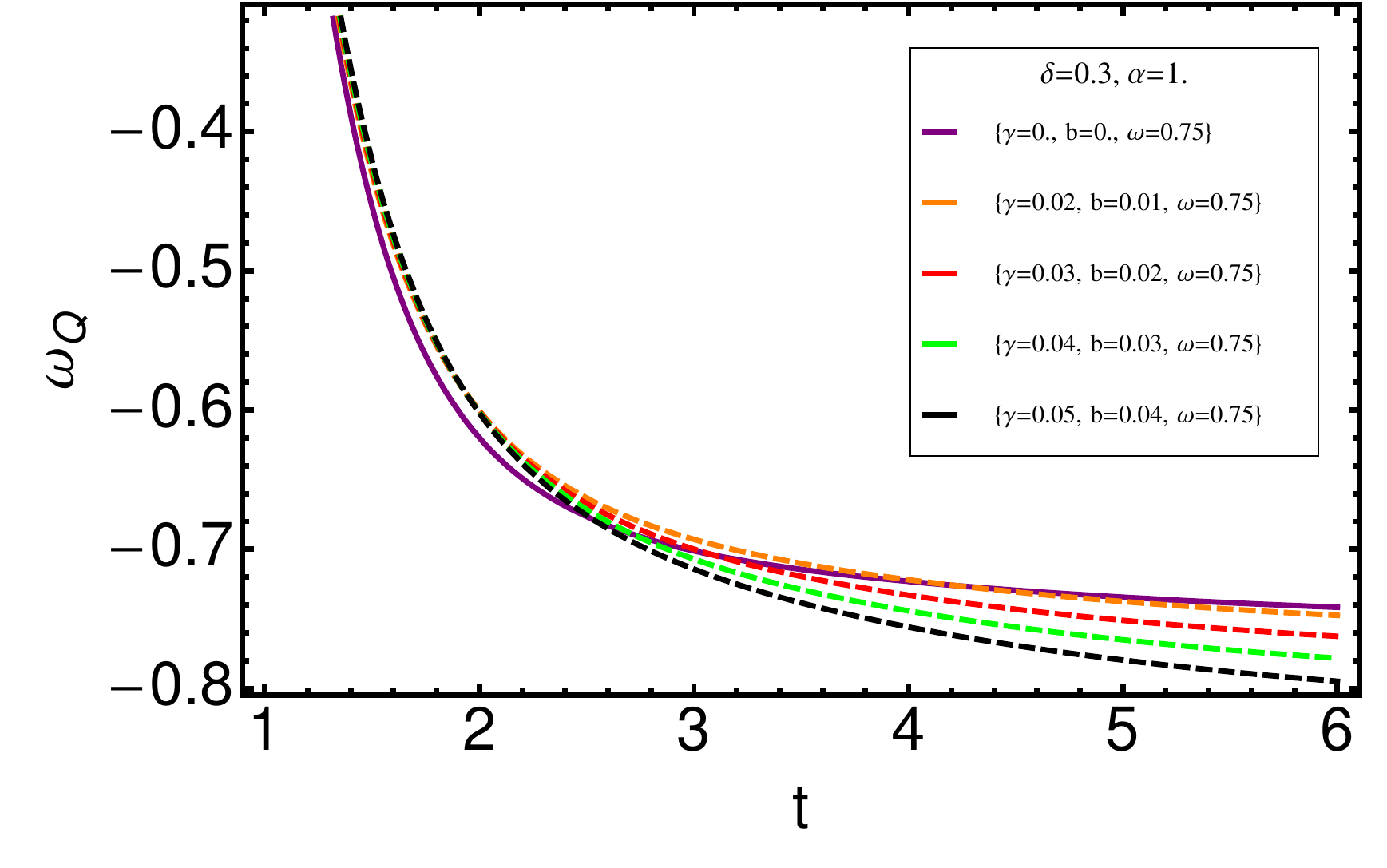}&
\includegraphics[width=50 mm]{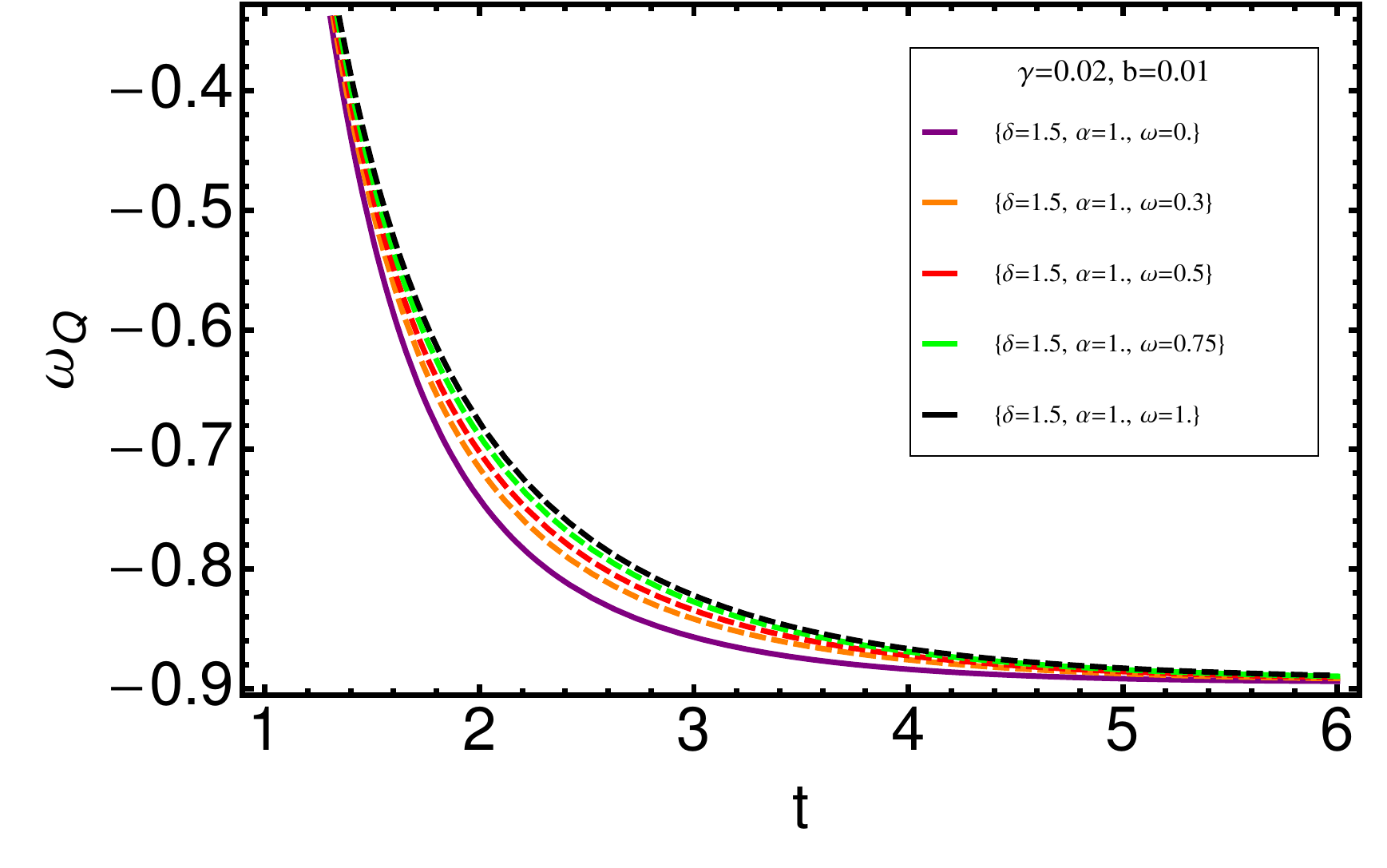}\\
 \end{array}$
 \end{center}
\caption{Behavior of EoS parameter $\omega_{tot}$ and $\omega_{Q}$ against $t$ for Model 5.}
 \label{fig:4}
\end{figure}

\subsection{\large{Model 6: $Q=bH^{1-2\gamma}\rho_{b}^{\gamma}\dot{\phi}^{2}$}}
\begin{figure}[h!]
 \begin{center}$
 \begin{array}{cccc}
\includegraphics[width=50 mm]{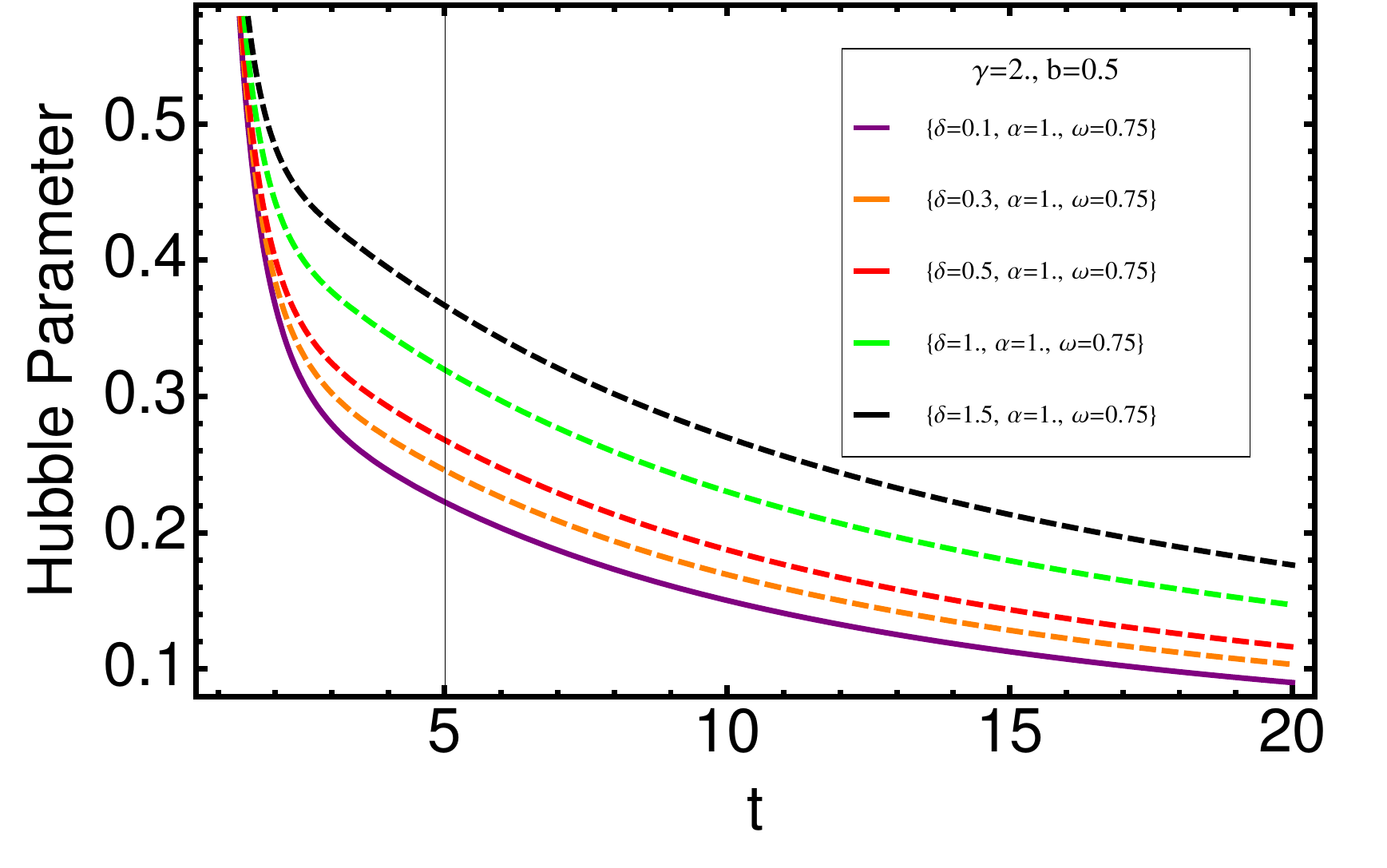} &
\includegraphics[width=50 mm]{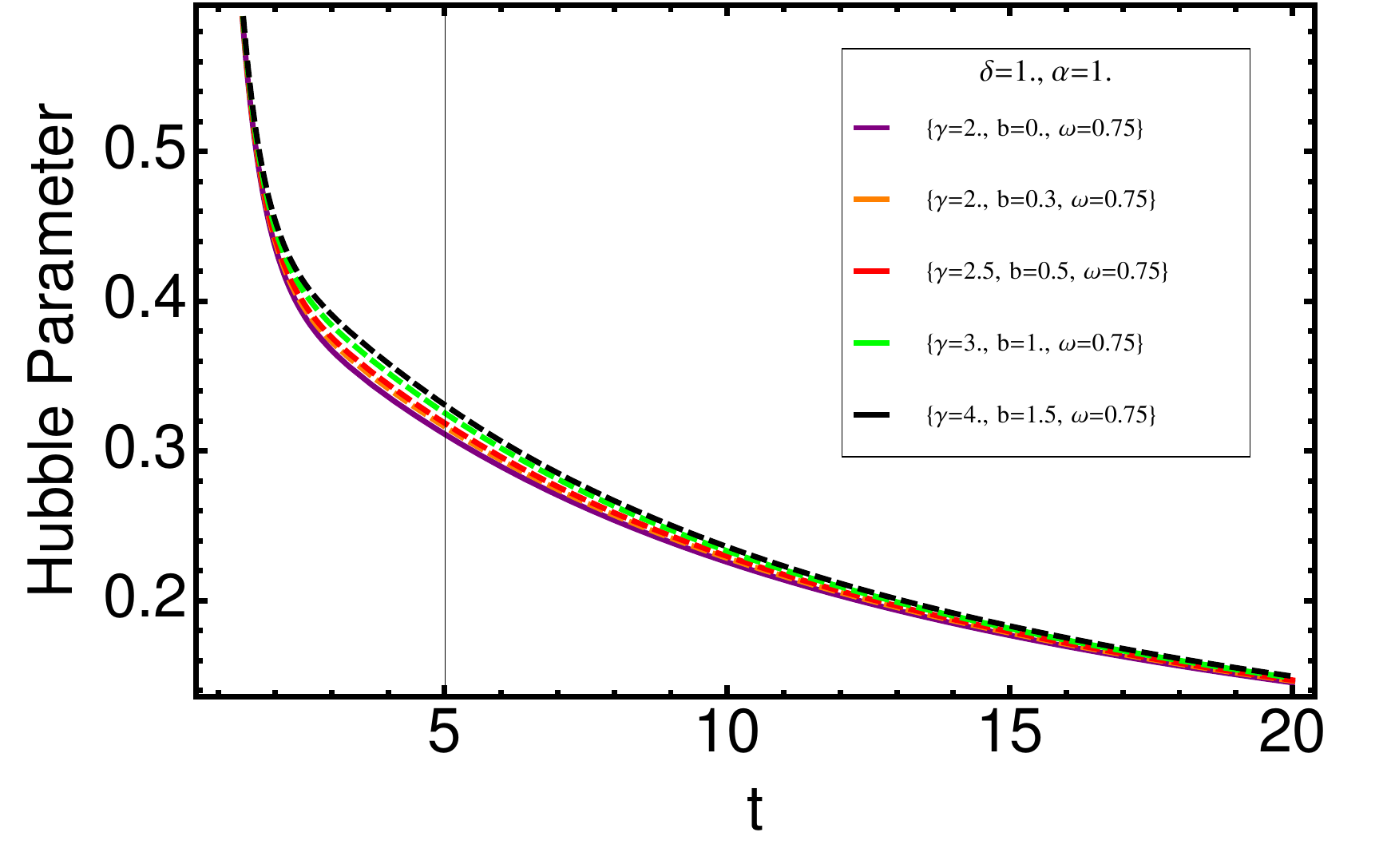}&
\includegraphics[width=50 mm]{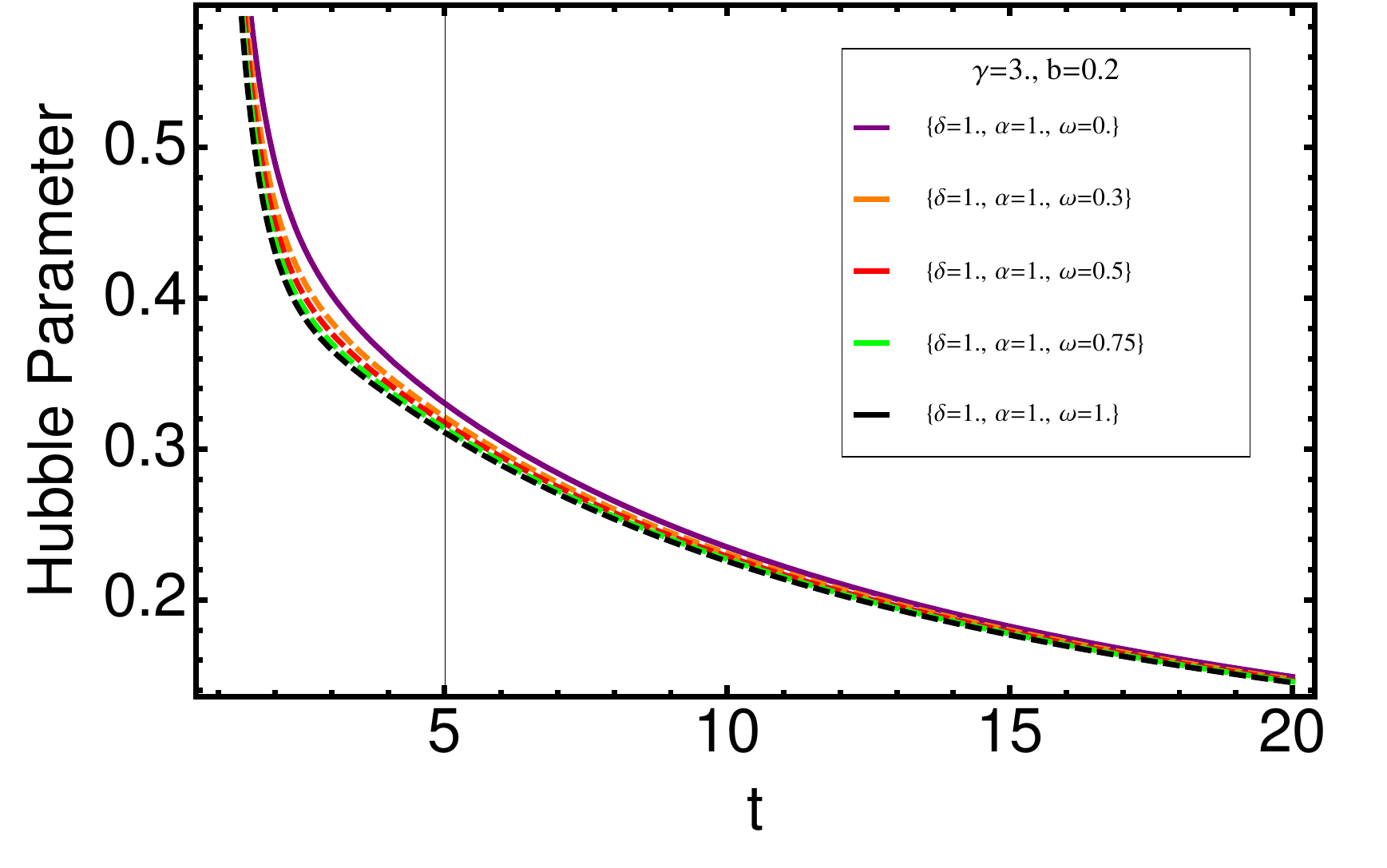}\\
\includegraphics[width=50 mm]{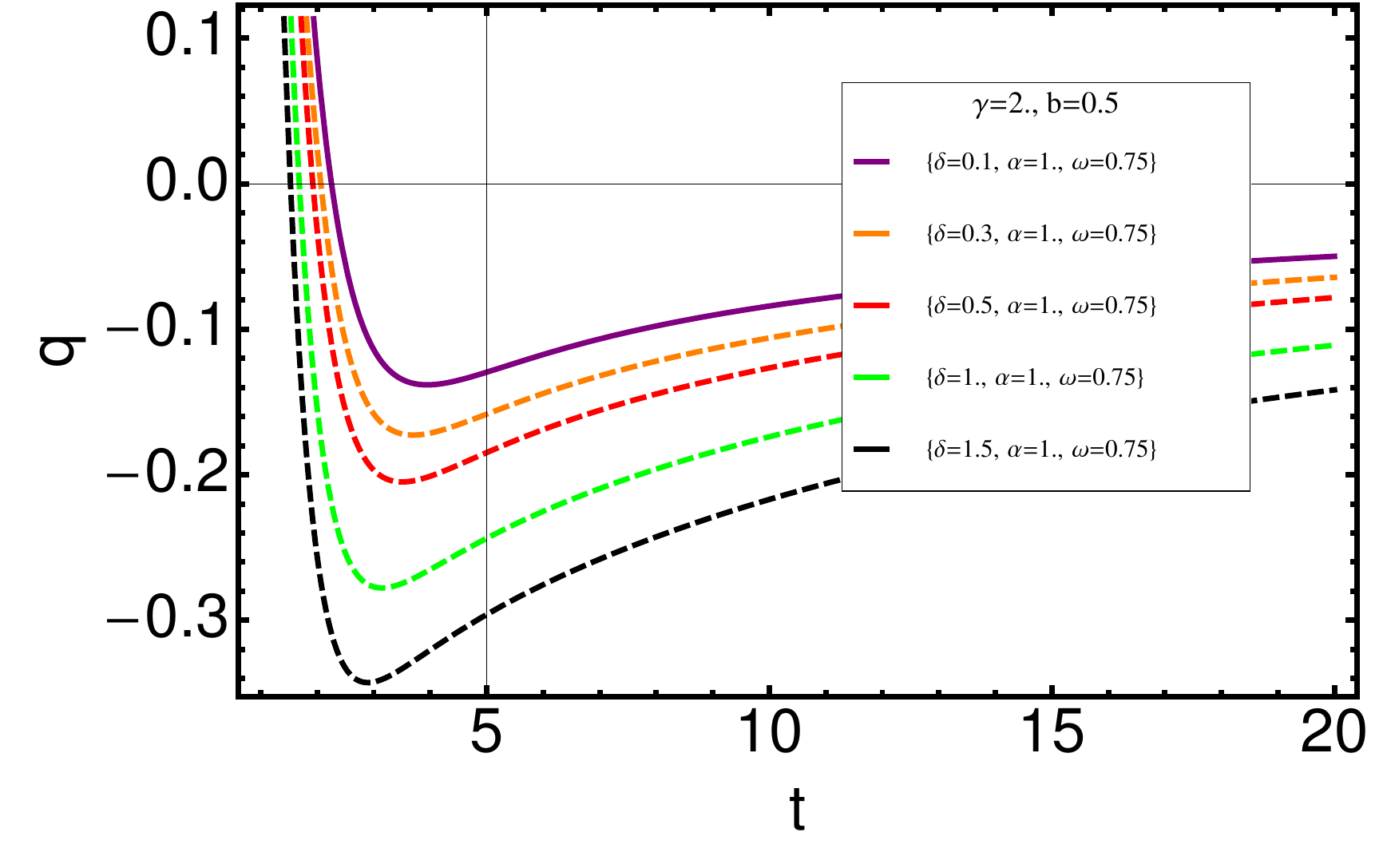} &
\includegraphics[width=50 mm]{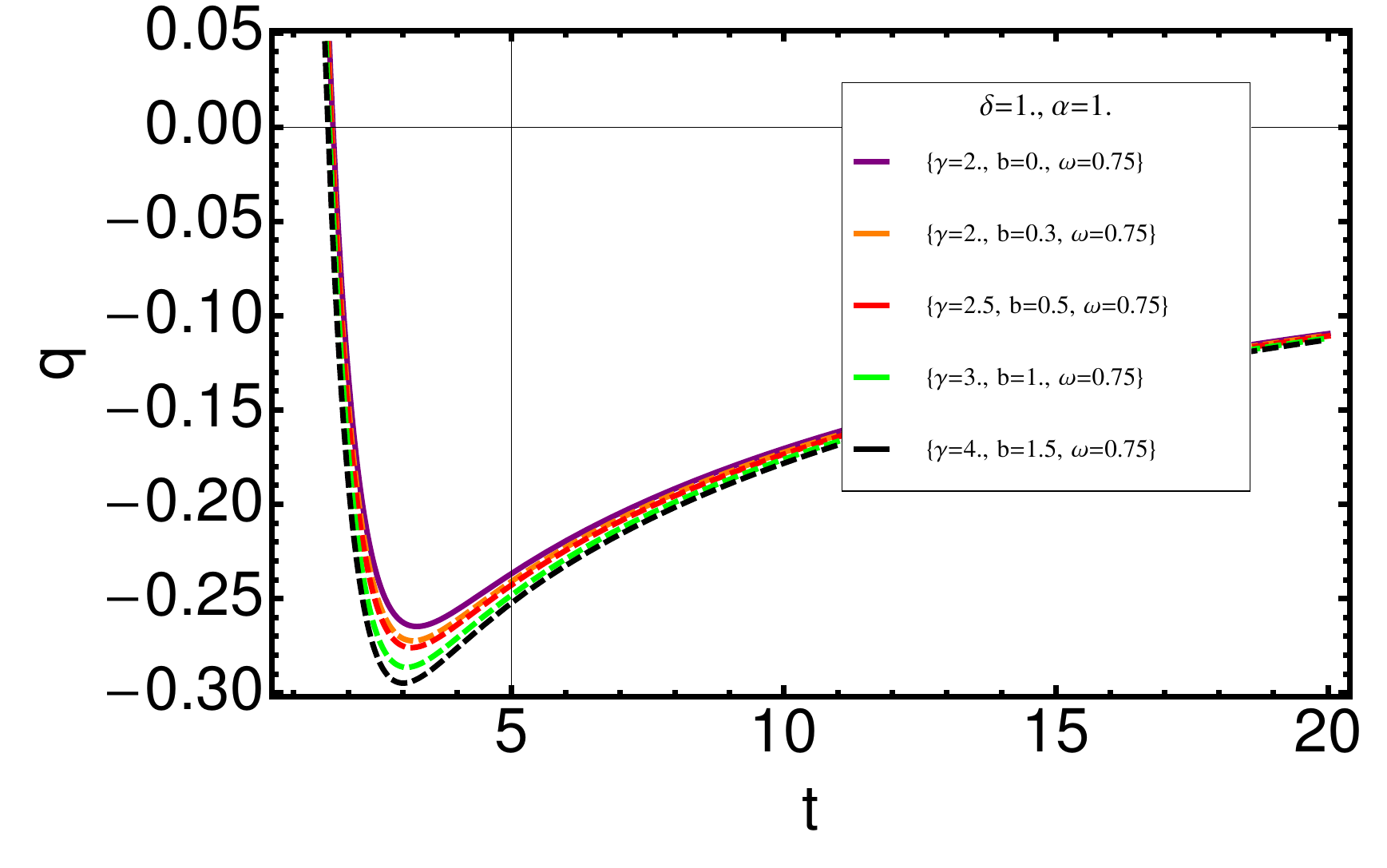}&
\includegraphics[width=50 mm]{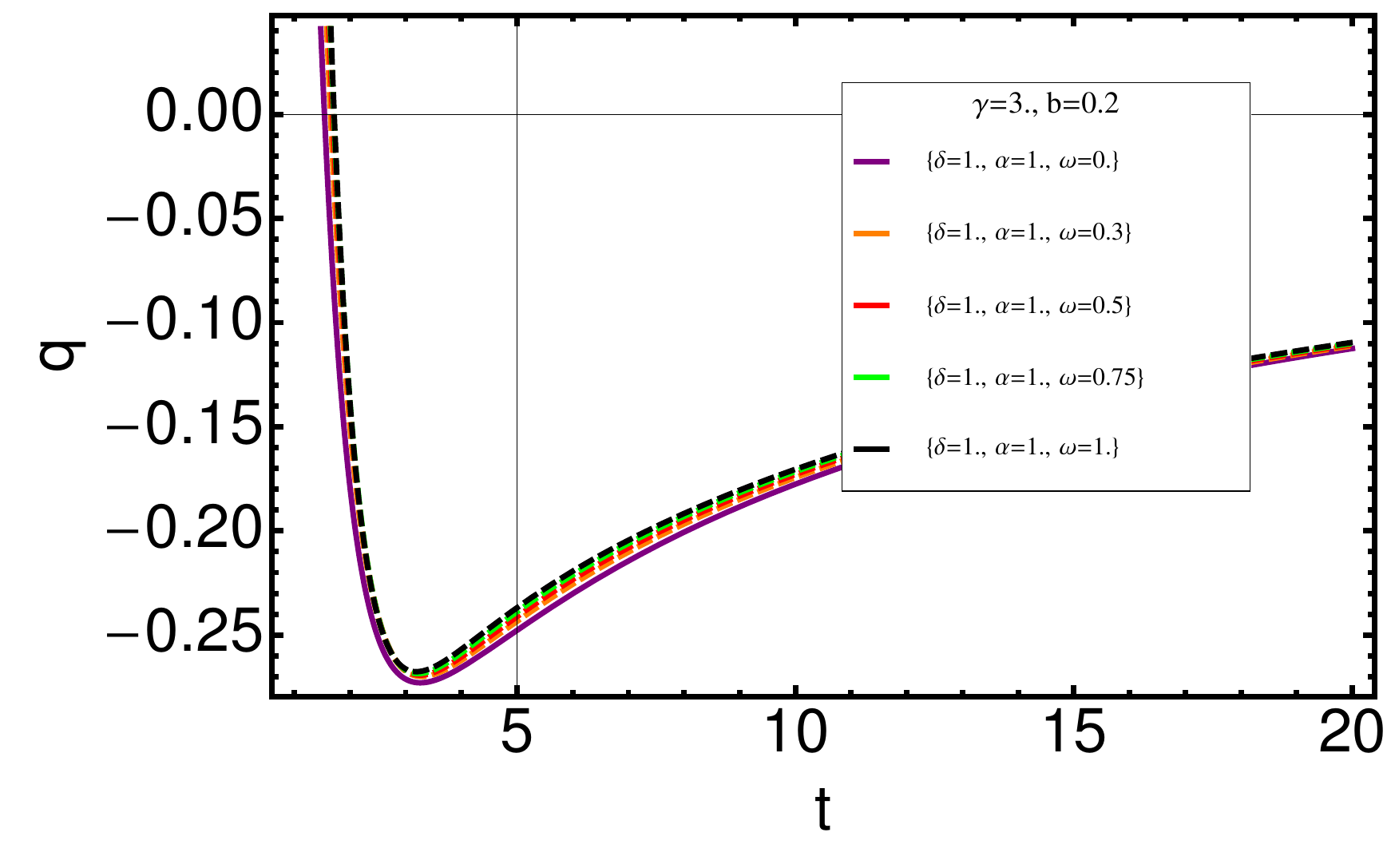}
 \end{array}$
 \end{center}
\caption{Behavior of Hubble parameter $H$ and deceleration parameter $q$ against $t$ for Model 6.}
 \label{fig:5}
\end{figure}

\begin{figure}[h!]
 \begin{center}$
 \begin{array}{cccc}
\includegraphics[width=50 mm]{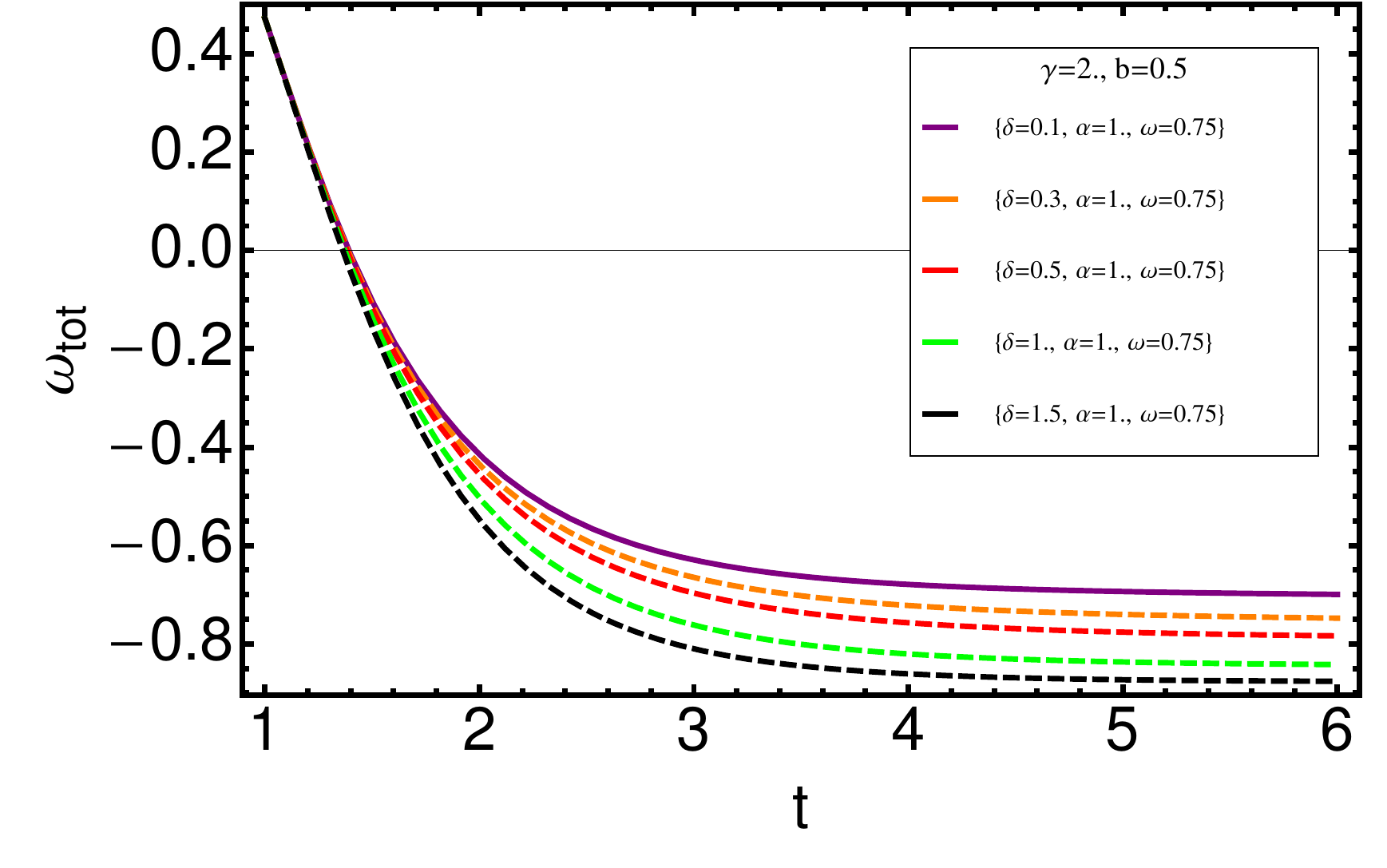} &
\includegraphics[width=50 mm]{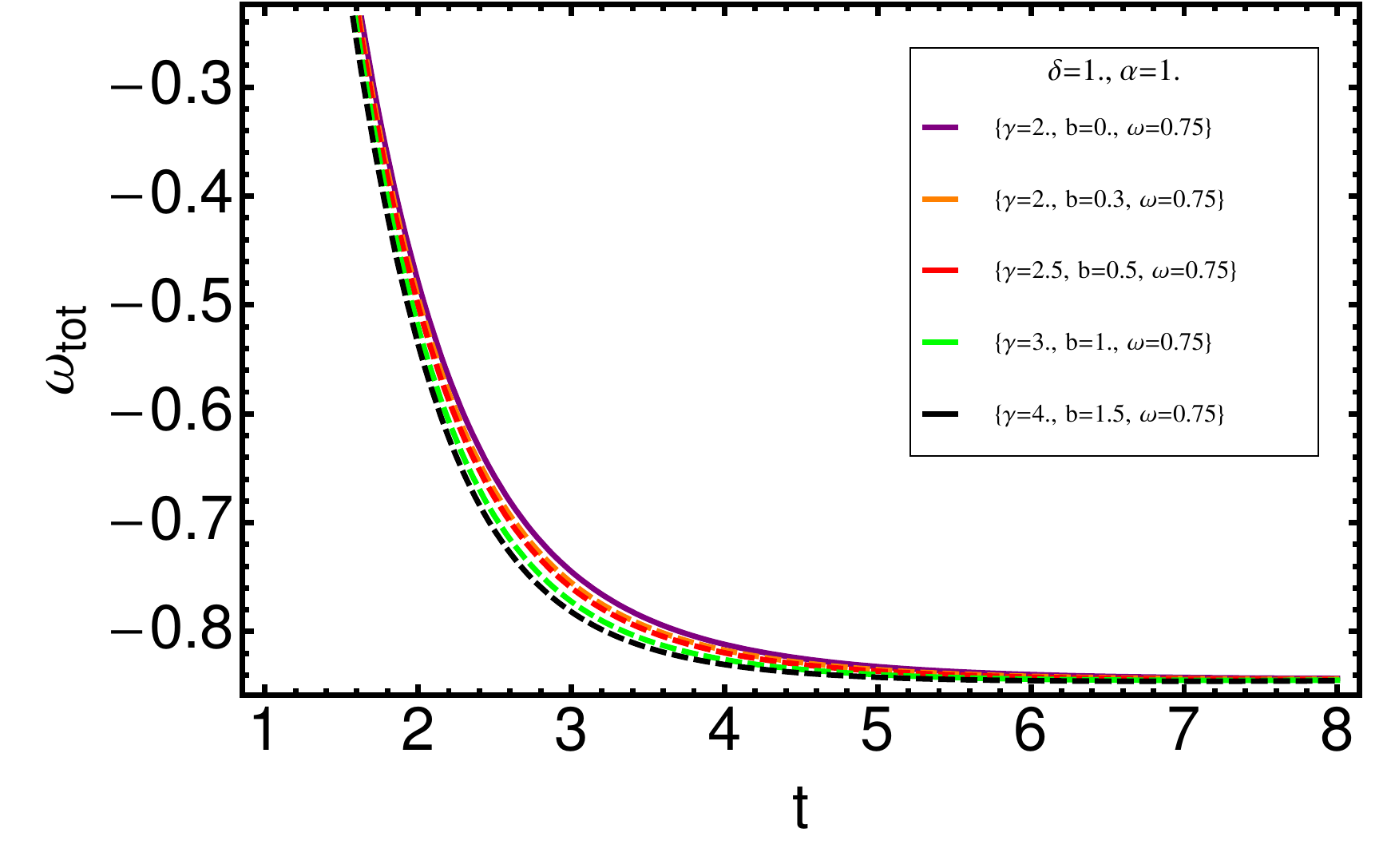}&
\includegraphics[width=50 mm]{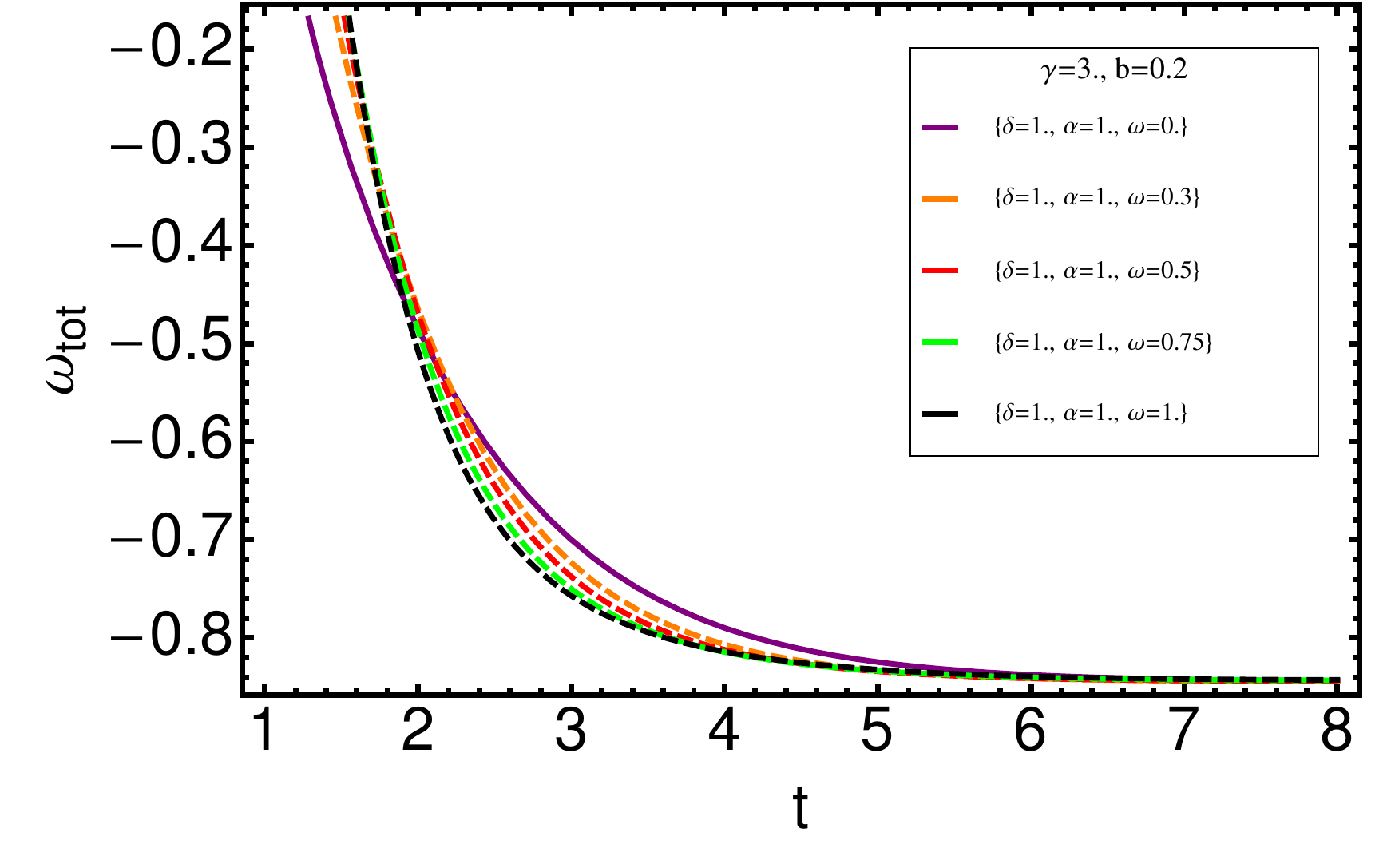}\\
\includegraphics[width=50 mm]{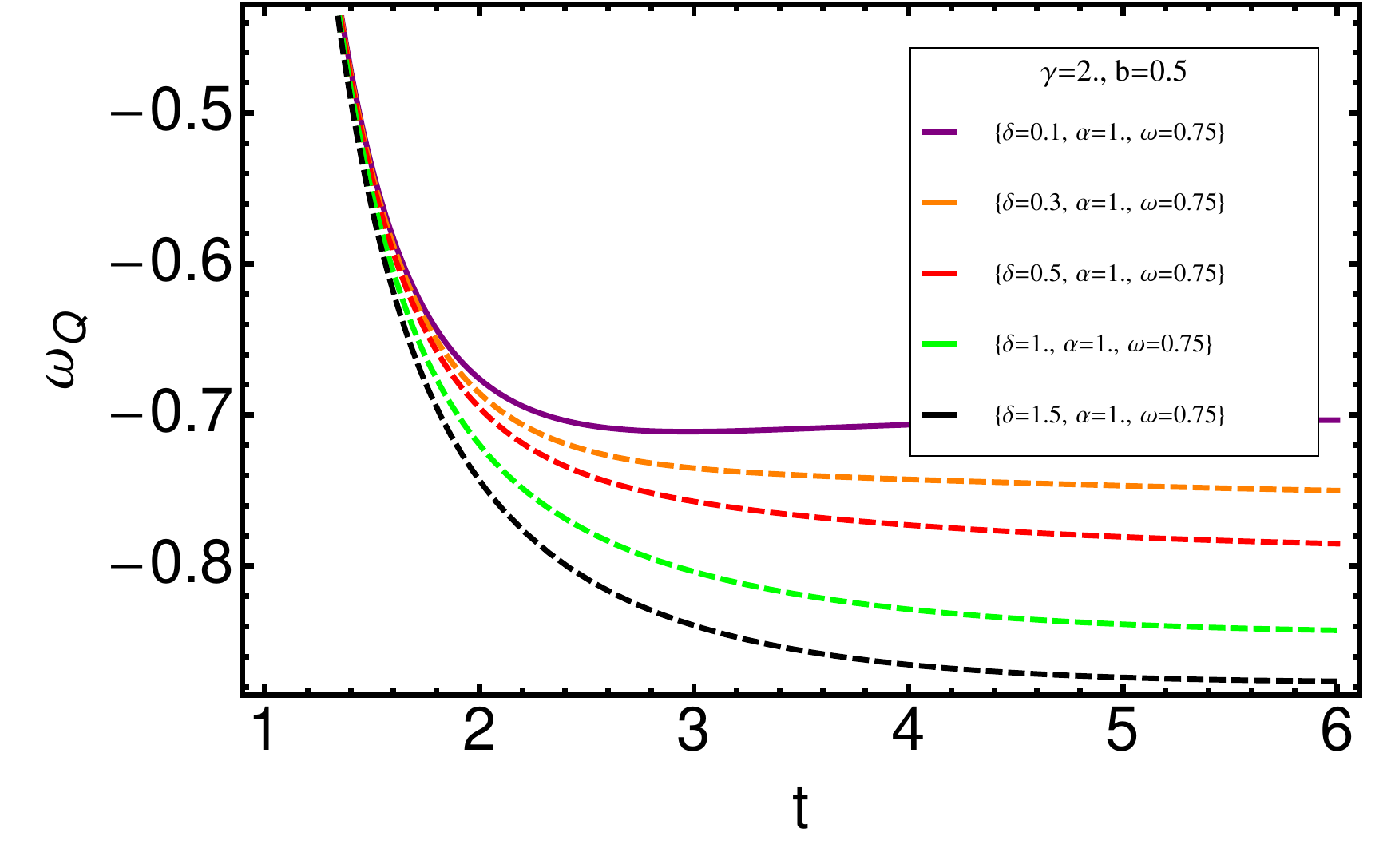} &
\includegraphics[width=50 mm]{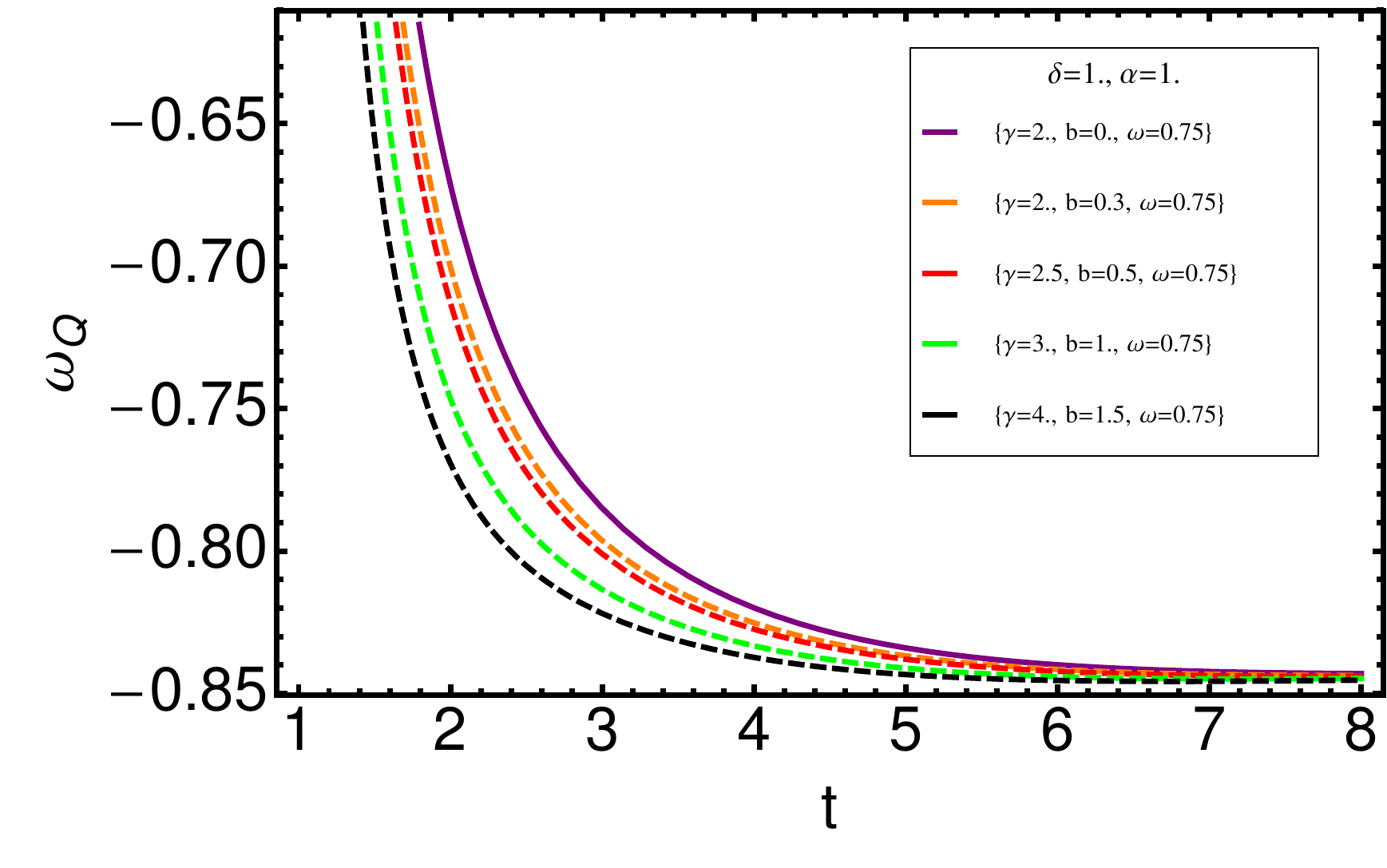}&
\includegraphics[width=50 mm]{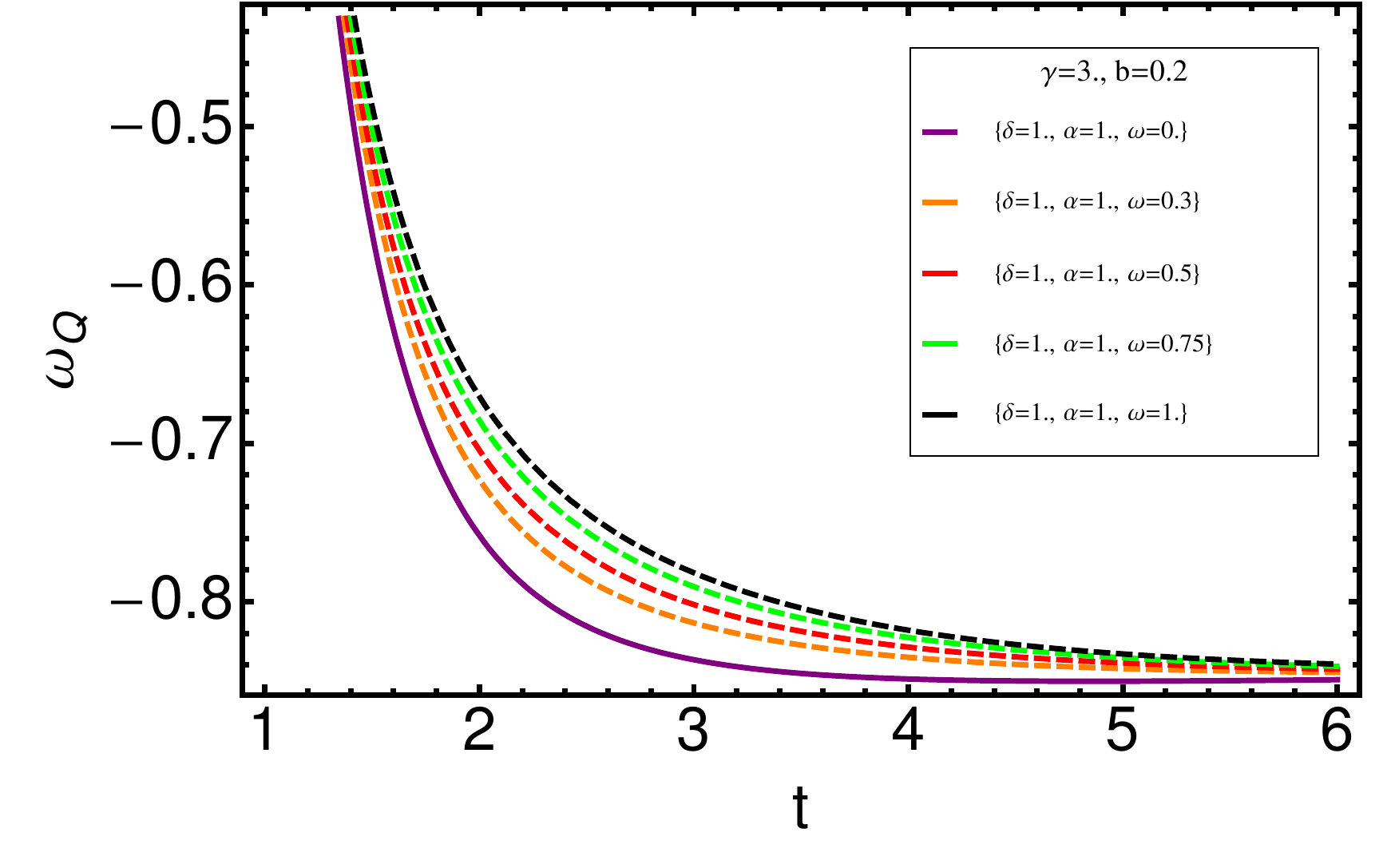}\\
 \end{array}$
 \end{center}
\caption{Behavior of EoS parameter $\omega_{tot}$ and $\omega_{Q}$ against $t$ for Model 6.}
 \label{fig:6}
\end{figure}

Analysis of the cosmological parameters for the varying $\Lambda(t)$ and interaction term $Q=bH^{1-2\gamma}\rho_{b}^{\gamma}\dot{\phi}^{2}$ presented in Fig. \ref{fig:5} and Fig. \ref{fig:6}. We conclude that this model as almost the same characters as all other models, therefore we conclude that this model is also a good model.  We conclude our work in the next section with some thoughts and finalizing obtained result, some observational data information is also given to prove our conclusions.

\section{Discussion}

In this work we have considered six diffrent models of interacting quintessence DE models, which is one of the well studied scalar field model of Dark Energy. We consider a Cosmology with varying $\Lambda(t)$ in Lyra manifold. We take into account modified field equations when $\Lambda(t)$ is considered. Within this background we started to analyse three forms of the interaction terms between DM and DE. We also assume that DM can be modeled as a barotropic fluid with $P_{b}=\omega_{b}\rho_{b}$. One of the forms of the interaction within the form of $\Lambda$ we already considered  in GR with varying $G(t)$ and $\Lambda(t)$. The second form of the interaction can be considered as one of the classical forms of the interaction considered in literature and it is a function from total energy density and its time derivative. While the last interaction $Q$ by its form is also a relatively new form. The construction of the third interaqtion term unit analysis is taken into account. It is a function of $H$, $\rho_{b}$ and $\dot{\phi}^{2}$. According to the $1\sigma$ level from $H(z)$ data $q\approx -0.3$ and $H_{0}=68.43\pm 2.8 \frac{Km}{sMpc}$ \cite{Kumar}. On the other hand from data of $SNe Ia$ we have $q \approx -0.43$ and $H_{0}=69.18 \pm 0.55\frac{Km}{sMpc}$ \cite{Kumar}. Also joint test using $H(z)$ and $SNe Ia$ give $-0.39 \leq q \leq -0.29$ and $H_{0}=68.93 \pm 0.53\frac{Km}{sMpc}$ \cite{Kumar}. Recent astronomical data based on anew infrared camera on the $HST$ gives $H_{0}=73.8 \pm 2.4\frac{Km}{sMpc}$ \cite{Riess1}. The other prob using galactic clusters data suggest $H_{0} = 67 \pm 3.2\frac{Km}{sMpc}$ \cite{Beutler}. Finally, $\Lambda CDM$ model suggests $q \rightarrow -1$ and the best fitted parameters of the Ref. \cite{Visser} say that $q \approx -0.64$. Conclusion of the presented facts is that that  generally $q \geq-1$.
Performing analysis of our models we see clearly, that $q \geq-1$ condition is can be satisfied, moreover we see that the carefull choose of values of the model parametres observational facts can be recovered.

\end{document}